# Developing IncidentUI$_{droid}$

## A Ride Comfort and Disengagement Evaluation Application for Autonomous Vehicles

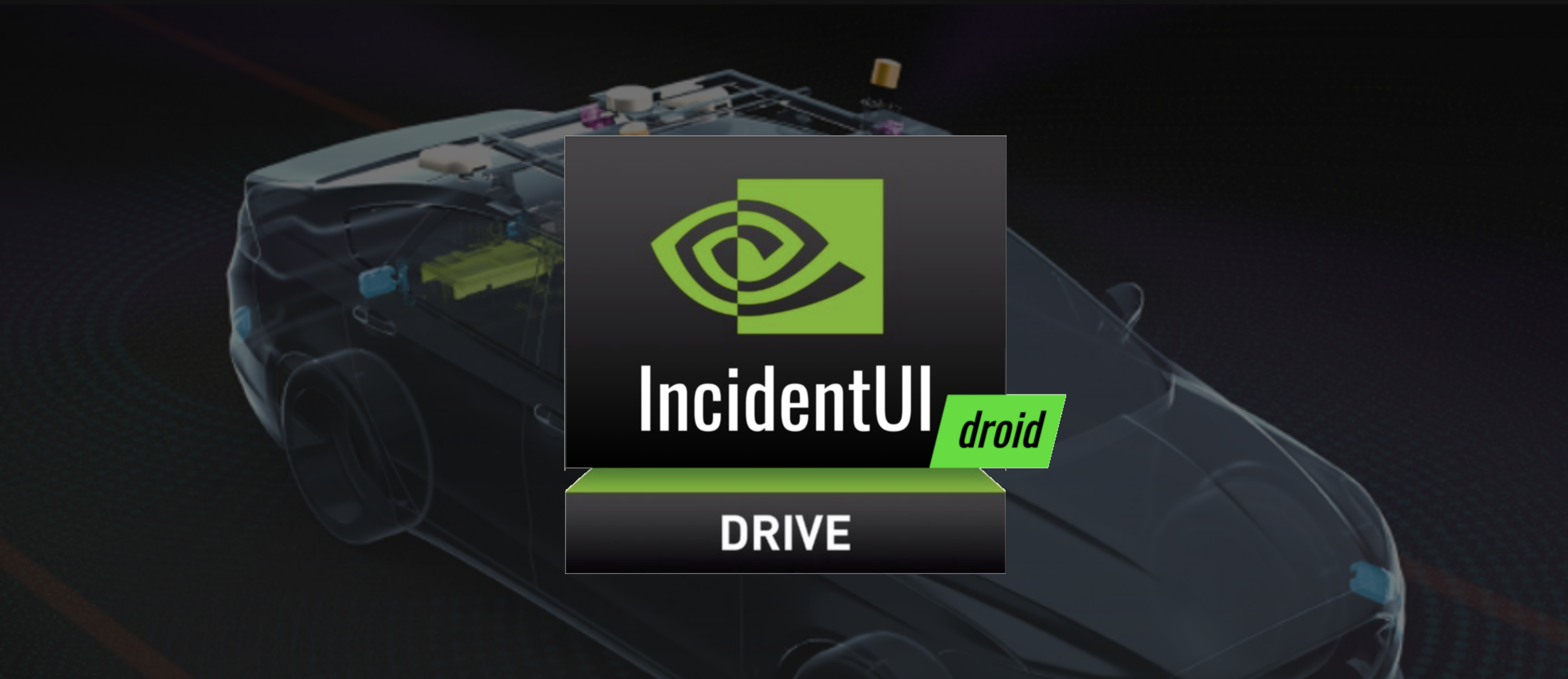

**Manas Mehta, Nugzar Chkhaidze, Yizhen Wang**

Worcester Polytechnic Institute

Nvidia Corporation

**Major Qualifying Project, C'20**

**Silicon Valley Project Center**

WPI

NVIDIA

Project Number: ND20

# Developing IncidentUI$_{droid}$: A Ride Comfort and Disengagement Evaluation Application for Autonomous Vehicles

A Major Qualifying Project Report:

Submitted in partial fulfillment of the requirements for the

Degree of Bachelor of Science

Submitted by:

**Manas Mehta, Nugzar Chkhaidze, Yizhen Wang**

Submitted to:

**Project Advisor: Mark L. Claypool**

**Nvidia Sponsor Liaison: Raymond Poudrier**

# Abstract

This report details the design, development, and implementation of $\text{IncidentUI}_{droid}$, an Android tablet application designed to measure user-experienced ride comfort and record disengagement data for autonomous vehicles (AV) during test drives. The goal of our project was to develop an Android application to run on a peripheral tablet and communicate with the Drive Pegasus AGX, the AI Computing Platform for Nvidia's AV Level 2 Autonomy Solution Architecture [1], to detect AV disengagements and report ride comfort. We designed and developed an Android XML-based intuitive user interface for $\text{IncidentUI}_{droid}$. The development of $\text{IncidentUI}_{droid}$ required a redesign of the system architecture by redeveloping the system communications protocol in Java and implementing the Protocol Buffers (Protobufs) in Java using the existing system Protobuf definitions. The final iteration of $\text{IncidentUI}_{droid}$ yielded the desired functionality while testing on an AV test drive. We also received positive feedback from Nvidia's AV Platform Team during our final $\text{IncidentUI}_{droid}$ demonstration.

# Acknowledgments

We would like to thank everyone who assisted in the development of our project. Without their help, our project would not have been possible.

First, we would like to thank our project sponsor, Mr. Raymond Poudrier from Nvidia, along with the rest of the Nvidia Drive Team for providing us with the opportunity to work for Nvidia in developing IncidentUI Android. We greatly appreciate their continuous support and involvement throughout the span of this project. We would like to specifically thank Mr. David Milewicz for his constant guidance during the development of IncidentUI Android.

Next, we would like to thank Professor Mark L. Claypool, our on-site advisor, for his continued guidance and support, throughout the length of this project. We really appreciate his guidance not only in the technical aspect of the project but also in team-building and in managing the sponsor-student relationship. Professor Claypool's advice and support during the preparatory term was critical to the success of the project in the following term and really enabled us to be better equipped for this project and Nvidia.

We would also like to thank our project sponsor, Nvidia for providing us with comfortable housing at Ironworks Apartments in Sunnyvale, California throughout the span of this project.

We would additionally like to mention and thank all the individuals and teams at Nvidia, particularly Mr. Bhanu Murthy along with the IFORV (End-to-End Product Testing) and the SQA (Quality Control) teams, who participated in our project demonstrations during multiple phases of our project and provided us with critical feedback that aided us greatly in the development of the project.

Finally, we would like to thank Worcester Polytechnic Institute for establishing and maintaining the Silicon Valley Project Center and providing us with this amazing opportunity to experience working in a team as part of a technology giant like Nvidia.

# Table of Contents

# 1. Introduction

Inspired by the dramatically expanding applications of artificial intelligence in the transport industry, the autonomous vehicle industry has transitioned into its apex development period. The global autonomous vehicle market revenue is expected to grow by 39.47% between 2019 and 2026 [2]. Many transport and technology companies have realized the potential of and invested heavily in the autonomous vehicle industry, including Nvidia, especially their Nvidia Drive technology. Nvidia Drive is a developer platform developed by Nvidia which has allowed them to make significant strides in the autonomous vehicle industry.

Nvidia participates in the autonomous vehicle industry with the Nvidia Hyperion Developer Kit, which is the autonomy solution for autonomous vehicle developers and the first commercially available Level 2+ autonomy developer kit. Nvidia Drive Hyperion itself is a reference architecture for Nvidia's second level autonomy solution [1]. The kit provides a complete sensor suite and (Artificial Intelligence) AI computing platform, along with a full software stack for self-driving development. Among the development kit, on the hardware side, is the core computing unit, Drive Pegasus AGX, the self-driving platform built on Nvidia Xavier, dedicatedly designed for autonomous driving; while on the software side, is the Nvidia DriveOS, the foundational software stack modules essential to for self-driving platform development and hardware drive including Real-Time OS, Nvidia CUDA libraries and more [3]. Based on all of those provided autonomous driving developer tools, Roadrunner is the actual Driving Application developed by Nvidia using the DriveOS SDK [4]. The current Roadrunner provides functionality to use sample sensor data for simulation driving on real roads for testing purposes.

The Hyperion Kit utilizes IncidentUI, an application developed to record disengagement and ride comfort data to improve the Drive Software. During the span of our project, we worked on developing $\text{IncidentUI}_{\text{droid}}$, as represented in Figure 1. $\text{IncidentUI}_{\text{droid}}$ is a tablet-based Android application that enables the co-pilot to evaluate and log ride comfort during a test vehicle run. The existing IncidentUI application, that runs on a touch screen attached to the Raspberry Pi in the center console of the Ego Vehicle, communicates with the Roadrunner deployed on the test vehicle's Pegasus AGX. Since there are several drawbacks of using Raspberry Pi and a front-end deployed in Qt for the IncidentUI, we aimed to redevelop the whole IncidentUI in Android, taking advantage of the intuitive user interface, versatile libraries and the wider accessibility of the Android platform.

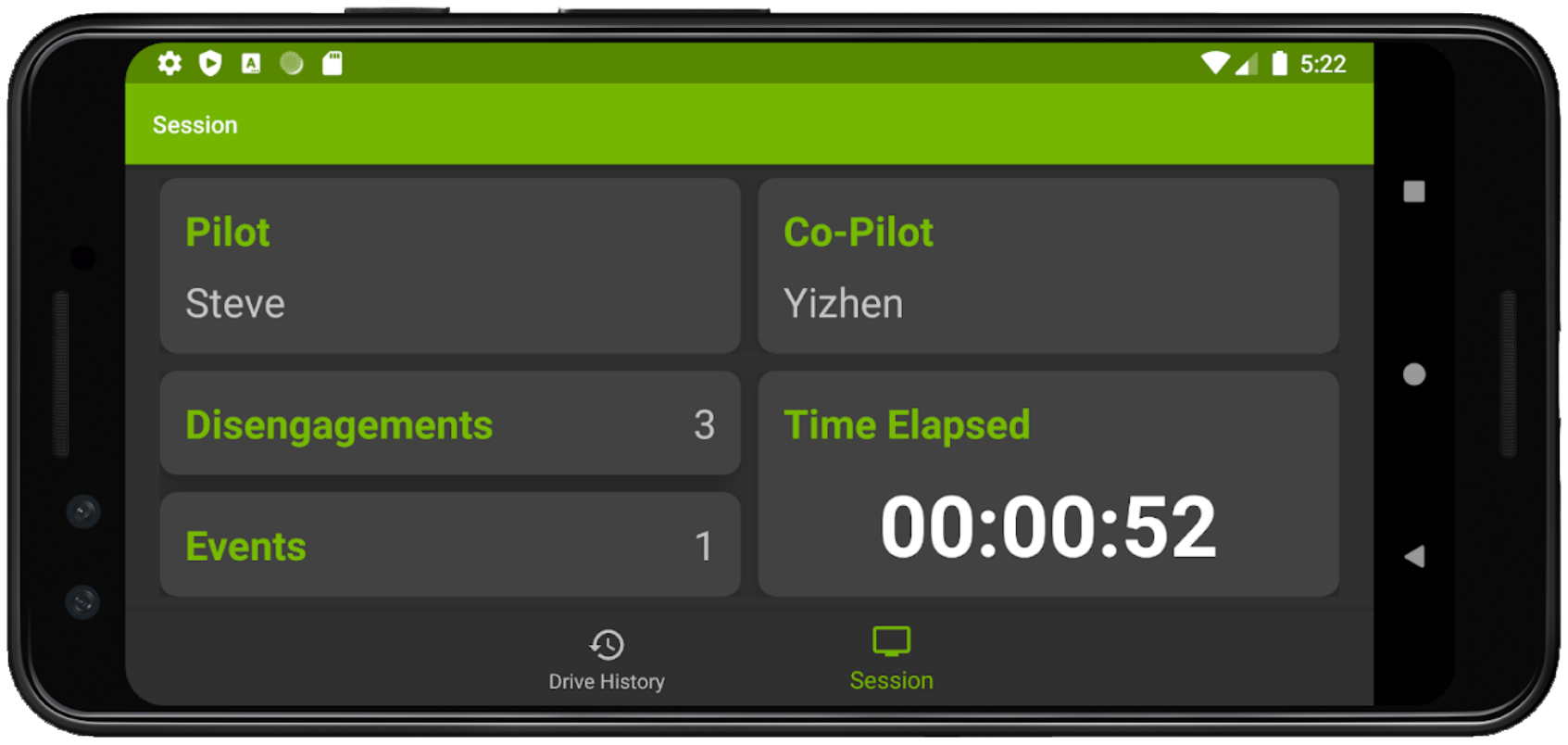

*Figure 1: Representation of IncidentUI$_{droid}$*

In our 7-weeks long project with the AV platform team at Nvidia, we redeveloped and transplanted the IncidentUI onto the Android platform. We developed the new system architecture and also redesigned the user interface and data flow. The user interface and the app features have been tailored to suit the needs of the co-pilot, to ensure that they can log ride comfort accurately and comprehensively, without being bogged down by an unintuitive UI, impractical features and software bugs. In our project, we worked with various technologies and tools such as Protocol Buffers, multiple network protocol connections [3], to ensure cohesive and interoperable sync between the tablet application and the Roadrunner which serves as a gateway to the test vehicle's infrastructure. We also tested and evaluated the functionality of IncidentUI$_{droid}$ to ensure its comprehensive ability to interact with the co-pilot and fulfill the data collection requirement for each driving session.

Our project successfully improved the performance and the interactivity of the IncidentUI. Not only did we transplant the application onto a more versatile Android platform with higher compatibility for future development, but we also reshaped the user experience with a much more intuitive user interface and working pattern. Additionally, we deployed a more robust system architecture design that utilizes a Java interface comprising Java implementations of Roadcast and Protobufs for communicating with Roadrunner. This improvement will help Nvidia to tweak and improve its self-driving technology using a more comfortable testing interface.

The remaining chapters of our project report are organized as follows: Chapter 2 elucidates the background research and defines some technical terms; Chapter 3 discusses the technical details of the methodology developed for the project; Chapter 4 delivers the final implementation of IncidentUI$_{droid}$; Chapter 5 summarizes the conclusion for the report; Chapter 6 recommends some ideas for future development; Chapter 7 discusses our experiences throughout the span of this project, and Chapter 8 concludes the report with a list of References.

# 2. Background Research

In preparation for this project, we familiarize ourselves with various concepts, tools, and technologies to better equip ourselves to develop IncidentUI$_{droid}$ and integrate it with the existing system architecture. This section elucidates these concepts and technologies in detail.

## 2.1 Nvidia Drive™

Nvidia Drive is the scalable AI platform providing hardware and software solutions for automakers with Nvidia's decades-long experience in AI. The Nvidia Drive Platform provides hardware solutions such as Drive AGX for high-performance and energy-efficient computing power in handling large numbers of applications, and software such as development toolkit, library, and photo-realistic simulation for testing and validating self-driving platforms. The Nvidia Drive development platform consists of the following four development platforms [5]:

**Drive Pegasus AGX** - an in-vehicle AI computing platform that enhances the performance of localization, mapping, and perception algorithms along with accelerated development of other AV development tools like Drive AV and Drive IX.

**Drive Hyperion** - Nvidia Drive Hyperion is a reference architecture for Nvidia's level 2+ autonomy solution consisting of a complete sensor suite and the AI computing platform Drive AGX, along with the full software stack under the umbrella of Drive Software [1].

**Drive Constellation** - Nvidia Drive Constellation, consisting of a Simulation Server and the Drive AGX AI computing platform, is the simulation platform used to evaluate AV algorithms in conjunction with the hardware using simulated real-world conditions like sensors, traffic and more [5]. In its development iteration, the simulation platform in the Nvidia Drive System is called Roadrunner. Roadrunner could virtually run on a desktop host or on Pegasus AGX with sampled car driving data to emulate a car driving session for testing purposes in the development iteration.

**Nvidia DGX** - DGX is a Deep Neural Net (DNN) training platform that comprises the Drive AGX AI computing platform and Nvidia's Deep Learning SDK, and is used to train neural networks to enhance AV perception.

## 2.2 Nvidia Drive Hyperion™

Nvidia Drive Hyperion is a reference architecture for Nvidia's level 2+ autonomy solution. The Drive Hyperion Developer Kit, as represented in Figure 2, consists of a complete sensor suite, a plethora of peripherals and the main AI computing platform: Drive Pegasus AGX, along with a full software stack under the umbrella of the Drive Software for autonomous driving, driver monitoring, and visualization [1].

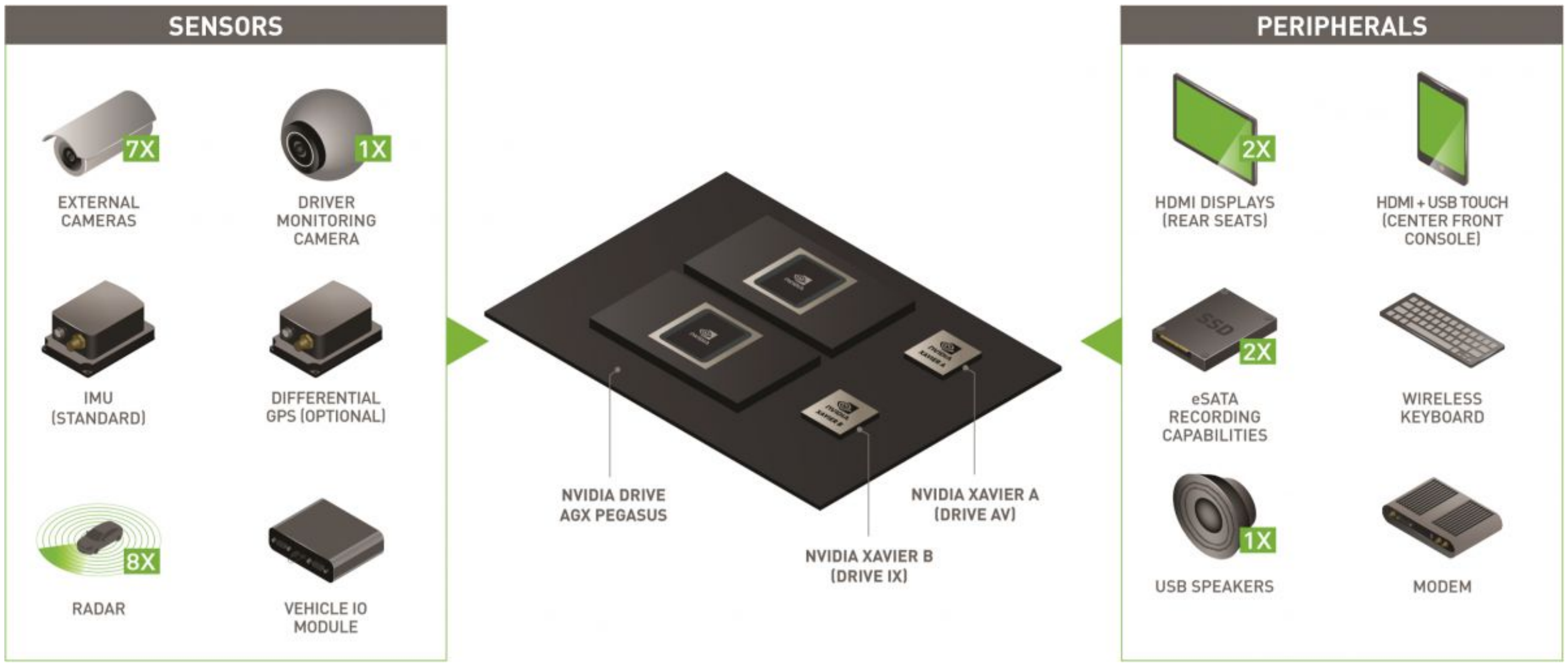

*Figure 2: Drive Hyperion Developer Kit*

The Hyperion kit enables developers to use the available hardware and the packaged software to develop and evaluate AV technology using simulated or on-road testing with test vehicle integration. The sensor suite, including cameras and radars, enables comprehensive data acquisition which is recorded and logged in the storage peripherals. The on-board displays visualize the sensor data and how it is being processed by Drive AV, while the modem allows the drive Software to be updated using the Drive OTA (over-the-air) update infrastructure.

## 2.3 IncidentUI

IncidentUI is an application designed to measure user-experienced ride comfort and record disengagement data for autonomous vehicles (AV) during test drives and is included with the Hyperion kit. The disengagement data and the user-generated discomfort event data is analyzed and used to evaluate the ride quality and the performance of the Drive AV Software during a test drive. This data is used to further enhance the performance of Drive AV's perception, mapping, and planning capabilities. Every time there is an AV Disengagement or an unpredictable AV maneuver resulting in the Co-pilot triggering an Event, IncidentUI displays a survey to allow the user to mention and describe the reason for the disengagement or event so it can be logged and analyzed to improve the performance of the autonomous driving platform. The comfort level during a disengagement or event is measured along two axes: Longitudinal (back and forth) and Lateral (sideways).

The current IncidentUI, as represented in Figure 3, is running on the Raspberry Pi which is a part of the Hyperion kit. The back end of the application is developed in C++ while the front end is designed and implemented using Qt and is displayed on a touch screen peripheral display attached to the Raspberry Pi residing in the front-center console of the Ego vehicle. Users can use the touch screen and the on-screen keyboard to record event and disengagement survey data. The display can also be used to fill out the login survey data (Pilot and Co-pilot information) that is requested in the Driver tab for each Roadrunner session (the duration between Roadrunner connecting to IncidentUI and recording data, and Roadrunner disconnecting from IncidentUI).

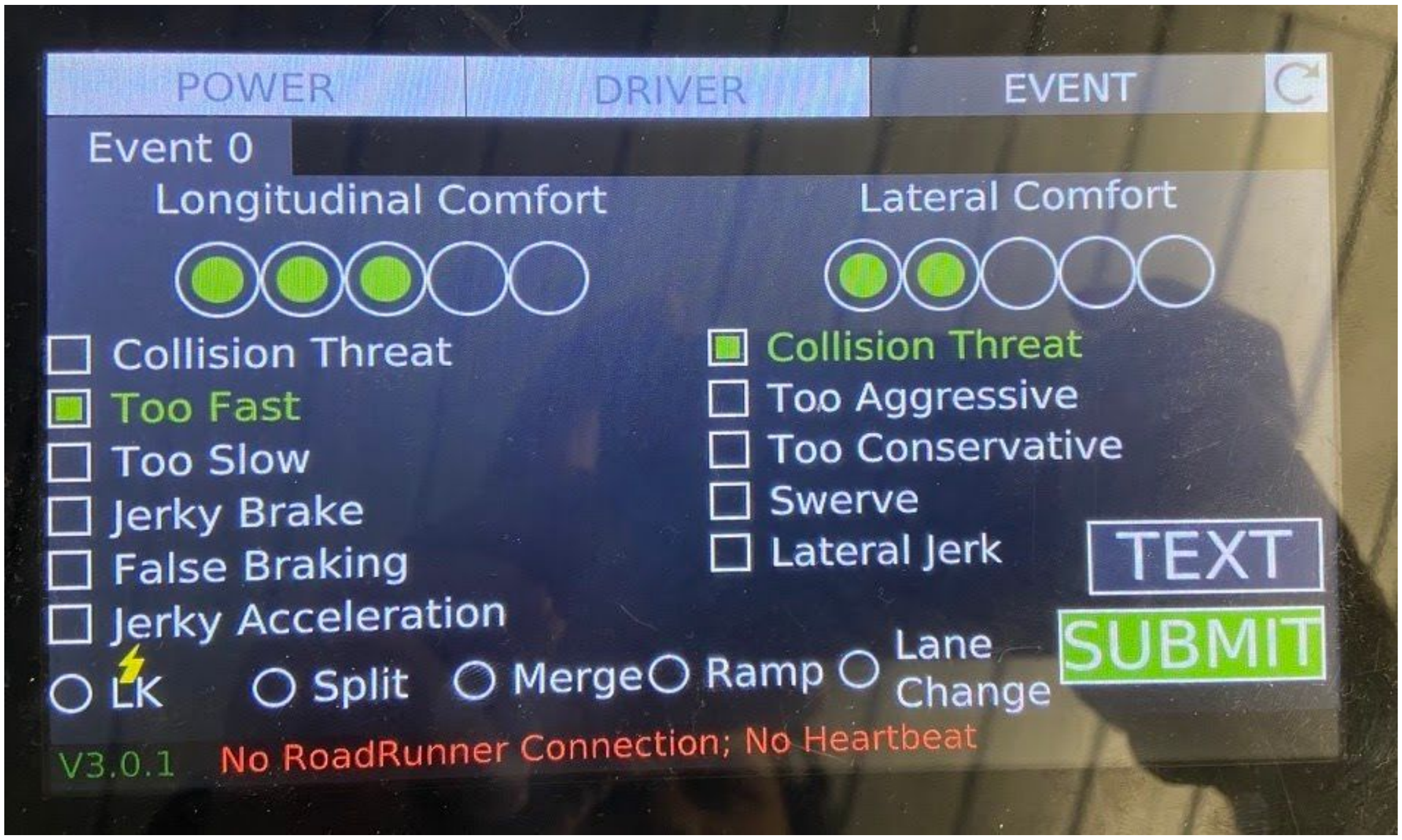

*Figure 3: IncidentUI Representation*

The current Incident UI running on the Raspberry Pi communicates with Roadrunner running on the Pegasus AGX through a proprietary communications protocol, Roadcast, which is developed in C++ and uses Protocol Buffers (protobufs) to serialize data that is transferred amongst the system architecture components. These protobufs are implemented in C++.

The Raspberry Pi is also connected to the ego vehicle speedometer and the Multiplexed Vehicle Electrical Center (mVEC), which dictates the flow of power to the system components using a Controller Area Network (CAN) bus interface. The mVEC can be controlled using the Power tab on IncidentUI. All the connections in the car are established through LAN using an ethernet connection and are part of a local static network set up in every test AV.

## 2.4 Current System Architecture: Test Bench

The current system architecture in the development environment includes the main AI Computing Platform: Pegasus AGX connected to the Raspberry Pi running IncidentUI via ethernet in a local static network. Roadrunner, the Autonomous Driving Application, runs on the Pegasus AGX and communicates with IncidentUI using a proprietary communications protocol developed in C++, Roadcast, which utilizes Protocol Buffers (protobufs) implemented in C++. The protobufs provide a unified format for serializing data in transmission and storage.

The Bench is an abbreviated term for the autonomous driving testing system used for testing various AV hardware and software components in the development iteration. The Bench, which mirrors the latest Hyperion architecture, consists of a Pegasus AGX driving platform as the core processing component and a Linux desktop as the host to update and flash the RoadRunner Docker image on the Pegasus AGX. Pegasus AGX is connected via ethernet to the Raspberry Pi running IncidentUI on it. The touch screen attached to the Raspberry Pi acts as the hardware interface for developers to interact with IncidentUI and communicate with Roadrunner during simulations and record sample disengagements and events during test sessions. Additionally, there is a dedicated set of peripherals like a screen, mouse, and keyboard that is used to interact with Xavier A, which is the component of Pegasus AGX running Roadrunner.

The Bench also has other architecture components like the mVEC that controls the power to all the system components and is controlled by the Power Tab on the IncidentUI running on the Raspberry Pi. In addition to all the sensors and peripherals, the Bench also has an external green button (a part of the dashboard in a test AV), which is used by the Co-pilot to trigger events and display the event survey on the IncidentUI touch screen.

The aforementioned architecture was used for the analysis of the Hyperion architecture and the current IncidentUI sub-architecture, and also for the development and evaluation of $\text{IncidentUI}_{\text{droid}}$. In the following subsections, we will mention in detail some of the major components of the test bench architecture that were used for the development of $\text{IncidentUI}_{\text{droid}}$.

### 2.4.1 Drive Pegasus AGX

Nvidia Drive Pegasus AGX (or just Pegasus) is the primary AI computing platform in the Hyperion Developer Kit. The design of the Pegasus is based on the Nvidia Xavier architecture and involves two Nvidia Xaviers: Xavier A which runs Roadrunner and deploys Drive AV, and Xavier B which deploys Drive IX to monitor in-car Pilot activity using an AI assistant. The Pegasus, as shown in Figure 4, is the core processing component in this autonomous driving infrastructure and also in the IncidentUI sub-architecture.

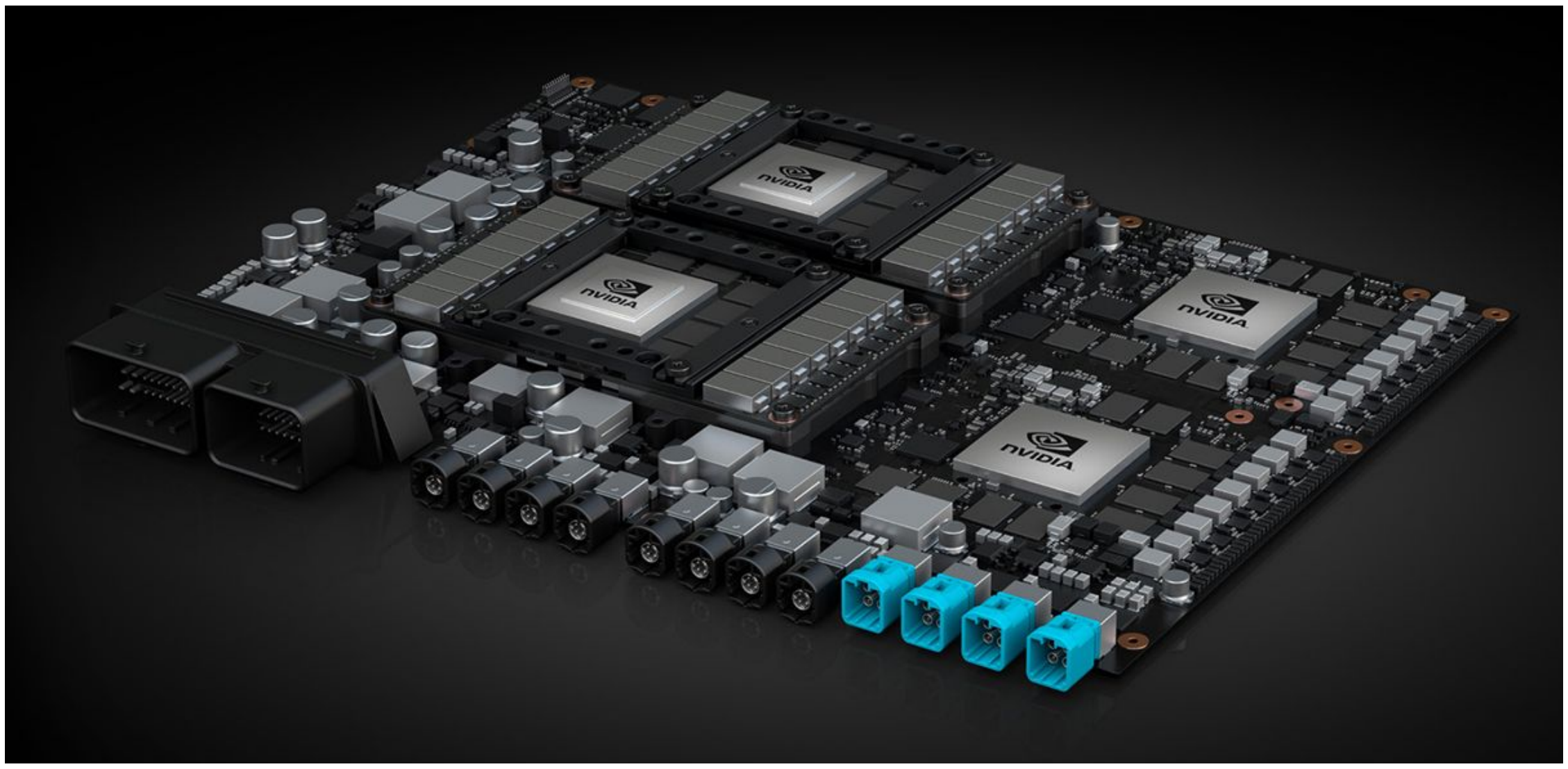

*Figure 4: Nvidia Drive Pegasus AGX*

The Pegasus is connected to the entire suite of sensors for comprehensive data acquisition which is recorded and logged in the storage peripherals and is used to ensure the robust functioning of each architecture component along with the enhancement of the Drive Software. The Pegasus also provides the hardware to run the autonomous driving simulation that is for developing and testing various hardware and software components. In the IncidentUI sub-architecture, the Pegasus runs Roadrunner and communicates with the IncidentUI using Roadcast. For the IncidentUI$_{droid}$ development iterations, the Pegasus was used as an emulator for sample driving session simulation, from which sampled driving data (survey requests and other signals) was used to test the functionality of IncidentUI$_{droid}$. The goal of the iterative testing was to establish and maintain robust two-way communication between Roadrunner on the Pegasus and IncidentUI$_{droid}$ on the Android tablet using Roadcast that utilized Protobufs for serialization.

### 2.4.2 Roadrunner

RoadRunner is the autonomous vehicle application that runs on the Drive Pegasus AGX and is developed using the DriveOS SDK. It is a frame for customized self-driving application development and testing. The current Roadrunner is able to collect data to guide the self-driving functionality with a suite of sensors connected to the Pegasus AGX and is also able to use sample sensor data to run self-driving simulations. Roadrunner itself is the name for the whole aggregated Nvidia autonomous driving application that deploys the self-driving algorithms, data processing, power manipulation, and performance analysis.

In the current system architecture, IncidentUI utilizes only a portion of the whole Roadrunner application. IncidentUI is only responsible for sending user-experienced disengagement and event survey data to Roadrunner for further analysis and Drive Software enhancement. In the Hyperion Developer Kit, RoadRunner sends a disengagement survey request to the IncidentUI via Roadcast when it detects a vehicle disengagement (pilot taking control of the AV) in the driving session. After the survey is filled, survey data is transmitted back to Roadrunner for the logs. Roadrunner uses a server to serialize and send login and disengagement survey requests to IncidentUI through Roadcast and runs a client that constantly listens for and processes the incoming event, login or disengagement survey data through Roadcast. Connections between Roadrunner and IncidentUI have to be established before each driving session by an exchange of heartbeats, which are bridged by Roadcast.

### 2.4.3 Raspberry Pi

A Raspberry Pi is a single-board computer developed by the Raspberry Pi Foundation, to educate people about computing [6]. Raspberry Pi, as shown in Figure 5, runs Linux and its primary operating system, Raspbian, is open source. The board itself has GPIO (General Purpose Input/Output) pins, sd card support, ethernet and USB-A ports, etc.

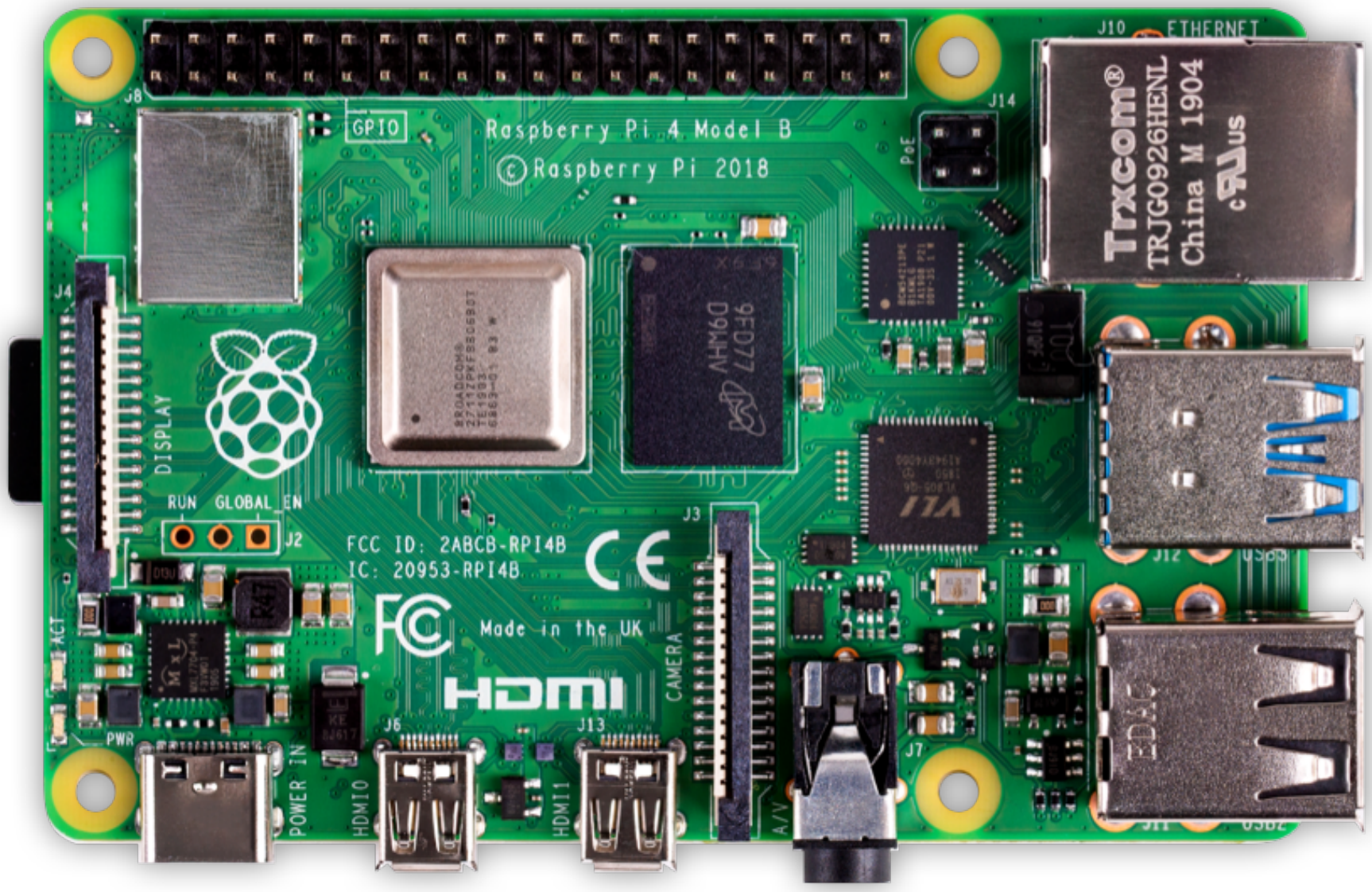

*Figure 5: Raspberry Pi 4 Model B*

Raspberry Pi is primarily used for the Internet of Things (IoT) and allows developers to control electronic components via physical computing [6]. The operating system can be installed in the Pi by writing the compressed file containing all the operating system files into an sd card and then booting up the Pi with the sd card installed. The Pi can be used in many diverse ways; it can be used to transmit files to devices connected to wirelessly (WiFi, Bluetooth) and using cables (USB or ethernet). The Pi can run various scripts and applications, and all it requires is a script (Python, C, or C++) that runs those applications on boot up, or a User Interface to interact with the files stored on the Pi.

The Raspberry Pi in the Hyperion architecture is connected to the Pegasus via Ethernet and has two major functions:

1) The Pi controls the power to the mVEC which manages the power to the architecture components through the CAN interface. The power relay control can be accessed through the Power Tab of the IncidentUI running on the Pi and displayed on the attached touch-screen.
2) IncidentUI, which is developed in C++ on its back end, runs on the Pi. The front-end of IncidentUI, implemented using Qt, is displayed on the touch-screen attached to the Pi. The Pi maintains communication between Roadrunner running on Pegasus and IncidentUI through Roadcast. The Pi packages survey data and heartbeats using the protobuf implementations, and sends them using roadcast to the Pegasus; similarly, it processes the survey requests and heartbeats received from Roadrunner via Roadcast and brings up corresponding surveys and on-screen messages.

#### 2.4.3.1 Qt

Qt is the open-source widget toolkit for creating graphical user interfaces on cross-platform applications in embedded systems [16]. In the existing IncidentUI, Qt is used to develop the GUI on the Raspberry Pi to display the user interface for IncidentUI on an attached touch screen. Qt on this version of IncidentUI is coded in C++ which is natively compatible with C++ code used for IncidentUI Raspberry Pi development.

### 2.4.4 Roadcast

Roadcast is the official DriveWorks module that is a communication protocol developed based on Protobufs for asynchronous data streaming [7]. The abstract data and transport layer of Roadcast gives users the flexibility to configure the transport layers. The communication between IncidentUI and RoadRunner is achieved by sending serialized data back and forth through the Roadcast pipeline that utilizes protobufs for serialization.

In the current Roadrunner system, Roadcast is used as the communication bridge amongst most system components. The survey data from IncidentUI is transmitted to Roadrunner for storage through Roadcast when surveys are filled out and submitted, and IncidentUI receives survey requests from Roadrunner through the same Roadcast pipeline. Either end has a server and client running. The server serializes and sends data to the other end when it is available and the client loops continuously to listen for and process incoming requests and data. All the survey data and vehicle information are already formatted in unified data structures.

## 2.5 Concepts and Technology

This section elucidates the various concepts, tools, and technologies that were used throughout the span of this project to design and develop IncidentUI$_{droid}$.

### 2.5.1 Android Development

Android is an open-source mobile operating system developed by Google and the OHA (Open Handset Alliance) for smartphones [8]. It makes use of mobile applications (apps) to cater to various needs like entertainment, music, communication, productivity, etc. Since Android is an open-source platform, it opens up many avenues for a plethora of software developers to develop their own apps.

The primary language used for mobile app development is Java, with Kotlin gaining more traction recently. Android Studio is the official integrated development environment (IDE) for Android application development and uses a Gradle-based build system, and an emulator [9]. Android Studio allows developers to design their own UI and add their own custom features to develop their own app which is then packaged into an APK (Android Package Kit) file, which can then be downloaded into an Android device directly from Android Studio or from the Google Play Store.

Each screen in an Android app is called an activity that has an XML layout file for designing the UI and a Java (or Kotlin) class that is used to add functionality to the elements defined in the corresponding XML layout file. Each activity can be further broken up into fragments with their corresponding XML layout and Java class to make the app more modular. Developers can make use of numerous UI elements (widgets, buttons, containers, etc.) and Java classes to develop features. Each application can be integrated with a wide range of APIs (Application Programming Interfaces) to make use of various services and features developed by other developers.

IncidentUI$_{droid}$ is entirely designed, developed and implemented in Android using Java. The application was deployed and iteratively tested on an Nvidia Shield tablet running Android version 7.0 with full root access. The version of IncidentUI$_{droid}$ was updated and evaluated over USB and WiFi using the Android Debug Bridge (adb).

### 2.5.2 Android Native Development Kit (NDK)

Android NDK is the Java Native Interface (JNI) toolset that is bundled with Android and enables developers to implement native C and C++ code in an Android Java application. NDK provides great compatibility between C++ and Java code at a native source code level and supports native libraries when implemented. The development of IncidentUI$_{droid}$ using one of the new proposed architecture design required Android NDK support to enable a smooth and robust flow of data between the Java end of the Android application and the C++ end of Roadcast and Protobufs. This mechanic required implementation of a layer of code translation which could be perfectly achieved by NDK: it provides a compatible layer for reforming Java code into C++ code and vice versa. NDK implements interfaces between the Android Java code and the native Android code, which allows the data that is stored in Java classes and structures to be translated to data stored in native classes and structures and vice versa. The native end defines callback functions which can be invoked to obtain data from the Java end and repackage it into native structures which can then be serialized and transmitted to Roadrunner via Roadcast.

### 2.5.3 Protocol Buffers (Protobufs)

Protocol Buffers (Protobuf) are a language and platform-neutral method of serializing structured data. It is a mechanism for storing and interchanging various kinds of organized information. Defined data structures (called messages) are serialized into a binary wire. By encoding messages into byte streams, Protobufs boost the efficiency of exchanging data across applications, based on different languages and platforms [10]. Serialization converts language-specific data structure into a bytestream, while deserialization executes the inverse operation. Both of these transformations are CPU-intensive and might become the bottleneck in data interchange [10]. A protoc executable is used to generate classes and methods from the protobuf definitions which are .proto files containing specific instructions on how data should be structured and serialized. These classes and methods can then be used for serialization and deserialization within the program.

Protobufs are the mechanism that Roadcast is built upon and they ensure a unified data structure formatting across the system for asynchronous data streaming. IncidentUI uses C++ implementations of protobuf definitions (.proto files) to integrate with the native C++ implementations of IncidentUI and Roadcast. Development of IncidentUI$_{droid}$ involved reusing the C++ implementations to integrate with the NDK interface of the Android application and also reimplementation of the same protobuf definitions in Java to develop a Java implementation of Roadcast.

### 2.5.4 Android OS Systems Engineering and Networking

$\text{IncidentUI}_{\text{droid}}$ is developed and deployed using the Android Operating System. The Nvidia Shield Tablet running Android version 7.0 provides an easy to use and compatible environment for $\text{IncidentUI}_{\text{droid}}$'s intuitive user interface. The Tablet was configured to run $\text{IncidentUI}_{\text{droid}}$ as the default application launcher on booting up to turn the Shield into a dedicated $\text{IncidentUI}_{\text{droid}}$ endpoint. Since Roadcast and the Hyperion infrastructure requires communication to be established between components that are assigned static IP addresses, the network configuration for the Android tablet deploying $\text{IncidentUI}_{\text{droid}}$ was revised and the tablet was assigned a static IP address.

A static IP address associated with the tablet running $\text{IncidentUI}_{\text{droid}}$ allowed the application to communicate and exchange data streams with Roadrunner running on the Pegasus. The correct networking configuration is very crucial in a network infrastructure utilizing static IP addresses and USB tethering, ethernet or WiFi to communicate amongst components. Comprehensive Android OS Systems Engineering and Networking is crucial for seamless and robust system integration.

### 2.5.5 Development Platform: Linux

All the development process of IncidentUI, operation of virtual vehicle simulation and testing was implemented on a Linux machine as the host because the Pegasus AGX runs on Linux. Linux is an operating system that utilizes a Linux kernel and is distributed under an open-source license [11]. The host machine that is a component of the Bench is a Linux desktop which is connected to the Pegasus AGX. The Pegasus is also connected to the suite of sensors, peripherals, and the Raspberry Pi. The host is responsible for deploying the Roadrunner application code as a docker image flashed onto the Pegasus AGX which implements the Driving Application. Code updates changing the functionality of Roadrunner are all compiled on the host and then flashed onto the Pegasus AGX.

For the development of $\text{IncidentUI}_{\text{droid}}$, we built a development environment which consists of the Pegasus AGX and a Linux host machine. The Linux Host machine is responsible for deploying Android code updates for the $\text{IncidentUI}_{\text{droid}}$ to the Nvidia Shield tablet. The code for $\text{IncidentUI}_{\text{droid}}$ was also developed on Android Studio that was set up on the Linux host machine and was evaluated on the emulator and the Nvidia Shield tablet over USB and WiFi using the Android Debug Bridge (adb).

## 2.6 Important Terms and Phrases

This section mentions and defines some of the important terms and phrases that the reader might be unfamiliar with that will be used throughout the report. These are mostly terms included in the AV vernacular and terms used to describe scenarios within Nvidia's AV division.

### 2.6.1 Autonomous Vehicle (AV)

Autonomous Vehicles are self-driving vehicles that have some level of autonomy or advanced driver assistance and perception. The Society of Automotive Engineers (SAE) has defined six levels of autonomy from Level 0 (No Driving Automation) to Level 5 (Full Driving Automation) [12]. All the autonomous vehicles mentioned in this report have Level 2 autonomy (Partial Driving Automation) or above, i.e. the vehicle can control both the steering wheel and the acceleration. Level 2 autonomy is not fully autonomous as it requires a human driver to sit in the driver seat and take control of the AV (disengage) if the vehicle performs an undesirable maneuver.

### 2.6.2 Ego Vehicle

An Ego vehicle or object refers to the main actor in an environment, i.e. the environment around the ego object is perceived from its perspective [13]. In the real world, a test AV is an example of an Ego object and all the vehicles and objects around it are considered non-ego objects. In this report, a reference to an Ego vehicle is referring to an AV undergoing a test run.

### 2.6.3 Disengagement

A Disengagement is defined by the Department of Motor Vehicles as "deactivation of the autonomous mode when a failure of the autonomous technology is detected or when the safe operation of the vehicle requires that the autonomous vehicle test driver disengage the autonomous mode and take immediate manual control of the vehicle." [14] The state of California requires the autonomous mode of an AV to include lane keep and cruise control; the moment either of these conditions is violated because of driver interference, a disengagement occurs. The reasons for each disengagement are essential in analyzing and evaluating the self-driving application and are required to be submitted to the California DMV annually in the form of a disengagement report [15].

At Nvidia, every time a disengagement occurs, the IncidentUI users are required to fill out a disengagement survey detailing the reasons for the disengagement, the lateral (sideways) and longitudinal (back and forth) comfort, the position of the ego vehicle and other additional information highlighting any discomfort caused by an unpredictable AV maneuver. Following a disengagement, the AV's actuation is blocked and the AV is not allowed to re-engage into autonomous mode until the disengagement survey displayed on IncidentUI is filled out and submitted.

### 2.6.4 Event

Events refer to scenarios where the test AV performs a maneuver that does not result in a disengagement but still causes enough discomfort to warrant a log of the scenario. Such events are manually invoked by the co-pilot or any IncidentUI users (by clicking a button) during a driving session and involve filling out an event survey that records the time of the event via an Event flag and the cause of the event along with the Lateral and Longitudinal discomfort. This event data is a useful evaluation metric that takes user experience (data that cannot be collected by the AV sensors) into account and thus supports the development of a more robust self-driving application. Events are not attributed to blocked actuators since they do not result in disengagement. Some examples of events include ride discomfort, engine failure, vehicle stalling, and animal crossing. Every scenario (something as minor as ride discomfort or something as major as engine failure) that is not detected as a disengagement but is determined to be significant enough to be logged and needs further analysis is recorded as an event.

# 3. Methodology

We divided our methodology into six phases to ensure a modular, comprehensive, and reliable project design and development. The first couple of phases were based around research, current application and system architecture analysis, User Interface (UI) design, and requirements gathering. The next few phases involved the actual design and development of the application, with development of two betas (including front-end Android application development, system integration, and hardware networking) corresponding to two proposed system architectures, application, and system architecture evaluation, and a final stable release development using the more robust and viable of the two proposed system architectures. The final phase involved IncidentUI$_{droid}$ in-car testing and evaluation during an AV test drive, final project and application demonstration, and feedback analysis for producing future work and feature recommendations. We followed the Agile Software Development Methodology for the development of IncidentUI$_{droid}$. We utilized Jira, as portrayed in Figure 6, to track the progress of every sprint.

Issues in Epic

| | | | |
|---|---|---|---|
| AVPSW-683 | Display & fill in disengagement survey on disengagement | CLOSED | Unassigned |
| AVPSW-684 | Tablet Internet connection to Pegasus | CLOSED | Nugzar Chkhaidze |
| AVPSW-685 | Connect tablet to in-car network | CLOSED | Nugzar Chkhaidze |
| AVPSW-686 | Document the design of the tablet system | CLOSED | Yizhen Wang |
| AVPSW-688 | Define data path for surevy request to survey pop-up | OPEN | Unassigned |
| AVPSW-689 | Receive roadcast messages on tablet | CLOSED | Manas Mehta |
| AVPSW-692 | Run Roadrunner and tablet app on same machine | CLOSED | Yizhen Wang |
| AVPSW-693 | Research testing framework for the UI | OPEN | Unassigned |
| AVPSW-695 | Compile RoadCast for Android | CLOSED | Nugzar Chkhaidze |
| AVPSW-701 | Set up a Functional Beta of the App on Android | CLOSED | Manas Mehta |
| AVPSW-702 | Research, set up and document Android NDK interface to send survey data to roadcast | CLOSED | Manas Mehta |
| AVPSW-703 | Add 'Test Track' button to front screen of disengagement survey | CLOSED | Manas Mehta |
| AVPSW-694 | Document the UI screens & UI flow | CLOSED | Nugzar Chkhaidze |
| AVPSW-715 | Tablet Ethernet connection to Pegasus | CLOSED | Nugzar Chkhaidze |
| AVPSW-709 | Setting Up RoadRunner on the system | CLOSED | Yizhen Wang |
| AVPSW-691 | Initialize roadcast connection between roadrunner and tablet | CLOSED | Nugzar Chkhaidze |
| AVPSW-712 | Demo Beta to SQA and IFORV Teams and get feedback | CLOSED | Manas Mehta |
| AVPSW-690 | Document & design NDK/Roadcast interface | CLOSED | Manas Mehta |
| AVPSW-717 | Set up Disengagement Survey UI, Methods, Model and NDK | CLOSED | Manas Mehta |
| AVPSW-738 | Set up static IP address for Ethernet Connection | CLOSED | Yizhen Wang |
| AVPSW-761 | Build Roadrunner emulator for connection testing | CLOSED | Yizhen Wang |
| AVPSW-771 | Create UI Design Flow & Sequence Diagrams | CLOSED | Nugzar Chkhaidze |
| AVPSW-772 | Reverse Tethering connection to Tablet through USB | CLOSED | Nugzar Chkhaidze |
| AVPSW-773 | Launch IncidentUI on start up on Tablet | CLOSED | Nugzar Chkhaidze |

*Figure 6: Jira Board for IncidentUI$_{droid}$ Development*

We ran 2-week long sprints with Sprint 1 covering Phases 1 and 2, Sprint 2 covering Phase 3, Sprint 3 covering Phase 4, and Sprint 4 covering Phases 5 and 6. For every sprint, we developed user stories, tasks, and sub-tasks to tackle the development of application features. We divided these sub-tasks amongst us and tracked them using Jira. In addition to the Jira Board, we tracked the progress of a sprint by Scrumming daily as a team and participating in AV Platform Team Stand-up Meetings twice a week. This section describes the methods and steps involved in each phase of the project development along with the results of implementing those methods.

## 3.1 Phase 1: Requirements and Feature Analysis

The first phase of the project focused on analyzing the current version of the IncidentUI application for features and requirements gathering for $\text{IncidentUI}_{\text{droid}}$. The IncidentUI was analyzed and features were acquired using a code analysis of the IncidentUI application, a build of IncidentUI set up, deployed and run on a Linux machine using the Qt Application Manager, and resources available online like Confluence pages detailing the IncidentUI architecture, data flow, sequence diagrams, and UI flow.

### 3.1.1 Current IncidentUI Data Flow and Feature Analysis

The first step in Phase 1 involved analyzing the components of the User Interface of IncidentUI and determining the UI flow between these components along with how data flows between these components and between IncidentUI and Roadrunner. To implement this step, we read through the documentation of IncidentUI available on Nvidia's Confluence pages describing the architecture and the sequence and data flow for IncidentUI. We also ran IncidentUI on a Linux machine after setting the machine up with the required packages and configurations. Running IncidentUI on a machine and playing around with the User Interface gave us some insight into the UI components and how the UI flow worked. We also analyzed the code for IncidentUI to understand the flow of logic and data better and how IncidentUI interacted with Roadcast. The existing version of IncidentUI was running on a low-resolution screen attached as a peripheral to a Raspberry Pi that comes packaged with the Hyperion Developer Kit. The Raspberry Pi controls the power relay and also dictates how the IncidentUI communicates with the Pegasus. IncidentUI is developed in C++ with Qt as the UI development tool. IncidentUI, as shown in Figure 3, involves three tabs: Power, Driver, and Survey.

The Power tab manages the Multiplexed Vehicle Electrical Center (mVEC) system power relay. The mVEC controls power to the architecture components using a Controller Area Network (CAN) bus interface.

The Driver tab is displayed every time there is a request for a Login Survey which is requested by Roadrunner when it starts and involves filling out the Pilot and Co-pilot information for the session of Roadrunner. This information is used by the Nvidia Driveworks Recorder to relate session logs to the Pilot and Co-pilot.

The Event tab is used to display an Event Survey when an Event is triggered by the co-pilot by pressing a green button attached to the dashboard of the AV, which sends an Event Flag to Roadrunner recording the actual time of the event in the logs. When the Event Survey, which contains information about the event and its cause, is filled out, it can be sent to Roadrunner by clicking the Send Button on the screen. If multiple events are triggered by the co-pilot, then a tab is created for each event. The co-pilot can switch between these tabs and access or fill out corresponding event surveys.

A Disengagement Survey is displayed when disengagement is detected by Roadrunner and a Disengagement Survey Request sent to IncidentUI. The Disengagement Survey involves information about the disengagement, for instance, the cause of the disengagement like End of Drive or Safety Issue, location of the Ego Vehicle, and other additional information describing the disengagement. Once the disengagement survey has been filled out by the user, it can be sent to Roadrunner by clicking the Send Button on the screen.

When a connection is established and maintained between Roadrunner and IncidentUI, heartbeat messages are sent back and forth. A lack of heartbeat messages is inferred as a lack of a Roadrunner connection and is displayed as an error message on the IncidentUI screen.

### 3.1.2 UI Screen and Flow Mockup Development

Based on the design of IncidentUI, we developed a rough mockup of what the screens would look like for IncidentUI$_{droid}$. We redesigned the data displayed on the screens and how the screens interacted with each other for a more intuitive UI flow. We decided on three main screens (Activities) for the application.

#### Initiate Drive

As shown in Figure 7, we decided to make a dedicated screen for Pilot and Copilot login information instead of a tab to force users to fill out the login survey information before being able to access the disengagement surveys and event trigger button, which will ensure that the Pilot and Co-pilot is filled out for the session logs.

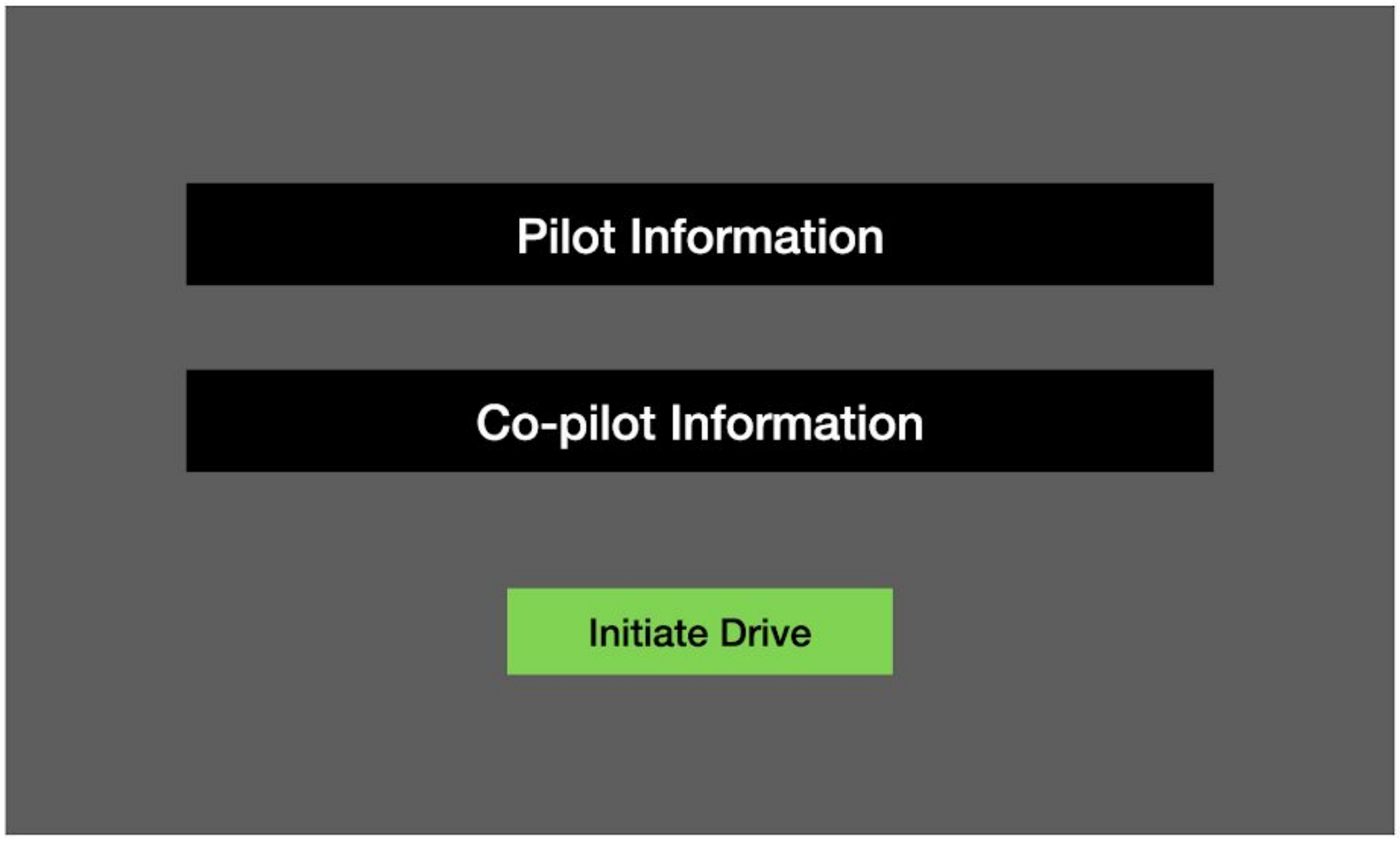

*Figure 7: Initiate Drive Screen Mockup*

After the login information is filled out, the users can click on the Initiate Drive button to start a new drive and load the dashboard screen.

### Dashboard

We designed a dashboard, as shown in Figure 8, that stores information for the current drive. We defined a drive as the duration of app use between user login and logout. The dashboard is displayed as a standby buffer screen between multiple sessions. The design of the application accommodates multiple drive sessions within a single drive owing to the common situation where the same users logged in disconnect and connect roadrunner multiple times for testing purposes and require data collected for each session while they are logged in. The use of a dashboard gives a more concrete boundary between when Roadrunner is connected and a session is underway, and when Roadrunner is disconnected and no data is being collected.

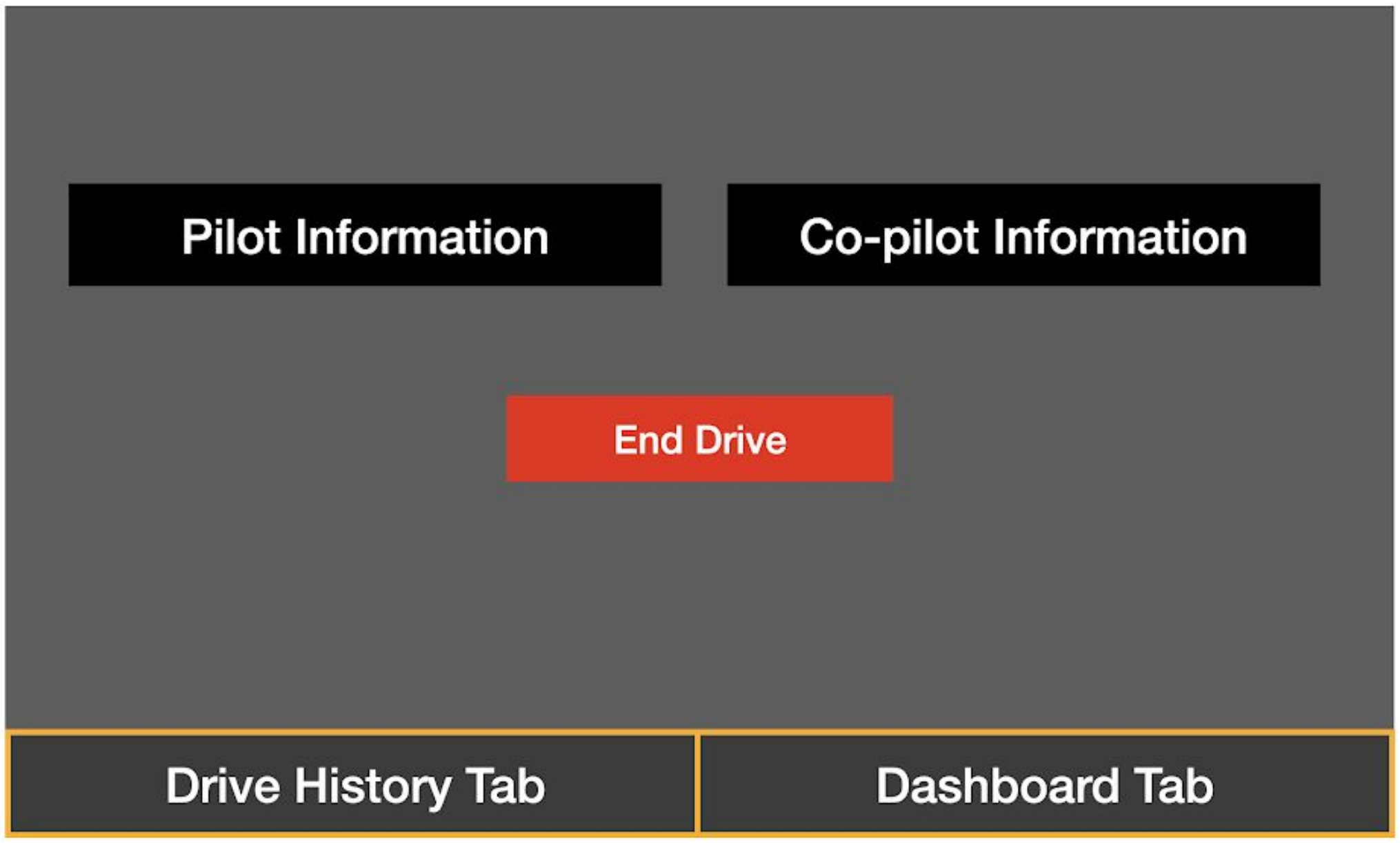

*Figure 8: Dashboard Screen Mockup*

A tab on the dashboard screen allows the user to view information about any past sessions. To end the drive and go back to the Initiate Drive login screen, users can click on the End Drive button.

### Session

The original IncidentUI did not have a primary screen that would be displayed while Roadrunner is connected and a session is active, as a result, we designed a session screen, as shown in Figure 9, to be displayed throughout the duration of a session.

We defined a session as the duration of app use during which Roadrunner is connected to $\text{IncidentUI}_{droid}$ and there is an exchange of heartbeats, i.e. when a connection is initiated with Roadrunner, a session starts and when a Roadrunner disconnects, a session ends. The sessions screen shows information about the current session like session duration, number of events, and number of disengagements. Another goal of the session screen was to display the Disengagement Survey which appears as a pop-up when Disengagement is detected by Roadrunner. We also decided to add a New Event button to trigger Events and display an Event Survey pop-up when it is clicked to eliminate the need for the physical green button in the Ego car dashboard.

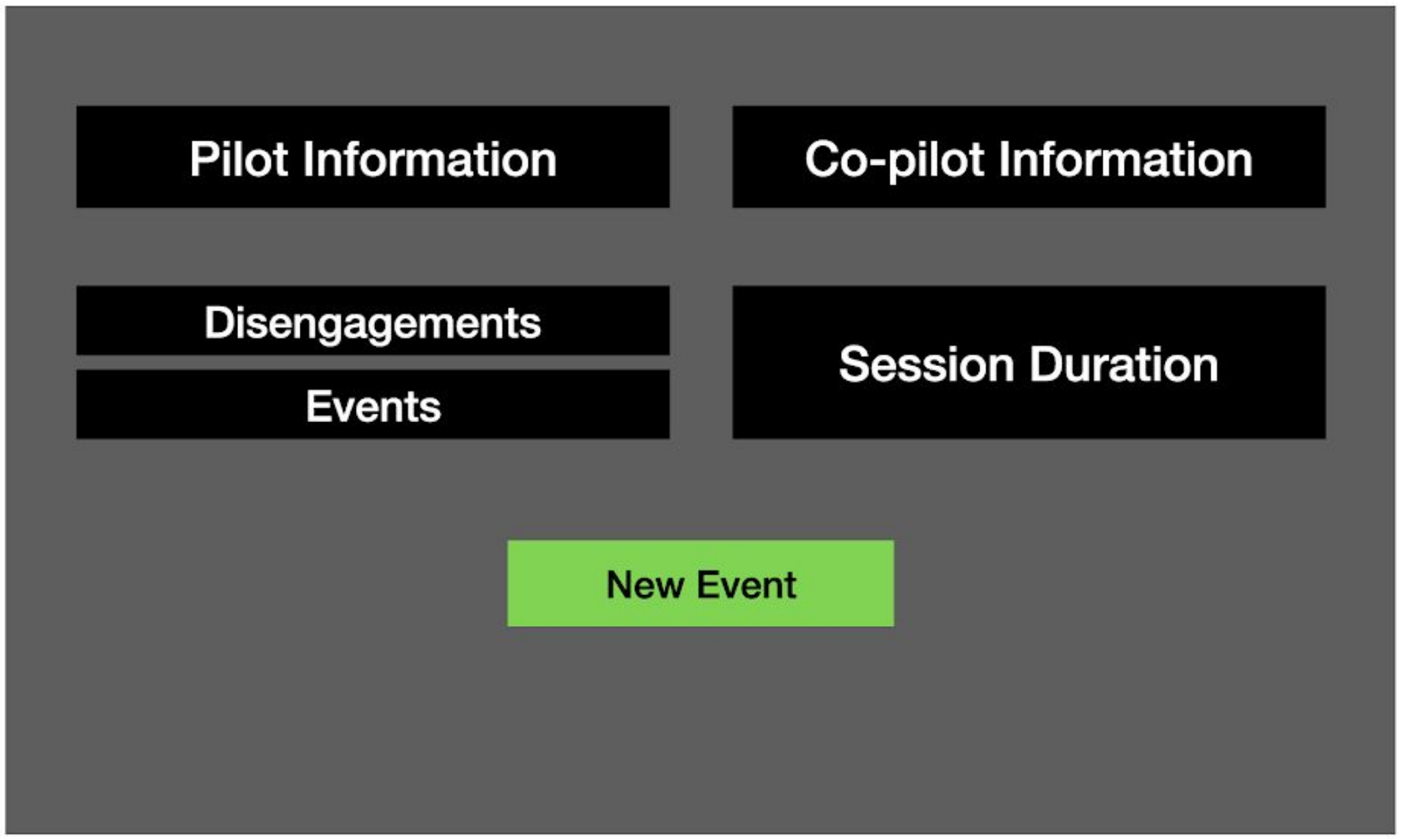

*Figure 9: Session Screen Mockup*

To ensure that the start and end of a session, i.e. Roadrunner connection, entirely dictated the display of the session screen, we did not add any buttons to allow the user to toggle the session screen. The session screen is displayed automatically when a Roadrunner connection is established and the dashboard is automatically loaded when the Roadrunner connection ends.

### 3.1.3 Requirements Inference

After analyzing the IncidentUI application and designing a UI mockup for $\text{IncidentUI}_{\text{droid}}$, we inferred the requirements for the project development and developed a project timeline. We needed to design the front-end of the application using the aforementioned screens along with additional functional application features and reconstruct a more intuitive UI flow between them. We also needed to redesign the system architecture to ensure communication between an Android application running on a peripheral tablet and the Pegasus. We also needed to debate the design and implementation of various aspects of the application, the integration techniques, and system architecture components to ensure robust and viable system integration and data flow. In order to ensure application viability and optimum user experience, we needed to develop application betas and evaluate them. Using the suggestions provided during application demonstrations, we needed to make changes to the application and finally develop a stable release for the application and determine its efficacy by an in-car test.

## 3.2 Phase 2: System Architecture Analysis and Design

During the second phase, with a mock-up of the front end of the application designed, we analyzed and redesigned the system architecture to ensure a robust and viable system integration between the peripheral tablet running $\text{IncidentUI}_{\text{droid}}$ and the Pegasus. This phase involved analyzing the existing system architecture design and the nature of data flow between the components of the system architecture, primarily how IncidentUI running on the Raspberry Pi communicated with Roadrunner running on Pegasus using Protocol Buffers (Protobufs) and a proprietary communications protocol, Roadcast. The system architecture analysis was done by examining the Testing Bench which was built using the components and current system architecture for Hyperion and used to evaluate different components of the Hyperion Kit. Code analysis of IncidentUI, Roadrunner and Roadcast, and research into online resources like Confluence pages detailing the system and IncidentUI architectures, data flow, sequence diagrams, and UI flow gave us a better understanding of the architecture.

With the system architecture and data flow analyzed, we designed two new system architecture plans to establish and maintain a robust communication and flow of data between the Android tablet running $\text{IncidentUI}_{\text{droid}}$ and the Pegasus. These system architecture designs were developed by using varying system integration interfaces which were researched and varied based on which components of the architecture were redesigned, what tools were used to integrate the different architecture components and how data and logic flowed amongst these components. We developed two architecture designs:

Plan A involved the use of the Android Native Development Kit (NDK) to interface between the Java Android Front-end of $\text{IncidentUI}_{\text{droid}}$ and the existing communications protocol, Roadcast written in C++. This interface design allowed the use of existing Roadcast and existing C++ protobuf implementations for communication between Roadrunner and $\text{IncidentUI}_{\text{droid}}$.

Plan B involved a redevelopment of Roadcast in Java on the $\text{IncidentUI}_{\text{droid}}$ side and implementation of existing protobuf definitions in Java, thus eliminating the need for the Android NDK interface, resulting in a more robust and viable design. The Java implementations of Roadcast and protobufs on the $\text{IncidentUI}_{\text{droid}}$ end processed and manipulated data identically to the C++ implementations of Roadcast and protobufs on the Roadrunner end.

This section details the methods implemented for the analysis of the existing system architecture and the design and development of the two aforementioned system architecture designs for $\text{IncidentUI}_{\text{droid}}$, along with the results obtained from the implementation of those methods.

### 3.2.1 Current System Architecture Analysis

The first stage of phase 2 involved the analysis of the existing system architecture. The analysis of the system architecture was done using code analysis of IncidentUI, Roadrunner and Roadcast, and research into online resources like Confluence pages detailing the system and IncidentUI architectures, data flow, sequence diagrams, and UI flow. To gain even deeper insight into the system architecture and how IncidentUI integrated and interacted with other components of the system like the CAN bus, the green button, and primarily the Pegasus running Roadrunner, we analyzed and interacted with the current system architecture set up on the Test Bench next to us. The current system architecture for communication between IncidentUI running on a Raspberry Pi and Roadrunner running on the Pegasus is represented in Figure 10.

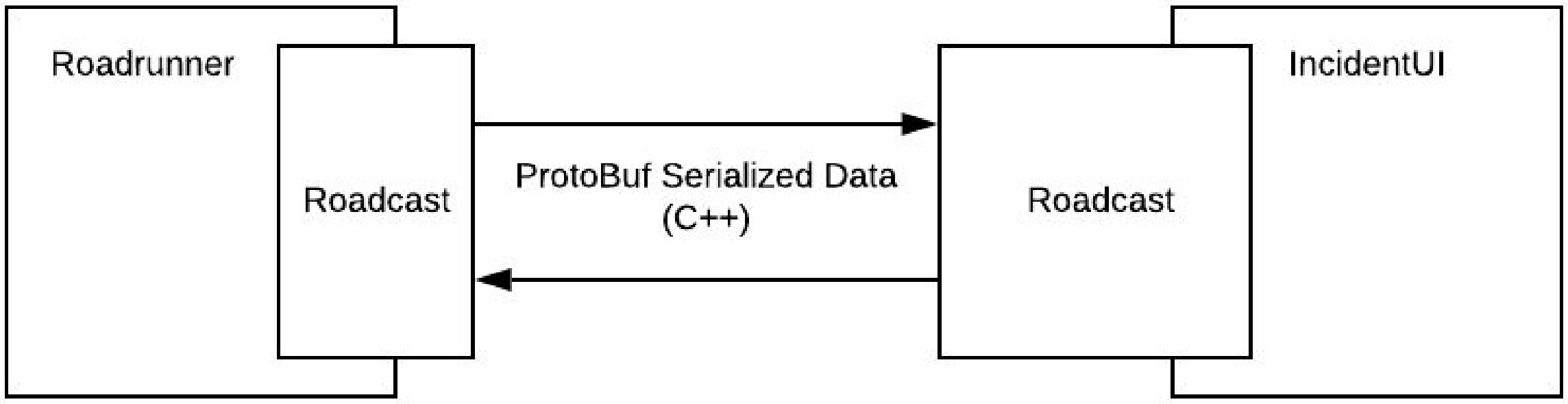

*Figure 10: IncidentUI Hyperion System Architecture: Test Bench*

The current IncidentUI system sub-architecture primarily comprises the Raspberry Pi running IncidentUI and the Pegasus AGX running Roadrunner. The Raspberry Pi and the Pegasus are part of a local static IP network set up inside the Ego vehicle. The components of this network communicate via ethernet. The primary communications protocol used for transferring data between IncidentUI and Roadrunner is called Roadcast. Roadcast is developed in C++ and utilizes protocol buffers (protobufs) for serializing data. As shown in Figure 9, Roadcast runs on either end (Raspberry Pi and Pegasus) of the sub-architecture and regulates the flow of data between the IncidentUI and Roadrunner. Roadcast packages survey requests and heartbeats from Roadrunner and deserialize them on the IncidentUI end; similarly, it packages event and disengagement data along with heartbeats sent by IncidentUI and deserializes them on the Roadrunner end. The protobufs used to serialize data sent via products are implemented in C++ to integrate with the existing C++ written infrastructure containing the Roadrunner, Roadcast, and IncidentUI. The system architecture also consists of a peripheral green button that is a part of the central console of the Ego vehicle and is used to trigger events.

When a connection is established between Roadrunner and IncidentUI, heartbeat messages are sent back and forth. These heartbeats are sent both ways during the entire time the connection between Roadrunner and IncidentUI is maintained. When Roadrunner stops running and recording data, the connection is closed and heartbeats are not detected anymore. This lack of heartbeat is displayed as an error message on the IncidentUI screen and means that Roadrunner has stopped collecting data for that session. When a connection is reestablished with Roadrunner, heartbeats are sent and detected on either end and Roadrunner requests a new Login Survey for this new session followed by the same flow of data and scenarios.

When disengagement is detected, Roadrunner sends a Disengagement Survey Request to IncidentUI which brings up the Disengagement Survey. The time of the disengagement is recorded in the Recorder. When the Disengagement Survey is displayed and being filled out, the actuators are blocked, i.e. The AV cannot reengage and drive autonomously until the filled out disengagement survey is sent to Roadrunner. After the Disengagement Data is sent to the Roadrunner, the AV can re-engage. An Event Survey is displayed when an event is triggered by the co-pilot by pressing the green button. Pressing the green button sends an Event Flag to Roadrunner recording the actual time of the event in the logs. An event is not the result of disengagement, so during an event survey is triggered or displayed, the AV is engaged. After the event survey is filled out it can be submitted to the Roadrunner.

In order to implement an Android tablet running $\text{IncidentUI}_{\text{droid}}$ and integrate it with the system architecture, we had to redesign the architecture. We came up with two plans for the new system architecture design: Plan A revolved around integrating the Java-coded $\text{IncidentUI}_{\text{droid}}$ with the C++-based system using the Native Development Kit and reusing the existing communications and serialization protocols; Plan B involved redesigning the communications and serialization schemes in Java and utilizing the Java-implemented Roadcast and protobufs to integrate $\text{IncidentUI}_{\text{droid}}$ with the Java-C++ hybrid system architecture.

### 3.2.2 System Architecture Design Plan A

Plan A for redesigning the system architecture to integrate $\text{IncidentUI}_{\text{droid}}$ involved utilizing Android's Native Development Kit (NDK), which is the Java Native Interface (JNI) toolset that comes bundled with Android Studio and can be used to design and deploy Android applications in native C++. This plan, as represented in Figure 11, involved using NDK as an interface between the Java-coded $\text{IncidentUI}_{\text{droid}}$ and Roadcast, which is developed using C++. This plan allowed the use of the existing C++-based implementations of Roadcast and Protobufs since the NDK interface allowed the construction of methods that convert data stored in Java classes on the Android end into data stored in C data structures on the native end. These data structures are identical to the ones packaged and transmitted back and forth between the current IncidentUI and Roadrunner via Roadcast, thus allowing for seamless integration with the current system architecture. This form of architecture design allowed the current version of Roadcast to run on the Android Tablet and use the C++ implementations of protobufs to serialize data that is being transmitted between $\text{IncidentUI}_{\text{droid}}$ and Roadrunner. The development of a robust native interface between Roadcast and the Android application required the protobufs and Roadcast libraries to be built in the Android environment as a part of the application apk (Android Package) and run on Android OS.

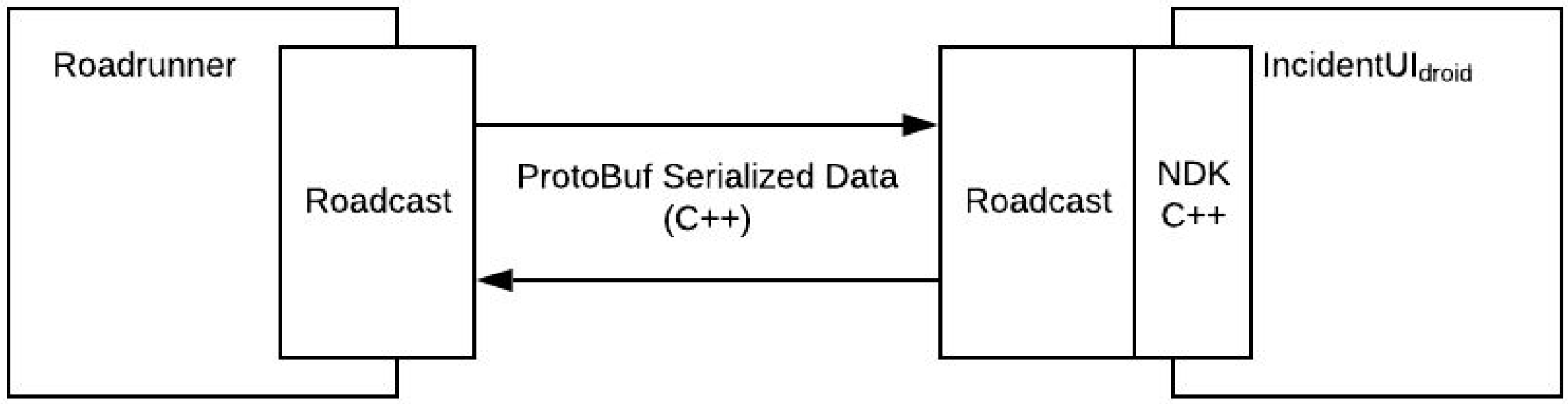

*Figure 11: $\text{IncidentUI}_{droid}$ System Architecture Design Plan A*

In plan A, Roadrunner sends heartbeats and survey requests via Roadcast that serializes the data using protobufs. When these messages from Roadrunner are received by Roadcast on the $\text{IncidentUI}_{\text{droid}}$ end, the data is deserialized and converted into C structures. The NDK interface converts these C structures into Java classes which are then utilized by the Android front end to access the data sent by Roadrunner and execute UI changes accordingly. Similarly, when a survey is filled out by the users, the survey data is packaged into a java class, which is then converted into C structures using the NDK interface. These C structures are then serialized and transmitted to Roadrunner using the C++ implementation of Roadcast. This data is finally deserialized by Roadcast on the Roadrunner end and stored in the Pegasus.

### 3.2.3 System Architecture Design Plan B

Plan B, as depicted in Figure 12, involved a more dramatic redesign of the system architecture, and revolved around the redevelopment of Roadcast and Protobufs in Java on the $\text{IncidentUI}_{\text{droid}}$ end. $\text{IncidentUI}_{\text{droid}}$ still communicated with Roadrunner using Roadcast and Protobufs were still being used for serialization, however, the NDK interface between the Java Android front end and C++-coded Roadcast was discarded for this plan, which eliminated the capability to integrate $\text{IncidentUI}_{\text{droid}}$ with the C++-based Roadcast and Protobufs seamlessly. Consequently, the plan involved the development of a new communications protocol coded purely in Java on the $\text{IncidentUI}_{\text{droid}}$ end. This new communications protocol, termed $\text{Roadcast}_{\text{Java}}$, mimicked the functionality of Roadcast written in C++. In order to serialize and deserialize data on the IncidentUI$_{droid}$ end using Roadcast$_{Java}$, the Protobufs needed to be implemented in Java so that they could be utilized in Roadcast$_{Java}$. To ensure that the scheme that is used to serialize data by Roadcast$_{Java}$ is compatible with Roadcast on the Roadrunner end, existing protobuf definitions were used. This compatibility between data formats processed by Roadcast$_{Java}$ and Roadcast was necessary to ensure that when the data that is serialized by Roadcast$_{Java}$ on IncidentUI$_{droid}$ end is deserialized by Roadcast on the Roadrunner end, the data format will match the existing data structures on that end and vice versa to ensure accurate data extraction. Implementing these existing protobuf definitions in Java generated classes that enable serialization of survey data and other messages by Roadcast$_{Java}$ into a data format that, on deserialization, can be processed by Roadcast and extracted into an existing data structure scheme that can be used by Roadrunner. These protobuf java classes also enable the deserialization and extraction of incoming Roadrunner data directly into Android usable Java classes. So, this plan involved communication between Roadrunner running C++-coded Roadcast and protobufs, and IncidentUI$_{droid}$ deploying Roadcast$_{Java}$ and Java implementations of protobufs to ensure a robust Java-C++ hybrid system architecture design.

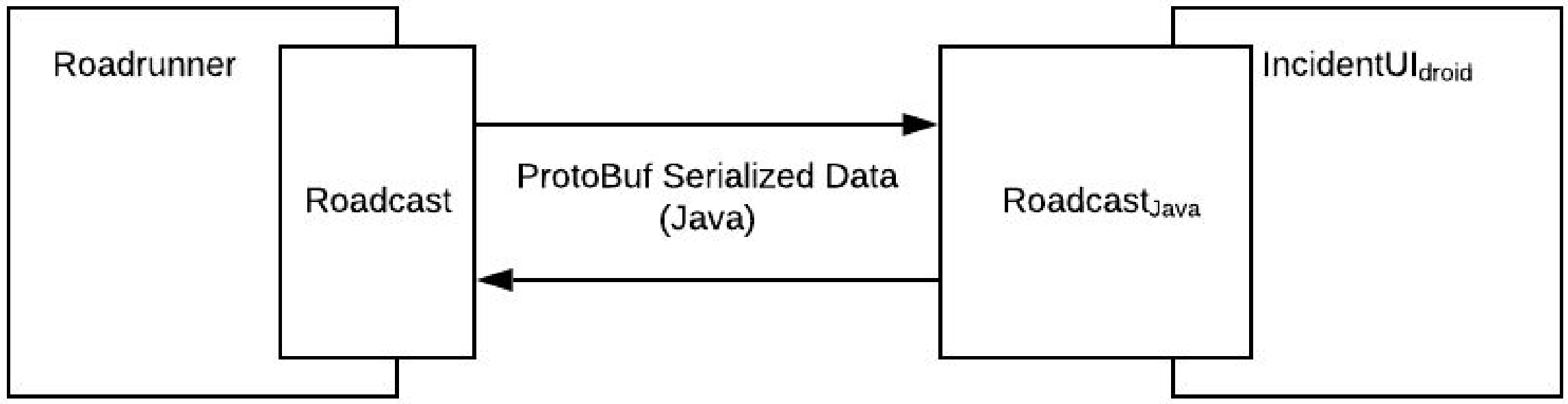

*Figure 12: IncidentUI$_{droid}$ System Architecture Design Plan B*

In plan B, Roadrunner sends heartbeats and survey requests via Roadcast that serializes the data using protobufs. When these messages from Roadrunner are received by Roadcast$_{Java}$ on the IncidentUI$_{droid}$ end, the data is deserialized directly into Java classes which are then utilized by the Android front end to access the data sent by Roadrunner and execute UI changes accordingly. Similarly, when a survey is filled out by the users, the survey data is packaged into a java class, which is then serialized and transmitted to Roadrunner using Roadcast$_{Java}$. This data is finally deserialized by Roadcast on the Roadrunner end and stored in the Pegasus.

## 3.3 Phase 3: IncidentUI$_{droid}$ Beta 1.0 Development

With the current system architecture analyzed, UI mockups constructed, and architecture redesign plans devised, we proceeded to develop Beta 1.0 for IncidentUI$_{droid}$ during Phase 3. Beta 1.0 involved the development of the Android front end of the application which was coded in Java and deployed using Android Studio. Android XML (Extensible Markup Language) was used to design the UI, and the data, UI, and functional flow were coded using Java. Beta 1.0 implemented Plan A for the system design architecture and thus involved the development of a Java Native Interface (JNI) between Java and C++ which was developed using the Native Development Kit (NDK) that comes bundled with Android Studio. Beta 1.0 was developed, deployed, and tested iteratively on Nvidia Shield Android Tablets running Android version 7.0. With the system integration not implemented yet, we utilized the tablets' hardware buttons to emulate Roadrunner messages. This section details the steps carried out and the implementation actualized during the development of Beta 1.0 for IncidentUI$_{droid}$.

### 3.3.1 Front End User Interface (UI) Design

The UI mockup developed during the first phase inspired the UI design and flow for the development of the front end for IncidentUI$_{droid}$ Beta 1.0. The UI consisted of three main screens (activities) with multiple tabs (fragments) within each screen. The UI was designed and developed using Android XML Layouts, and the functionality of the UI elements was coded in Java. The application UI went through multiple iterations and testing phases before the final Beta 1.0 UI was deployed. Android studio was used to develop Beta 1.0 and the application apk was deployed and tested iteratively on multiple Nvidia Shield Tablets. IncidentUI$_{droid}$ is an application designed to collect surveys detailing the causes of ego AV disengagements and involves an interface to track events detailing any ride discomfort experienced in the ego vehicle. This section details the design of the three main application screens for the IncidentUI$_{droid}$ Beta 1.0 - Initiate Drive, Dashboard, and Session along with the sub-tabs for each; it also describes the design flow, i.e., the way these screens interact with each other.

#### Initiate Drive Screen

The first screen that is displayed when the application is launched after getting into the ego vehicle is the Initiate Drive screen. A screenshot of the screen is shown in Figure 13. The screen has text fields to enter the Pilot's and Co-pilot's Nvidia IDs. It also has a button "*Start Drive*" to initiate the drive and launch the dashboard screen. The pilot and copilot information are collected for each drive and used to relate AV disengagements, events, ride comfort, session information, and other test data to the test users.

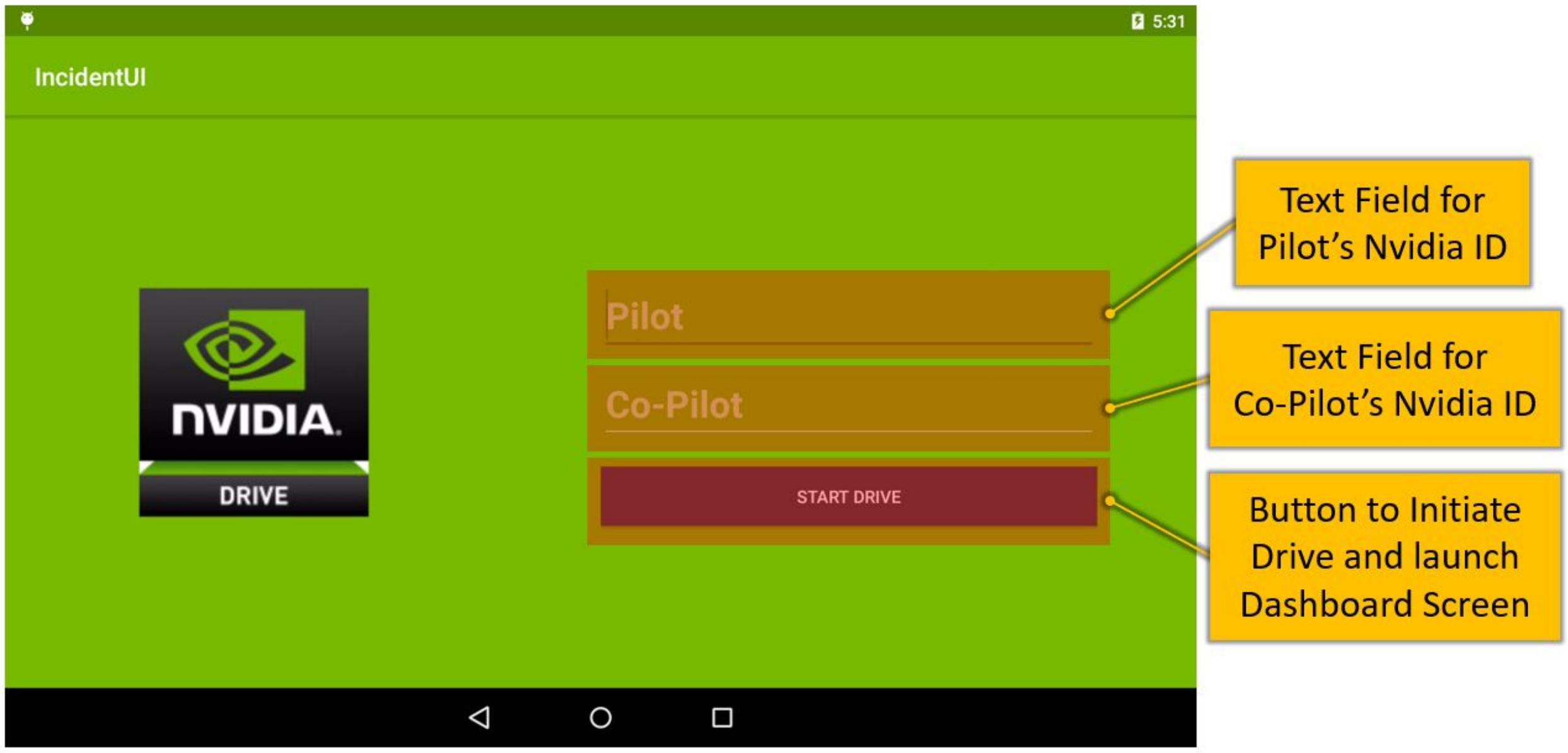

*Figure 13: Initiate Drive Screen*

### Dashboard Screen

The dashboard screen is displayed when a drive is initiated and in between active sessions. It has the following three tabs under it.

#### *Drive History Tab*

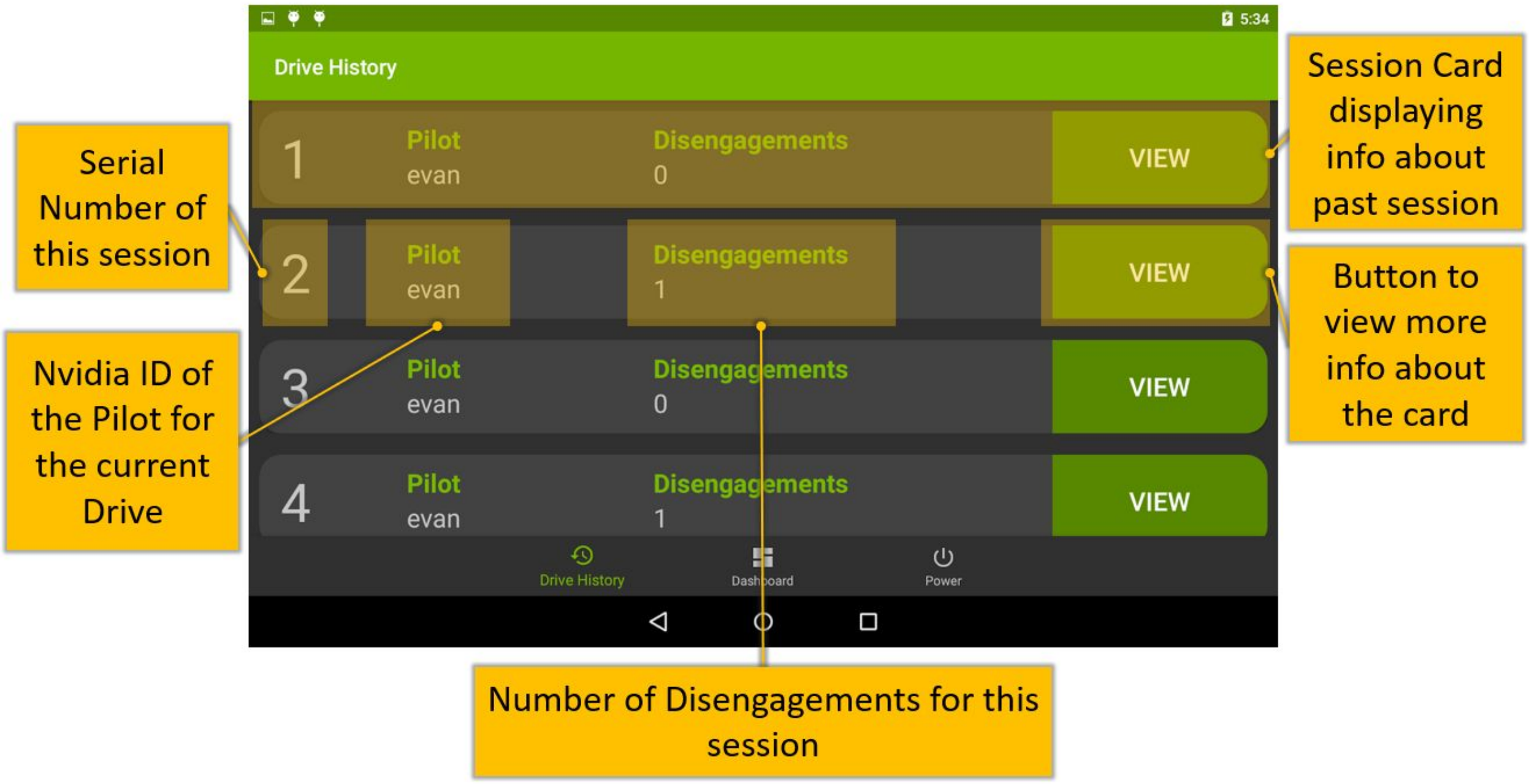

*Figure 14: Drive History Tab*

The Drive History Tab, as shown in Figure 14, stores information about every past session after it has ended. This tab generates a Session Card for every past session and contains information about that session. These cards appear in a vertical list in this tab that can be scrolled through. This tab is also available in the Session Screen, allowing the user to view the drive history during an active session. This tab contains the cards for previous sessions. Each Session Card has the sequence number of the session. It also displays the Nvidia ID of the Pilot for the current Drive and the number of Disengagements for this session.

There is a "*View*" button on each Session Card that, on being clicked, brings up a dialog window that displays more information about that session, as shown in Figure 15. This Session Information dialog window displays the Pilot's and Co-pilot's Nvidia IDs for the current drive. It also displays the number of disengagement and events generated during that session. The dialog window also shows the Session Duration, i.e. the length of time that session lasted. At the bottom of the dialog window, there is a "*Back*" button that dismisses this Session Information dialog and goes back to the Drive History tab.

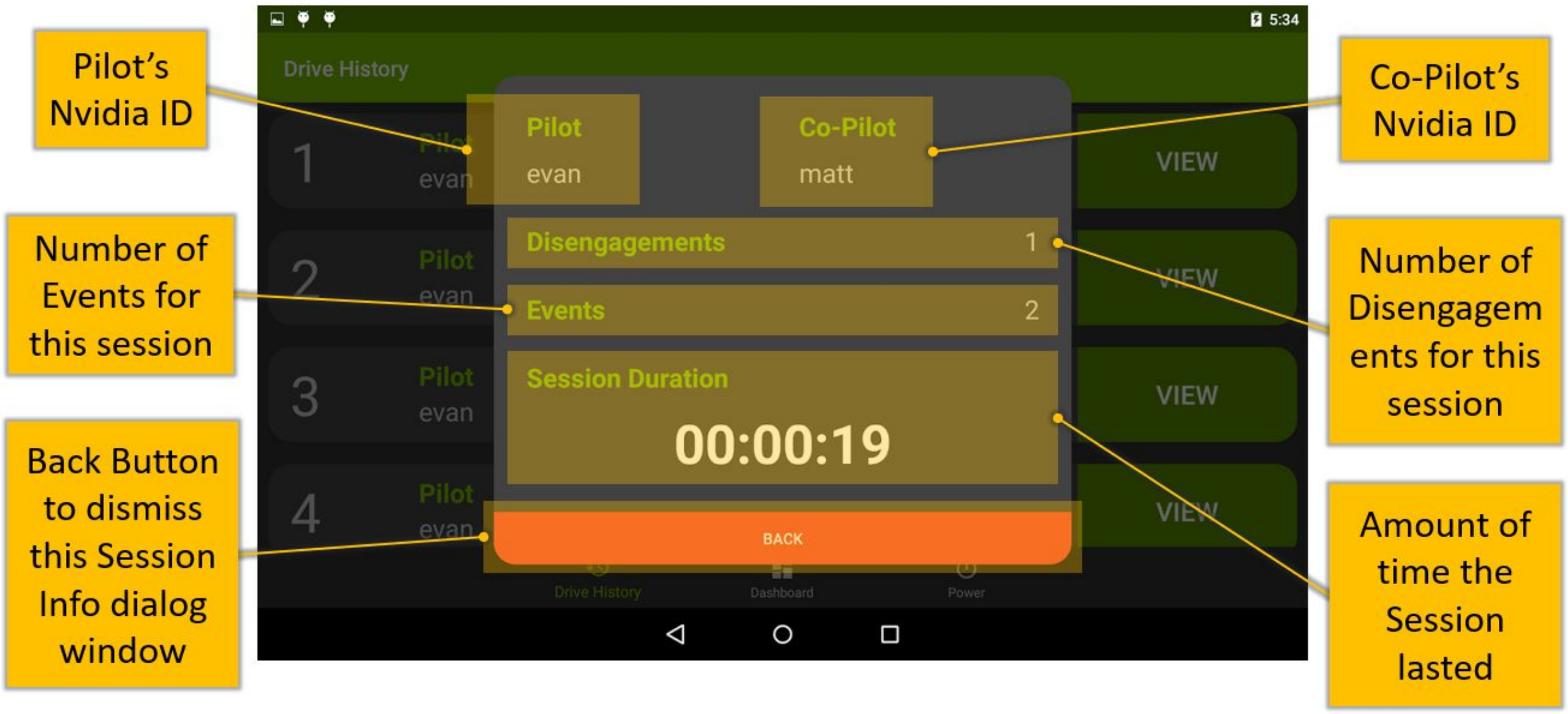

*Figure 15: Session Information Window*

### *Dashboard Tab*

The dashboard tab, as shown in Figure 16, is displayed by default whenever the dashboard screen is launched. This tab displays the Nvidia IDs of the Pilot and the Co-pilot on the current drive. It has an "*End Drive*" Button that terminates the current drive and takes the user back to the Initiate Drive Screen. At the top right corner of the screen, there is an indicator that shows the current Roadrunner connection status, i.e. if the client running on IncidentUI$_{droid}$ detects a heartbeat from Roadrunner or not.

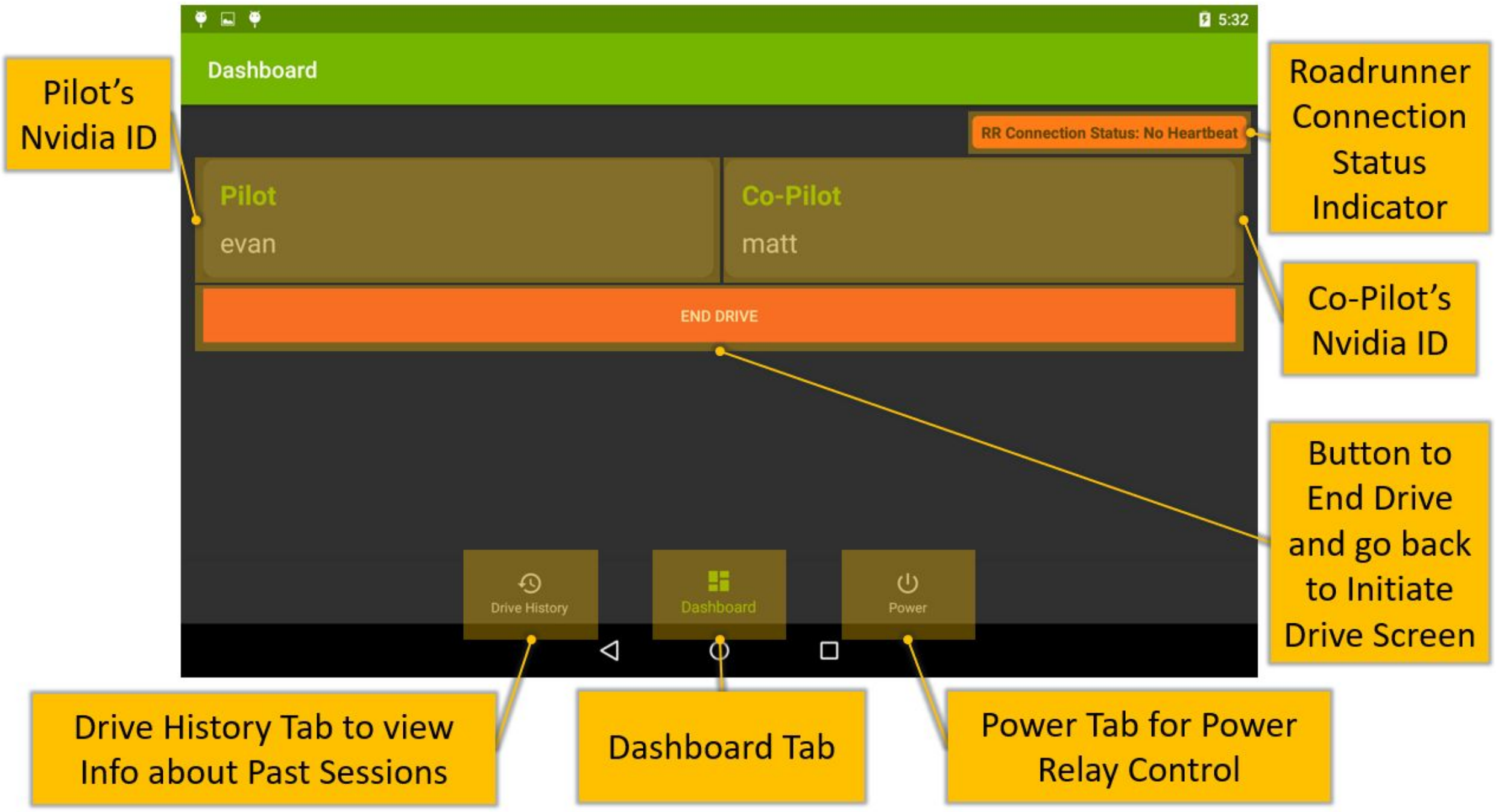

*Figure 16: Dashboard Tab*

At the bottom of the screen, there is a navigation menu that allows users to switch between the three tabs in the dashboard screen, i.e. Drive History Tab, Dashboard Tab, and Power Tab.

### *Power Tab*

The power tab is currently empty but has been included as a tab in the UI in case the application is further developed to include the power relay control functionality, thereby completely eliminating the need for the Raspberry Pi in the Hyperion system architecture.

## Session Screen

During an active session, the session screen is displayed. The ability to initiate or end a session is not under the control of the user; instead, it is controlled by the status of the Roadrunner connection, i.e. the moment Roadrunner connects, a session is initiated, and a session ends when Roadrunner is disconnected. The session screen has the following two tabs:

*Drive History Tab*

The Drive History tab is also available in the Session screen to view information about past sessions during an active session. The User Interface elements of this tab and their functions are the same as explained previously for the Dashboard screen.

*Session Tab*

The session tab, as shown in Figure 17, is the main tab that is displayed during an active session. It displays the current Pilot's and Co-pilot's Nvidia IDs, along with the number of disengagements and events encountered in this session. There is a "*New Event*" button that records a new event by bringing up the Event Survey. The number of events and disengagements are updated in the session tab in real-time. There is also a chronometer that tracks the amount of time that has elapsed since the current session started. At the top right corner of the screen, there is an indicator that shows the current Roadrunner connection status, i.e. if the client running on $\text{IncidentUI}_{\text{droid}}$ detects a heartbeat from Roadrunner or not.

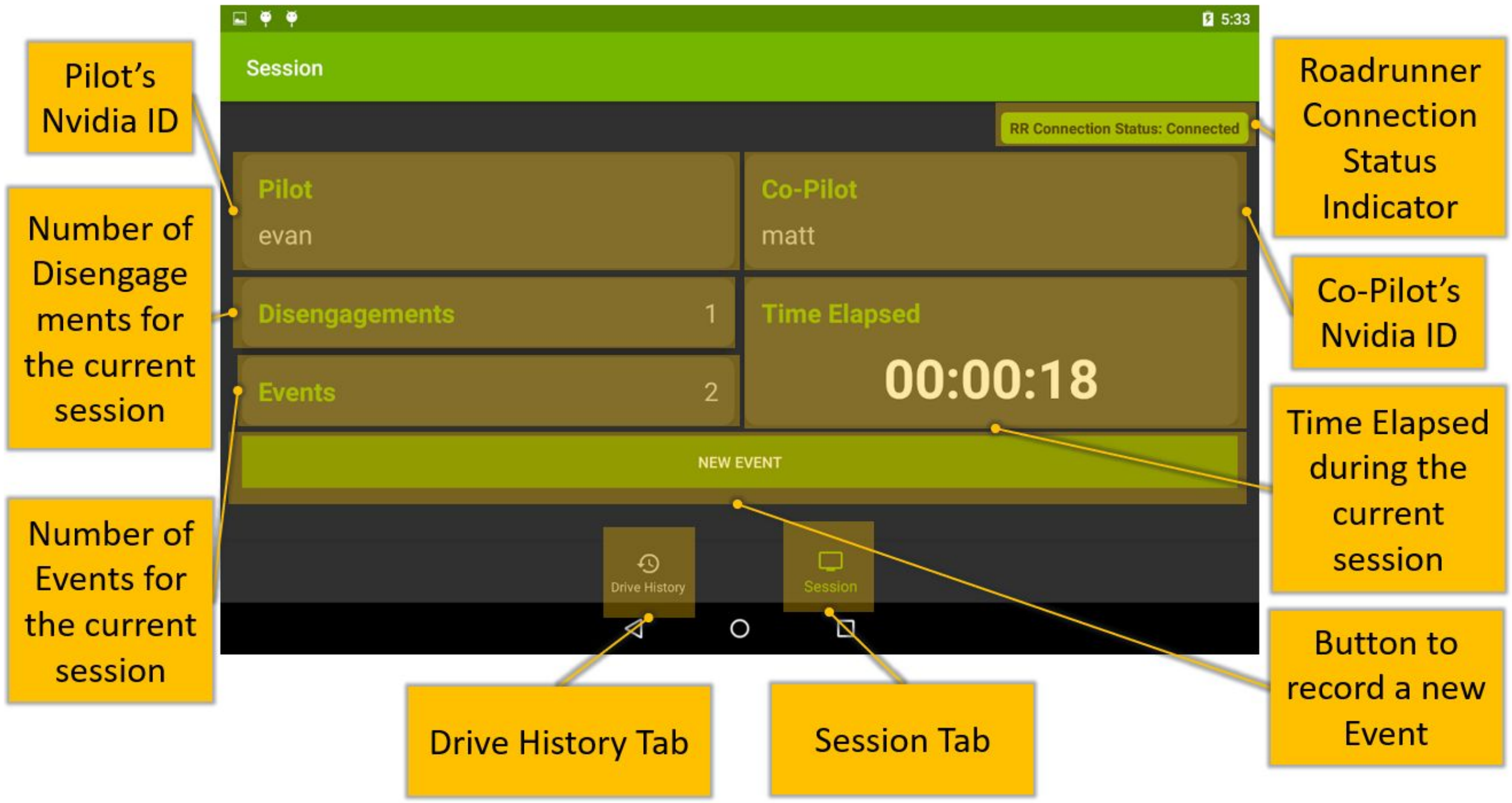

*Figure 17: Session Tab*

The disengagement survey is brought up automatically on the screen when Roadrunner detects an ego vehicle disengagement and sends a disengagement survey request to $\text{IncidentUI}_{\text{droid}}$. During the development and testing of Beta 1.0, Roadrunner messages sent to $\text{IncidentUI}_{\text{droid}}$ were emulated using hardware buttons on the Nvidia Shield tablet; clicking the Volume-Down hardware button on the tablet mimicked a disengagement survey request from Roadrunner and displayed the Disengagement Survey on the tablet screen. At the bottom of the screen, there is a navigation menu that allows users to switch between the two tabs in the session screen, i.e. Drive History Tab and Session Tab.

### Surveys

The main function of IncidentUI is to track events and disengagements and their causes using surveys. They make use of pre-existing parameters to measure the comfort of the ego vehicle's maneuvers across 2 axes - lateral and longitudinal and use these parameters or other additional information to report the reasoning for the lateral/longitudinal comfort rating assigned by the co-pilot for the event/disengagement. These surveys are displayed and can be filled out only while a session is active, i.e. Roadrunner connection is present. The survey responses are sent to the Roadrunner using Roadcast and then stored in the Pegasus. There are two kinds of surveys used for recording data:

#### *Event Survey*

The Event Survey is used to track an event. An event is recognized as a scenario where the ego vehicle's maneuver causes lateral and/or longitudinal discomfort without requiring a disengagement from the pilot. Whenever the ego vehicle does a maneuver that the co-pilot believes violates the comfort parameters, they can generate an event. To do so, they can click on the "New Event" on the Session tab, which brings up the "Event Survey" as shown in Figure 18. Ride comfort is measured across 2 axes using sliders on a scale of 0 to 5. Checkboxes let the user choose potential causes for Lateral and Longitudinal Discomfort, with the option to select multiple causes simultaneously. The possible causes of Longitudinal Discomfort are as follows:

- Collision Threat
- Too Fast
- Too Slow
- Jerky Acceleration
- Jerky Break
- False Breaking

The causes of Lateral Discomfort are as follows:

- Collision Threat
- Swerve
- Lateral Jerk
- Too Aggressive
- Too Conservative

The Event Survey also allows the Co-pilot to define the position of the Ego-vehicle at the time of the discomfort event to give locational context to the cause of the event. This locational context improves the quality of the event data, which can then be used to enhance the Drive Software further. The Ego-Vehicle's position is selected from radio buttons for the following options:

- Lane Keep
- Merge
- Lane Change
- Split
- Ramp

The Event Survey also allows the user to provide additional information related to the event that is not covered by the predefined options. The user (usually the co-pilot) can fill out the "*Additional Information*" in a dedicated text box.

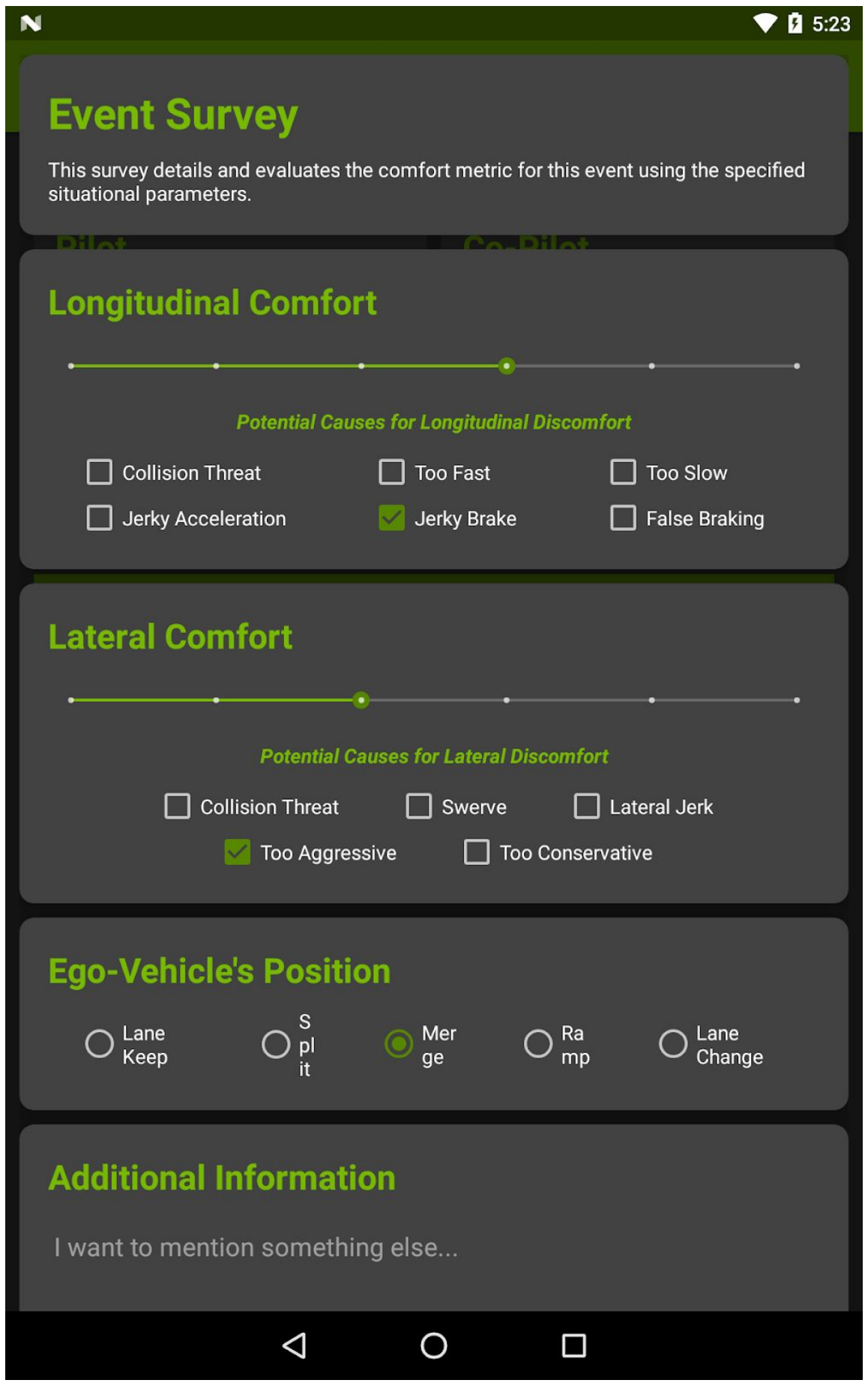

*Figure 18: Event Survey*

### *Disengagement Survey*

The Disengagement survey, as shown in Figure 19, is used to record disengagement data, i.e. lateral and longitudinal ride comfort, cause of the disengagement, the explanations for the cause of the disengagement, and any additional information describing the disengagement. In case of an ego vehicle disengagement, Roadrunner sends a disengagement survey request to $\text{IncidentUI}_{droid}$ using Roadcast, and consequently, the disengagement survey is displayed on the tablet screen.

The ride comfort is measured across 2 axes: longitudinal and lateral, and can be selected using sliders from a range of 0 to 5, where 0 is least comfortable and 5 is most comfortable.

The Cause of Disengagement can be selected from four possible options using Radio Buttons. The possible causes of disengagement are as follows :

- Safety Issue
- Intended and safe
- End of Drive
- Other

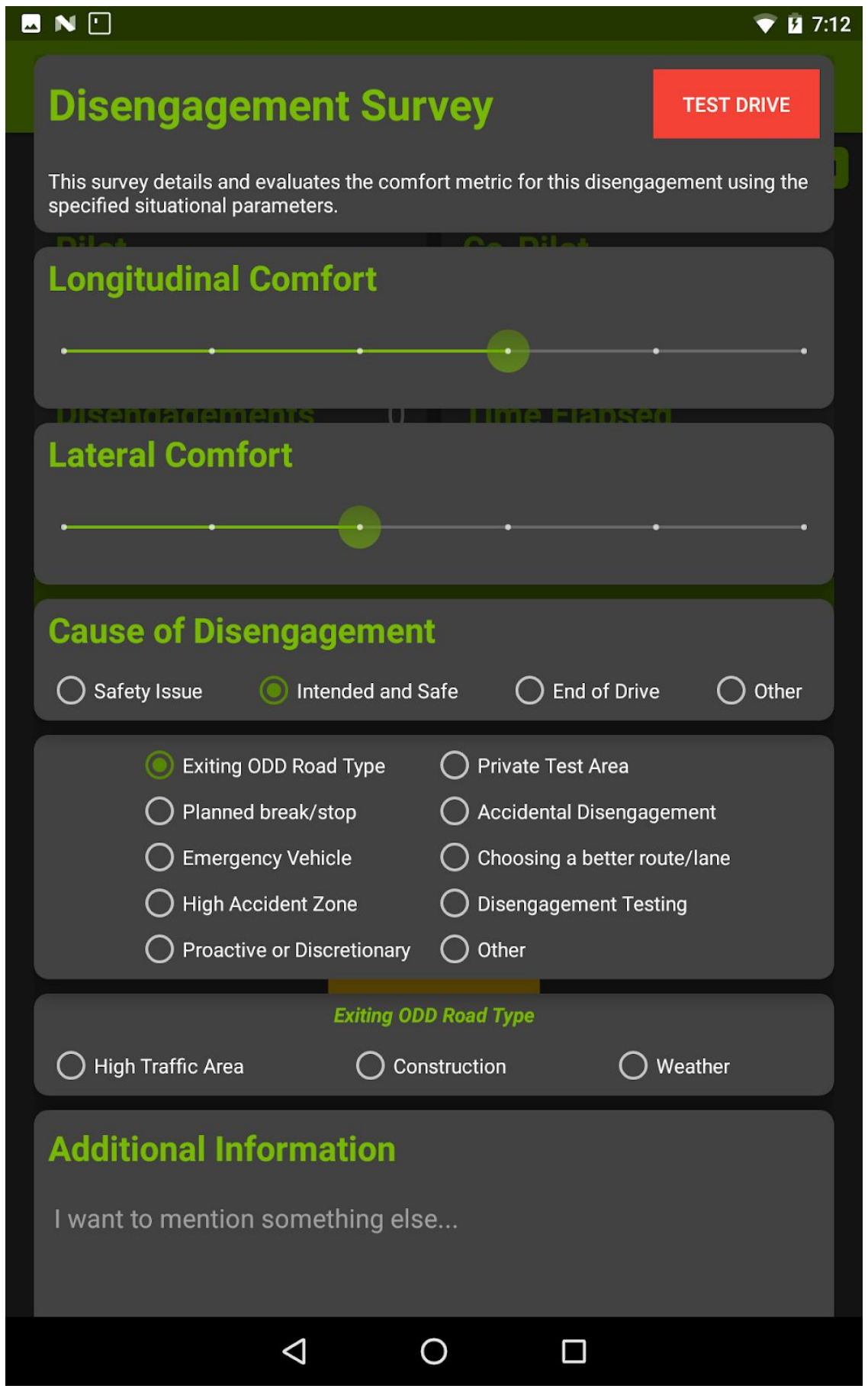

*Figure 19: Disengagement Survey*

If the cause of the disengagement is some safety issue caused by the nature of the maneuver performed by the ego vehicle that forced the pilot to take control and trigger a disengagement, then the "*Safety Issue*" option is selected as the Cause of Disengagement on the Disengagement Survey. Selecting this option displays additional fields to provide explanations for the Safety Issue. These fields include potential causes for Lateral and Longitudinal Discomfort and the location of the Ego Vehicle at the time of the disengagement. The layout of the fields is identical to that of the Event Survey, as described in the previous section.

If the disengagement is triggered by the pilot intentionally because of varying reasons and it is not the result of a safety issue, i.e., the disengagement is the result of some reason external to the ego vehicle misbehaving, then the "*Intended and Safe*" option is selected as the Cause of Disengagement on the Disengagement Survey. Choosing the "*Intended and Safe*" option prompts the user to provide more context to the disengagement cause by selecting one of the following explanations using radio buttons:

- Exiting ODD Road Type
- Planned break/stop
- Emergency Vehicle
- High Accident Zone
- Proactive or Discretionary
- Private Test Area
- Accidental Disengagement
- Choosing a better route/lane
- Disengagement Testing
- Other

From these aforementioned options, selecting "*Exiting ODD Road Type*" prompts the user to provide more environmental context to the explanation for the disengagement like a High Traffic Area, Weather Conditions or, a Construction Zone. The rest of the options are self-explanatory and do not require additional context. If none of the predefined options give an accurate explanation for the "*Intended and Safe*" cause of disengagement, then the "*Other*" option can be selected, which displays a text box that can be used to provide an explanation for the disengagement that is not covered by the predefined options.

At the end of an AV drive, the pilot takes control of the ego vehicle to mark the end of the test session and the data collection process, which triggers a disengagement. This scenario can be recorded by selecting the "*End of Drive*" option as the Cause of Disengagement on the Disengagement Survey. If none of the aforementioned options accurately define the cause of disengagement, then the "*Other*" option can be selected on the Disengagement Survey, which displays a text box that can be used to provide a cause for the disengagement that is not covered by the predefined options. The Disengagement Survey also allows the user to provide additional information related to the disengagement that is not covered by the predefined options. The user (usually the co-pilot) can fill out the "*Additional Information*" in a dedicated text box.

During an AV test drive at Nvidia's private test track, multiple disengagements and events are triggered to test and tweak the Hyperion Software and Hardware. The Disengagement Survey has a "*Test Drive*" button on the top-right corner that automatically submits a disengagement survey with the following information filled out:

- Longitudinal Comfort: 0
- Lateral Comfort: 0
- Cause of Disengagement: Intended and Safe
- Explanation: Private Test Area
- Additional Information: None

This feature is set up to ensure that the test users are not bogged down by having to fill out the disengagement survey with duplicate information every time an intentional disengagement is triggered as part of the test session on the private test track.

### 3.3.2 Android Native Development Kit (NDK) Interface Design

With the front end Android application set up, an NDK interface needed to be developed to enable robust data transfer between the Java-coded Android application and the system libraries and protocols like Roadcast, which were developed using C++. Android Native Development Kit (NDK), which is a Java Native Interface package that comes bundled with Android Studio, was used to develop an interface consisting of methods that repackage data in Java objects into C++ structure instances and vice versa.

When Roadrunner sends a survey request message to $\text{IncidentUI}_{\text{droid}}$ via Roadcast, the message is received by Roadcast running on the $\text{IncidentUI}_{\text{droid}}$ end and it is deserialized using Protobufs to generate a C structure instance. The data stored in this C structure is processed by a method in the NDK interface and converted to a Java class object. Based on the contents of this Java class object, UI elements are triggered; for instance, a survey pops up on the $\text{IncidentUI}_{\text{droid}}$ screen. After the survey is filled out and submitted, a Java object is generated with its data derived from the fields filled out in the survey. This Java object is then processed by the NDK interface and converted into a C structure that is serializable by the Protobufs that are utilized by Roadcast and matches the data scheme on the Roadrunner end. This serialized data is sent to Roadrunner, where it is deserialized by Roadcast and stored in the Pegasus.

The NDK interface comprises two primary methods, one to handle messages from Roadrunner to $\text{IncidentUI}_{\text{droid}}$ and the other for messages from $\text{IncidentUI}_{\text{droid}}$ to Roadrunner. Each of these methods utilizes a unique method for every type of message transmitted between Roadrunner and $\text{IncidentUI}_{\text{droid}}$, i.e. a disengagement survey request will be translated from C++ to Java using a method different from the one used to translate disengagement data from Java to C++. The NDK interface is developed entirely in C++ with seamless access to C++-coded system libraries and protocols and involves calls to JNI methods that access Java methods and objects that are a part of the Android end of the application.

### 3.3.3 Beta 1.0 Demonstration

The final stage of the development of Beta 1.0 for $IncidentUI_{droid}$ involved a demonstration of the beta to the end-users of $IncidentUI_{droid}$, i.e. the IFORV (End-to-End Product Testing) and the SQA (Quality Control) teams. The demonstration focused on the User Interface (UI) Design and Flow of the Android application with some insight into the system integration strategies and architecture design.

The demonstration began with a presentation highlighting the UI screens and tabs along with the functions of the various UI elements. This presentation was followed by a walkthrough of the actual Android application beta running on an Nvidia Shield Tablet. The walkthrough demonstrated the UI and data flow along with the way users could interact with the application. It also demonstrated the UI changes resulting from the interaction between $IncidentUI_{droid}$ and Roadrunner, where the messages from Roadrunner were simulated using the hardware buttons on the tablet. The walkthrough highlighted the changes in the UI design and flow along with the enhanced intuitiveness, compatibility, and robustness of the new Android $IncidentUI_{droid}$, compared to the existing IncidentUI deployed on the Raspberry Pi.

After the walkthrough, the tablet was passed down to the SQA and IFORV teams as part of a free user interaction and evaluation session. This session was followed by a Q&A that involved discussions about UI design and flow, application use, suggested features, and some general feedback. The Q&A also involved a discussion about system integration, architecture design, and platform compatibility, which also highlighted the versatility, efficacy, and enhanced robustness of deploying $IncidentUI_{droid}$ in an Android environment. The demonstration also displayed the ease of use and portability of a tablet application.

#### Feedback Analysis

The demonstration of Beta 1.0 for $IncidentUI_{droid}$ broadly received positive feedback from the SQA and IFORV teams. Some of the quotes from the discussions with the team members are as follows:

- *"The biggest issue with IncidentUI is typing on that small screen. It is much easier to type on this tablet."*
- *"I don't like the current UI. This UI is a lot better and easier to interact with."*
- *"The fixed display cannot be passed on to the people in the backseat. This tablet is much more portable."*
- *"Using the in-console display disturbs the pilot. This tablet will remain with the co-pilot so that's safer."*

- *"Android is more versatile. Like I can have this application running on my phone which means multiple people can use it at the same time."*

As can be devised from the aforementioned quotes, the teams really appreciated the redesigned User Interface and screen flow along with the enhanced user interactivity of $\text{IncidentUI}_{\text{droid}}$. They were in favor of Android as an environment for the application because of its enhanced robustness and greater compatibility with various platforms. Another aspect of the new design they appreciated was the deployment of the application on a tablet which is much more portable and easy to use than a fixed in-console display, allowing for feedback from multiple users. The teams had many questions about the system architecture design and integration strategies and were in favor of both system architecture design plans discussed previously.

In a nutshell, the SQA and IFORV teams really appreciated the transition of IncidentUI from a fixed Raspberry Pi display to a portable Android tablet and commented that using the Android Tablet-based $\text{IncidentUI}_{\text{droid}}$ would considerably improve their testing experience during the disengagement and ride comfort data collection process.

## 3.4 Phase 4: $\text{IncidentUI}_{\text{droid}}$ Beta 1.0 System Integration

With the $\text{IncidentUI}_{\text{droid}}$ Beta 1.0 application developed, deployed, and evaluated on the Nvidia Shield tablets and an NDK interface set up, we began integrating the application with the existing system architecture utilizing *Design Plan A*. The integration strategy under system architecture design Plan A involved utilizing the existing Protobuf and Roadcast C++ implementations in an Android environment and integrating them with $\text{IncidentUI}_{\text{droid}}$ via the NDK interface. The NDK interface, which is developed in C++, has methods that regulate data transmission between Roadcast and Android end of $\text{IncidentUI}_{\text{droid}}$. This section describes the steps taken to deploy the C++ implementations of Protobuf and Roadcast in an Android environment and couple Roadcast on the $\text{IncidentUI}_{\text{droid}}$ end with the NDK interface. This section also highlights the issues encountered during system integration under Architecture Design Plan A which resulted in a pivot to Design Plan B.

### 3.4.1 Protobufs Implementation in C++

Nvidia utilizes Protocol Buffers as a part of the Hyperion Kit to serialize data that is transmitted between Roadrunner and IncidentUI. System integration of $\text{IncidentUI}_{\text{droid}}$ under Architecture Design Plan A involved using the existing C++ implementation of Protobufs. Since Roadcast utilizes the protobuf classes and methods for serialization, the building of the Roadcast libraries is dependent upon the building of the protobufs library, which is termed "*avprotos*". This plan involved deploying the C++ protobuf classes generated from the existing "*.proto*" files in an Android environment and then calling the methods associated with these classes to serialize data on the $\text{IncidentUI}_{\text{droid}}$ NDK interface end. As part of this deployment, we configured and built the system Protobuf libraries using the same Protobuf compiler, i.e. "*protoc*" file used by Roadcast, to maintain compatibility between the serialization scheme implementations on the Roadrunner and $\text{IncidentUI}_{\text{droid}}$ end. Each "*protoc*" file is built using the Protobuf source files. This "*protoc*" file is then used to configure the "make scheme" used for building the system Protobuf libraries. This "*protoc*" file is also used to generate the required Protobuf classes from the "*.proto*" files.

The limited documentation available online for utilizing C++ implementations of protobufs in an Android environment pointed towards the generation of a cross-compiled system Protobuf library using cross-compilation standalone toolchains. This cross-compiled system Protobuf library enables the utilization of C++ implementations of Protobufs, i.e. the Protobuf classes generated from the "*.proto*" files, by the NDK interface in an Android environment. A cross-compilation standalone toolchain was developed for each of the 32-bit and 64-bit versions of ARM architecture, which is the instruction set architecture used by the Tegra K1 chipset in Nvidia Shield Tablets [17]. We also developed a cross-compilation standalone toolchain for each of the 32-bit and 64-bit versions of Linux-x86 architecture to enable the testing of the application in the Android Studio Emulator deployed on the Testing Host Machine running Ubuntu Linux. The "make scheme" was reconfigured for the Android environment using the newly generated cross-compilation standalone toolchains and the same "*protoc*" file. This reconfigured "make scheme" was then used to build and compile the cross-compiled Protobuf library. However, the generation of the cross-compiled system Protobuf library using the standalone toolchains was met with some Android NDK version compatibility issues as discussed in detail later in section 3.4.3.

Once the cross-compiled system Protobuf library is compiled, the "*protoc*" file is used to generate the Protobuf classes as part of the "*avprotos*" library. The "*avprotos*" files utilize the cross-compiled system Protobuf library to support robust data serialization in an Android environment. The NDK interface is coupled with Roadcast, which utilizes classes in the avprotos library to serialize data that is transmitted between $\text{IncidentUI}_{\text{droid}}$ and Roadrunner.

### 3.4.2 Roadcast Integration

Another aspect of system integration of $\text{IncidentUI}_{\text{droid}}$ is the deployment of Roadcast in an Android environment and the coupling of Roadcast with the NDK interface. The coupling of the NDK interface and Roadcast involved setting up a loop that listens for incoming survey requests and heartbeats sent to the $\text{IncidentUI}_{\text{droid}}$ Roadcast Client from Roadrunner. When a message is received and deserialized by the Roadcast client, the listener loop is alerted and it forwards these messages that are packaged as C++ structures to the NDK interface where they are processed and converted into Java objects. The loop also listens for survey data and heartbeat messages received from the NDK interface and forwards them to the Roadcast Server running on the $IncidentUI_{droid}$ end. When the survey data is processed by the NDK interface and converted into a C++ structure, the listener loop is alerted and it forwards the message to the Roadcast Server, which serializes the data and transmits it to Roadrunner.

To enable $IncidentUI_{droid}$ to utilize Roadcast, all of the libraries and files associated with Roadcast needed to be deployed in an Android environment as part of the system integration of $IncidentUI_{droid}$. This deployment involved building an Android apk consisting of the application itself, i.e. $IncidentUI_{droid}$, and the Roadcast library as a support library. However, the deployment of Roadcast in an Android environment resulted in some compatibility issues between the NDK system files and the Roadcast library files, which is discussed in detail in the next section.

### 3.4.3 Issues with Beta 1.0 System Integration

The process of integrating the Beta 1.0 for $IncidentUI_{droid}$ with the system architecture using Design Plan A was concluded prematurely owing to some insurmountable barriers encountered while deploying Protobufs and Roadcast in an Android environment.

#### Issues with Protobufs Integration

The primary issue with the deployment of C++ Protobufs implementations in an Android environment is the incompatibility between the standalone toolchains required for generating the cross-compiled Protobuf library and the Android NDK version required to accommodate the C++ system libraries required by Roadcast. The minimum NDK version required for the necessary C++ system libraries used by Roadcast is "*r21*", but the standalone toolchains become obsolete with version "*r19*" [18], which makes it impossible to accommodate a cross-compiled Protobuf library and Roadcast simultaneously. Lowering the NDK version below "*r19*" allows a cross-compiled Protobuf library to build with standalone toolchains but the C++ system libraries required by numerous Roadcast files aren't included in the NDK bundle for this version which results in numerous undeclaration and import errors. On the other hand, raising the NDK version to "*r21*" accommodates the C++ system libraries required by Roadcast, but obsoletes standalone toolchains, which eliminates the capability to build a cross-compiled Protobuf library. This inability to reconcile the two major facets of system integration: Roadcast and Protobufs, became a point of failure for the system integration of Beta 1.0 under Architecture Design Plan A.

#### Issues with Roadcast Integration

The deployment of Roadcast in an Android environment was met with numerous redeclaration and compatibility issues between the Roadcast library files and the NDK C++ system libraries. These file, class, function, and variable redeclaration issues arose from the fact that the existing Roadcast library comprises of customized redeclared versions of existing C++ system library files; however, these redeclaration issues are only resolved for a Linux environment, in which the current version of Roadcast is deployed. A redeployment of Roadcast libraries into an Android environment reproduced these redeclaration issues as a result of incompatibility with the identical C++ system library files in the NDK bundle. A resolution of these redeclaration errors required manually sifting through hundreds of thousands of lines of Roadcast and NDK bundle system code and rebuilding the entire Android apk with every minute code patch, which, after two weeks of debugging, still resulted in a failed apk build. The compatibility issues were amplified by the fact that Roadcast is compiled using GCC, but Android Studio relies on Clang for compiling native code, which resulted in further build errors. We predicted that debugging these errors would require an indefinite amount of time and ultimately result in an unstable system architecture, which resulted in another point of failure for the system integration of Beta 1.0 under Architecture Design Plan A.

Owing to the aforementioned issues, strong time constraints, and the lack of progress regardless of several attempts and help from software engineers at Nvidia, the team decided to conclude the development Beta 1.0 and pivot to the development of Beta 2.0 utilizing Design Plan B for integrating $\text{IncidentUI}_{\text{droid}}$, which involved a more drastic but robust system architecture redesign.

## 3.5 Phase 5: $\text{IncidentUI}_{\text{droid}}$ Beta 2.0 Development

With the overwhelming issues encountered during the development, and more specifically the system integration, of Beta 1.0, we decided to pivot to the development of Beta 2.0 for $\text{IncidentUI}_{\text{droid}}$. Beta 2.0 incorporates the same Java-coded Android front end design as Beta 1.0 but involves a more drastic redesign of the system architecture under Design Plan B, with the $\text{IncidentUI}_{\text{droid}}$ back end and system sub-architecture developed entirely in Java to build a robust system and ensure seamless data transmission throughout this system. In addition to the development and evaluation of Beta 2.0, this phase also involved extensive $\text{IncidentUI}_{\text{droid}}$ integration and functional testing using a Roadrunner Emulator. This section details the steps taken during the design, development, system integration, and evaluation of the Java-coded $\text{IncidentUI}_{\text{droid}}$ Beta 2.0.

### 3.5.1 Android Front End and Interface Redesign

Since the front end of the $\text{IncidentUI}_{\text{droid}}$ application developed during Beta 1.0 was deemed intuitive and effective on evaluation, Beta 2.0 incorporates the same front end design for the Android application with some minor UI changes to improve text readability and incorporate a practical screen design regardless of the screen orientation. To limit application use exclusively to $\text{IncidentUI}_{\text{droid}}$ during a test drive and ensure absolute user attention and participation, the application is configured as the default Android application launcher, which forces $\text{IncidentUI}_{\text{droid}}$ to be displayed on boot up and stay on-screen throughout the duration of device use. In order to actualize Java implementations of Protobufs and Roadcast on the back end and develop a system sub-architecture developed entirely in Java according to Architecture Design Plan B, the Android NDK interface was decoupled. This decoupling involved eliminating the NDK system libraries and the NDK interface developed to transform data from Java objects to native structures and vice versa. On the system end, the C++ Roadcast support libraries and the Protobufs C++ system libraries were also removed from the Android apk. This decoupling resulted in a Java-coded front end Android application identical to the one developed during the Front End Development of Beta 1.0 for $\text{IncidentUI}_{\text{droid}}$.

### 3.5.2 $\text{Roadcast}_{\text{Java}}$

With the front end developed for Beta 2.0 and the NDK interface and C++ system libraries decoupled, the system integration of $\text{IncidentUI}_{\text{droid}}$ under Architecture Design Plan B was initiated, which involved the development of $\text{Roadcast}_{\text{Java}}$: a new communications protocol coded purely in Java. $\text{Roadcast}_{\text{Java}}$ replicates the functionality of the C++-coded Roadcast and utilizes Java implementations of Protobufs to serialize and deserialize data on the $\text{IncidentUI}_{\text{droid}}$ end. Similar to Roadcast, $\text{Roadcast}_{\text{Java}}$ implements a Client and a Server to regulate the exchange of data with Roadrunner.

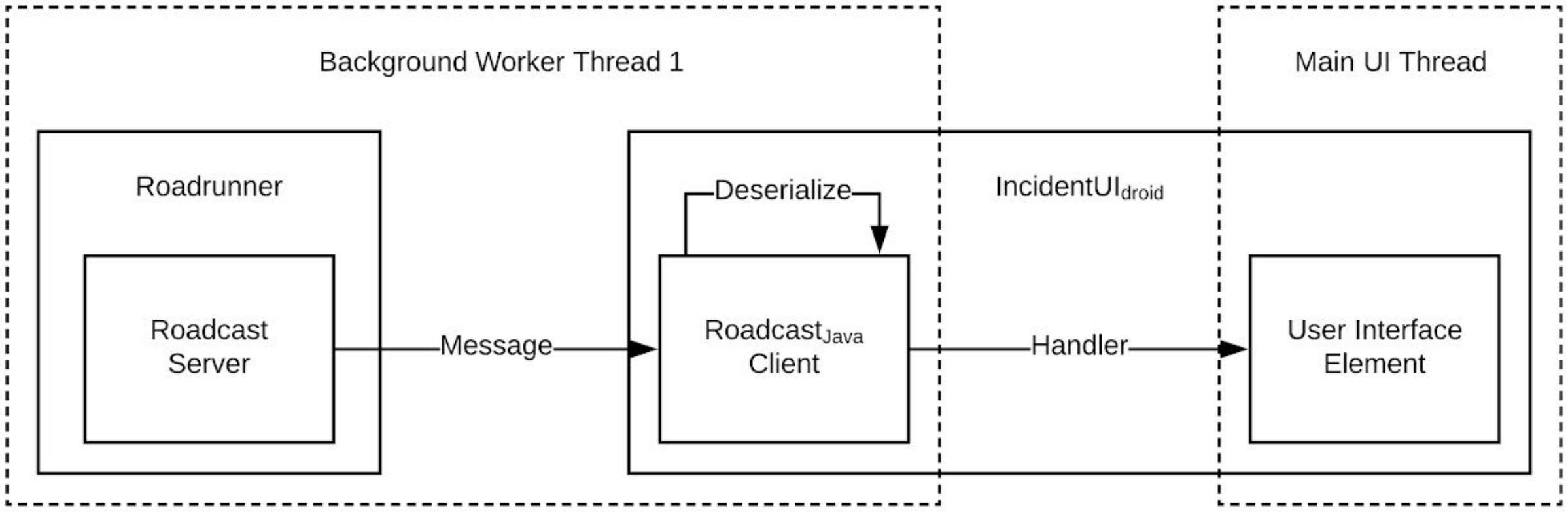

*Figure 20: $\text{Roadcast}_{\text{Java}}$ Client Sub-architecture*

The $\text{Roadcast}_{\text{Java}}$ Client, as shown in Figure 20, is responsible for receiving and processing survey requests and heartbeat messages sent by Roadrunner. The Client runs on a background worker thread as an "*IntentService*'' and listens for incoming data from Roadrunner in a loop. The $\text{Roadcast}_{\text{Java}}$ Client is configured to listen for messages from Roadrunner on a specific port and from a specific Server IP address that the Roadcast Server running on Roadrunner is bound to. Since the Client runs as a background service on a worker thread, the Android front end of the application running on the main UI thread remains interactive and functions parallel to the background listener Client. The $\text{Roadcast}_{\text{Java}}$ Client, on receiving a message from the Roadcast

Server running on Roadrunner, deserializes the message data directly into a Java object using the Java implementations of Protobufs, i.e. Java class serialization methods derived from the *"avprotos"* Protobufs definitions. The Client then sends this object or a corresponding signal to the UI thread using a "*Handler*". This "*Handler*" is utilized by the Android front end to access the data sent by Roadrunner and execute UI changes accordingly.

$\text{Roadcast}_{\text{Java}}$ utilizes a server to generate and send survey data and heartbeat messages to Roadrunner. The $\text{Roadcast}_{\text{Java}}$ Server, as shown in Figure 21, runs on another background worker thread as an "*IntentService*" and sends generated messages to Roadrunner in a loop. The Server is bound to a specific IP address and configured to accept client connections on a specific port. The Server also utilizes this port to send serialized messages to the Roadcast Client running on Roadrunner.

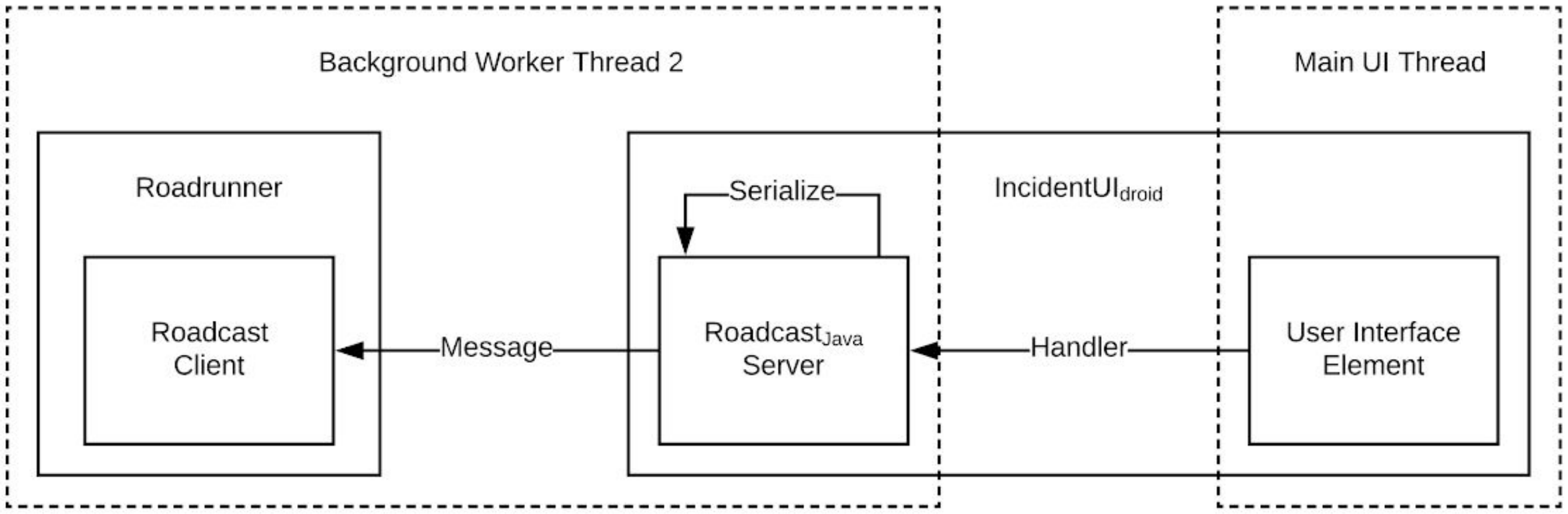

*Figure 21: $\text{Roadcast}_{\text{Java}}$ Server Sub-architecture*

When a survey is filled out and submitted on the main UI thread, the survey data is packaged into a Java object. The Server then sends this object or a corresponding signal to the background worker thread using a "*Handler*". On receiving the Java survey object and/or the signal from the "*Handler*", the $\text{Roadcast}_{\text{Java}}$ Server utilizes the Java class serialization methods to serialize the Java object and generate a message, which is then transmitted to the Roadcast Client running on Roadrunner.

On application startup, the $\text{Roadcast}_{\text{Java}}$ client attempts to connect to the Roadcast server in a loop using a specific port and Server IP address. On a successful connection, the $\text{Roadcast}_{\text{Java}}$ client receives and processes the heartbeat messages sent by the Roadcast server. Similarly, the $\text{Roadcast}_{\text{Java}}$ server gets initialized and bound to a specific IP address and starts listening for client connections on a specific port. On a successful Roadcast client connection, the $\text{Roadcast}_{\text{Java}}$ server generates and sends heartbeat messages to the connected Roadcast client. After a successful and robust two-way connection is established between Roadrunner and $\text{IncidentUI}_{droid}$, varying survey request messages and corresponding survey response messages are sent back and forth during an active test session. At the start of every active session, a Login Survey Request is sent by Roadrunner and the corresponding Login Survey Response containing information about the current Pilot and Copilot is sent back by $\text{IncidentUI}_{droid}$. This exchange is followed by a series of Event and Disengagement survey requests and responses sent back and forth between Roadrunner and $\text{IncidentUI}_{droid}$. Throughout the span of an active session, heartbeats are exchanged between Roadrunner and $\text{IncidentUI}_{droid}$ to indicate an active roadrunner connection status. After a test session ends and Roadrunner disconnects, the $\text{Roadcast}_{Java}$ client retries connecting to the Roadcast server in a loop, and simultaneously the $\text{Roadcast}_{Java}$ server continues to listen for incoming Roadcast client connections. After a successful two-way handshake is established, a new test session is generated.

The IP addresses and the ports used by the $\text{Roadcast}_{Java}$ server and client are modified depending upon the communication medium used to transmit data between Roadrunner and $\text{IncidentUI}_{droid}$. For instance, if $\text{IncidentUI}_{droid}$ is connected to Roadrunner over the Nvidia corporate WiFi network, then the $\text{Roadcast}_{Java}$ server and client are assigned dynamic IP addresses and corresponding ports, and if $\text{IncidentUI}_{droid}$ is connected to Roadrunner via USB tethering then the $\text{Roadcast}_{Java}$ server and client are assigned static IP address and corresponding ports as part of the local AV static network. $\text{Roadcast}_{Java}$ maintains a robust Java-based system sub-architecture for $\text{IncidentUI}_{droid}$ whilst also successfully replicating the functionality of the C++-coded Roadcast to ensure a robust integration of Roadrunner and $\text{IncidentUI}_{droid}$.

### 3.5.3 Protobufs Implementation in Java

Beta 2.0 still utilizes Protobufs for serializing data sent back and forth between Roadrunner and $\text{IncidentUI}_{droid}$. Under Architecture Design Plan B, the Protobufs are implemented in Java to enable $\text{Roadcast}_{Java}$ to utilize them and ensure robust serialization on the $\text{IncidentUI}_{droid}$ end. To maintain compatibility between the serialization scheme implementations on the Roadrunner and $\text{IncidentUI}_{droid}$ end, we configured and built the Java system Protobuf libraries and the Java-based "*avprotos*" library using the same Protobuf compiler, i.e. "*protoc*" file used by Roadcast. Since the system architecture under Design Plan B does not involve native code and libraries and the system Protobufs libraries are generated in Java and include support for Android, system integration under Design Plan B doesn't require a cross-compiled Protobufs library that accommodates standalone toolchains and native system libraries. We also used the existing Protobufs definitions in the "*.proto*" files to generate Java implementations of Protobufs, i.e. Java Protobufs serialization classes, to maintain compatibility between Roadrunner and $\text{IncidentUI}_{droid}$. The methods associated with these Java Protobufs serialization classes are utilized by Roadcast$_{Java}$ to deserialize incoming messages from Roadrunner into Android-usable Java objects. These methods are also utilized by Roadcast$_{Java}$ to serialize data stored in Java objects into messages that are transmitted by IncidentUI$_{droid}$ to Roadrunner. The Java implementations of Protobufs maintain a robust Java-based system sub-architecture for IncidentUI$_{droid}$ whilst also successfully integrating with the C++ implementations of Protobufs utilized by Roadrunner to ensure a robust Java-C++ hybrid system architecture integration.

### 3.5.4 Android Tablet Networking

The final aspect of system integration for IncidentUI$_{droid}$ Beta 2.0 involved configuring the hardware and network settings for the Nvidia Shield Tablet that was used for developing, deploying, and evaluating IncidentUI$_{droid}$. The three varying hardware network solutions provided to integrate the Nvidia Shield Tablet running IncidentUI$_{droid}$ with the Pegasus that deploys Roadrunner, are as follows:

- Tethering using USB On-The-Go (OTG)
- Ethernet LAN
- WiFi - Nvidia's Corporate Wireless Network

Each route to configure the hardware networking for the Nvidia Shield Tablet requires a unique setup. In order to debug the functionality of IncidentUI$_{droid}$ and ensure that the application is working as intended, we utilized the Android Debug Bridge (ADB). ADB is a command-line tool that enables communication with an Android device via a Unix shell [19]. The Unix shell is also used to execute a variety of commands to tweak the system and hardware settings for a connected Android device. We utilized ADB for IncidentUI$_{droid}$ debugging and Android Tablet hardware network and system configuration for all three aforementioned hardware network solutions.

To enable communication via USB Tethering and Ethernet LAN, we configured new network interfaces on the Pegasus. These network interfaces were configured to be in the same static local sub-network as the Nvidia Shield Tablet to allow steady communication between the two, without drastic modifications to the network configuration. The WiFi network solution was more straightforward and involved configuring the IP address assigned to the IncidentUI$_{droid}$ Android Tablet in the Pegasus' network configuration files.

Each proposed network solution has its pros and cons. USB Tethering involves a wired connection between the Pegasus and the Nvidia Shield Tablet, and contrary to the other proposed network solutions, it provides support for charging the tablet while it is connected to the Pegasus. Compared to the other network solutions, Ethernet LAN requires the least amount of time. The WiFi network solution requires the least amount of network configuration modifications and is the most practical to use. Even though future iterations of IncidentUI$_{droid}$ will depend heavily on the WiFi network solution, Nvidia's current AV infrastructure does not comprise a WiFi network owing to network security concerns.

### 3.5.5 Integration and Functional Testing with Roadrunner Emulator

In addition to the iterative user interface, hardware network configuration, and front end functional testing, this phase involved the development of a custom Roadrunner test build to emulate the functionality of the Hyperion Kit Roadrunner. The Roadrunner build is set up to test the functionality of $\text{IncidentUI}_{\text{droid}}$ and the integration stability of the Beta 2.0 system sub-architecture. The Roadrunner build is deployed on the Pegasus that is set up as a component of a test sub-architecture identical to the IncidentUI Test Bench sub-architecture. The custom build runs a sample test session recorded during a previous AV test drive.

The custom-built Roadrunner connects to $\text{IncidentUI}_{\text{droid}}$ and automatically exchanges Heartbeat messages throughout the span of an active test session. At the start of the test session, it sends a Login Survey Request to $\text{IncidentUI}_{\text{droid}}$. The custom build is set up to run in a Linux environment from the command line of the Jetson AGX Xavier A module in the Pegasus and allows the user to manually trigger Event, Disengagement and Login Survey Requests. The command line can also be used to manually alter the Lateral and Longitudinal Sequence Numbers on the Disengagement Survey Requests. In addition to the $\text{IncidentUI}_{\text{droid}}$ connection status, the Roadrunner build prints out the Message Type (Heartbeat, Event Flag, Event Survey, Disengagement Survey, or Login Survey) and the Message Content, i.e. the Survey Data, for every message it receives from $\text{IncidentUI}_{\text{droid}}$. The build is designed to emulate every possible Roadrunner scenario during an AV test drive and is used to ensure the proper functioning of $\text{IncidentUI}_{\text{droid}}$. It is also used to evaluate the quality of data transmission between Roadrunner and $\text{IncidentUI}_{\text{droid}}$ and the integration stability of the Beta 2.0 system sub-architecture under Design Plan B.

## 3.6 Phase 6: Evaluation and Stable Release Deployment

With a functional Beta 2.0 for $\text{IncidentUI}_{\text{droid}}$ deployed and integrated with the redesigned system sub-architecture, the sixth and final phase in the development of $\text{IncidentUI}_{\text{droid}}$ was initiated. This phase involved a Beta 2.0 application demonstration and the development, deployment, evaluation, and in-car testing of the stable release for $\text{IncidentUI}_{\text{droid}}$. This section details the steps taken during this phase to evaluate and enhance Beta 2.0 and deploy the final stable release for $\text{IncidentUI}_{\text{droid}}$.

### 3.6.1 $\text{IncidentUI}_{\text{droid}}$ Beta 2.0 Application Demonstration

This step of Phase 6 involved a demonstration of Beta 2.0 to the AV Platform Team that includes the IFORV and the SQA teams. The demonstration described the User Interface Design, System

Architecture Design and Integration Strategy, and the functionality of $IncidentUI_{droid}$ Beta 2.0. The demonstration began with a presentation highlighting the steps involved in the development of Beta 2.0, with an overview of the issues encountered during the development and system integration of Beta 1.0 using Architecture Design Plan A, and the consequent pivot to Beta 2.0 and Design Plan B. The presentation also mentioned the design and flow of the User Interface along with the development of the system integration strategies involving $Roadcast_{Java}$ and Java implementations of Protobufs.

This presentation was followed by a walkthrough of the actual Android application Beta 2.0 running on an Nvidia Shield Tablet. The walkthrough demonstrated the functionality of $IncidentUI_{droid}$ and the nature of data transmission between Roadrunner and $IncidentUI_{droid}$ by utilizing the test sub-architecture set up in the previous phase. We utilized the custom Roadrunner build to emulate an AV test drive and manually trigger disengagements and events using shell commands. Secure Shell (SSH) was used to remotely access the test Pegasus running the custom Roadrunner build. The test Pegasus running the Roadrunner build was connected to the Nvidia Shield tablet running $IncidentUI_{droid}$ over Nvidia's wireless corporate network by utilizing the WiFi hardware network solution mentioned previously.

SSH was used to access the Roadrunner build logs during the demonstration that displayed the Heartbeats and the Survey Data that Roadrunner received from $IncidentUI_{droid}$. The heartbeat messages sent and received by the Roadrunner build were displayed in the logs throughout an active session which indicated a continuous exchange of heartbeats between Roadrunner and $IncidentUI_{droid}$. Manually triggering a disengagement on the Roadrunner build sent a Disengagement Survey Request to $IncidentUI_{droid}$, which was recorded in the Roadrunner session logs. Receipt of this Disengagement Survey Request by $IncidentUI_{droid}$ triggered the disengagement survey, which was displayed on the $IncidentUI_{droid}$ screen. The disengagement survey, on being filled out and sent to the Roadrunner build, appeared in the session logs as well.

After the walkthrough, the tablet was passed down to the AV Platform team members as part of a free user interaction and evaluation session. This session was followed by a Q&A that involved discussions about the UI design and flow, application use, suggested features, and some general feedback. The Q&A also involved a discussion about system integration, architecture design, and platform compatibility, which also highlighted the versatility, efficacy, and enhanced robustness of deploying $IncidentUI_{droid}$ in an Android environment and developing a Java-based system sub-architecture. The demonstration also displayed the ease of use and portability of a tablet application.

### Feedback Analysis

The demonstration of Beta 2.0 for $IncidentUI_{droid}$ broadly received positive feedback from the AV Platform Team. The team members really appreciated the redesigned User Interface and the enhanced user interactivity of $IncidentUI_{droid}$. They were in favor of Android as an environment for the application because of its enhanced robustness and greater compatibility with various platforms. Another aspect of the new design they appreciated was the deployment of the application on a tablet which is much more portable and easy to use than a fixed in-console display, allowing for feedback from multiple users. The team members also appreciated the enhanced integration stability of a Java-based system sub-architecture and the deployment of $Roadcast_{Java}$ and Java implementations of Protobufs in an Android environment. As part of the discussion about the design, development, integration, and evaluation of $IncidentUI_{droid}$, the AV Platform Team provided some useful insight on the aspects of $IncidentUI_{droid}$ that could be further improved and suggested some additional features for the final stable release of $IncidentUI_{droid}$. The team also discussed the possible direction for the future development of $IncidentUI_{droid}$.

In a nutshell, the AV Platform Team team really appreciated the transition of IncidentUI from a fixed Raspberry Pi display to a portable Android tablet and commented that using the Android Tablet-based $IncidentUI_{droid}$ would considerably improve their testing experience during the disengagement and ride comfort data collection process. They also believe that the deployment of $IncidentUI_{droid}$ in an Android environment and the development of a purely java-based system sub-architecture would facilitate the maintenance and modification of the user interface, application features, and system architecture of $IncidentUI_{droid}$.

### 3.6.2 $IncidentUI_{droid}$ Stable Release Development

With the $IncidentUI_{droid}$ Beta 2.0 deployed, evaluated, and demonstrated, we concluded its development and pivoted to the development of a full stable release for $IncidentUI_{droid}$, which focused on the development of some new features suggested during the Beta 2.0 application demonstration. This step also involved some general debugging, UI cleanup, and code refactoring. This section describes the design, development, and testing of the four major $IncidentUI_{droid}$ stable release features suggested during the Beta 2.0 application demonstration. The development of these new features was focused on the Android application end and was done to improve user experience by implementing intuitive features that are designed to give the users more control of the application and enhance the data collection process. This step concluded with the deployment of a final Stable Release for $IncidentUI_{droid}$.

#### Feature 1: Display Active Session Data on User Login

One of the features implemented during the development of the $IncidentUI_{droid}$ stable release was designed to handle a functional edge case where a disengagement or Roadrunner connection is detected by $IncidentUI_{droid}$ before the user has entered the Pilot and Co-pilot Login Information. The application screen flow for $IncidentUI_{droid}$ has been designed to require the users to enter the Pilot and Co-pilot Login Information for every test drive to ensure that the disengagement and ride comfort data for every session can be related to a pilot and co-pilot. This requirement is enforced by forbidding the users from accessing the Session screen and the Event and Disengagement Surveys until the login information has been filled out on the Login screen and the Login Survey has been submitted to initiate the test drive. During normal operation, as depicted in Figure 22, the Login screen is displayed on application startup, where the user fills out the Pilot and Co-pilot Login Information and starts the drive to launch the Dashboard screen.

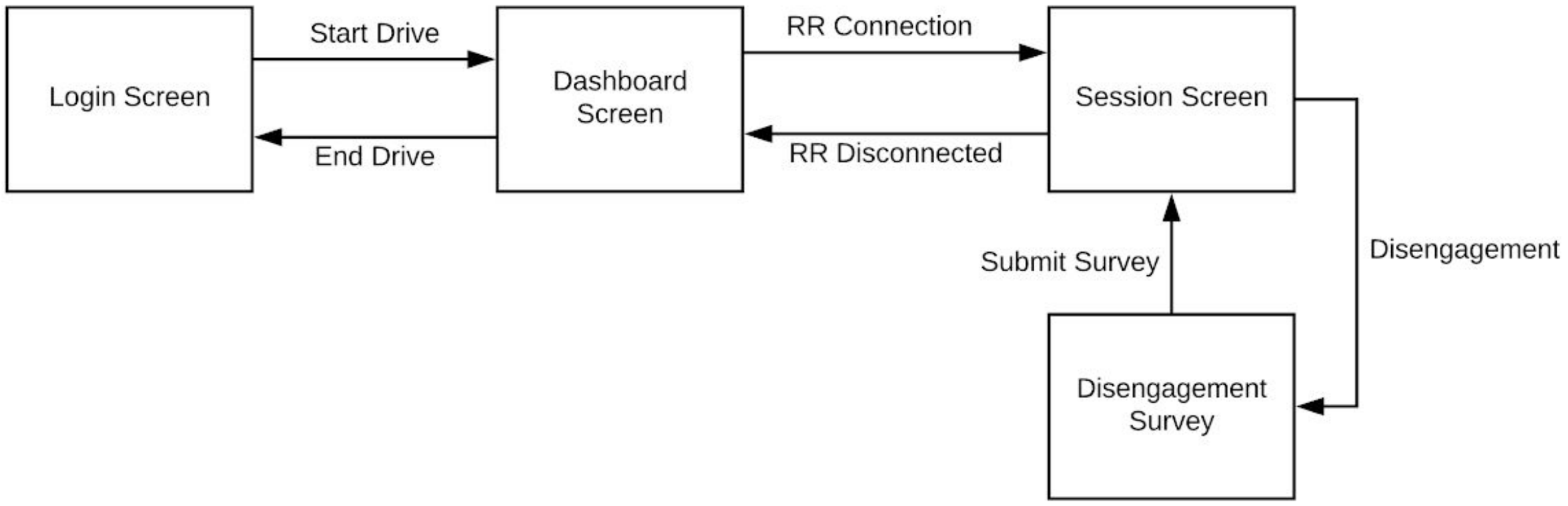

*Figure 22: Feature 1 - IncidentUI$_{droid}$ Normal Operation*

From the Dashboard screen, the Session screen is loaded and the login information is sent to Roadrunner when a Roadrunner connection is established. If a Disengagement is detected during this active session, Roadrunner sends a Disengagement Survey Request to IncidentUI$_{droid}$, which displays the Disengagement Survey on the IncidentUI$_{droid}$ screen. Once the Disengagement Survey is filled out and submitted, the Session screen is loaded again.

The aforementioned scenario takes place during the normal operation of IncidentUI$_{droid}$. However, this feature is set up to handle the edge case where a Roadrunner connection is established with IncidentUI$_{droid}$ before the user has entered the Pilot and Co-pilot Login Information. For this edge case, as depicted in Figure 23, when a Roadrunner connection is established with IncidentUI$_{droid}$, the Login screen is displayed because the Login Information has not been filled out and submitted to initiate the drive. In this scenario, once the user fills out and submits the login information by clicking the "*Initiate Drive*" button on the Login screen, the Dashboard screen is skipped and the Session screen is displayed directly to account for the ongoing session that was initiated when the Roadrunner connection was established on the Login screen. If a Disengagement is detected during this active session, Roadrunner sends a Disengagement Survey Request to $\text{IncidentUI}_{\text{droid}}$, which displays the Disengagement Survey on the $\text{IncidentUI}_{\text{droid}}$ screen. Once the Disengagement Survey is filled out and submitted, the Session screen is loaded again.

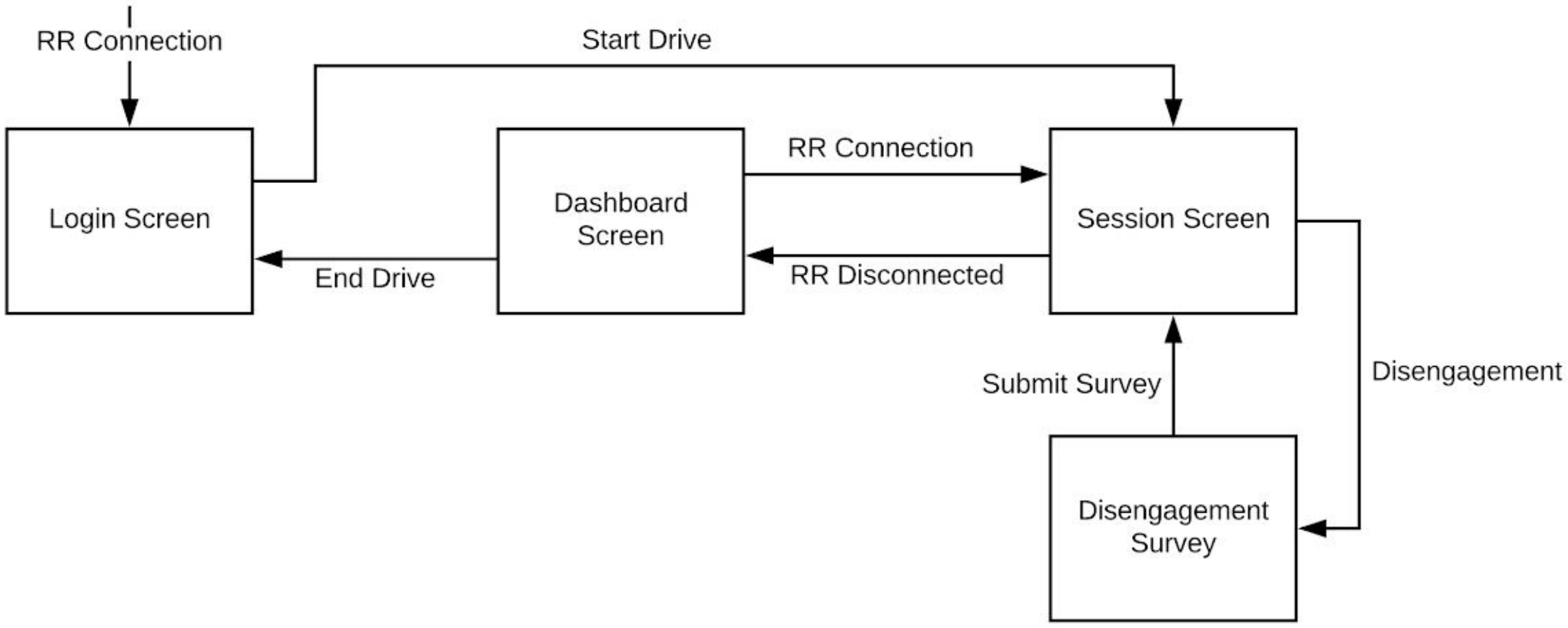

*Figure 23: Feature 1 - Roadrunner Connection Edge Case*

This feature is developed to handle another edge case where a Roadrunner connection is established and a disengagement is generated before the user has entered the Pilot and Co-pilot Login Information. For this edge case, as depicted in Figure 24, when a disengagement is detected by Roadrunner, it sends a Disengagement Survey Request to $\text{IncidentUI}_{\text{droid}}$, which continues to display the Login screen because the Login Information has not been filled out and submitted to initiate the drive. Since the AV remains disengaged until the Disengagement Survey is submitted to Roadrunner, the users are required to fill out the Disengagement Survey to re-engage the AV, and as the Login screen is displayed during an active disengagement in this scenario, the users are forced to fill out the Pilot and Co-pilot Login Information to gain access to the Disengagement Survey. Once the user fills out and submits the login information by clicking the "*Initiate Drive*" button on the Login screen, the Dashboard screen is skipped and the Disengagement Survey on the Session screen is displayed directly to account for the disengagement that was detected previously on the Login screen. Once the Disengagement Survey is filled out and submitted, the Session screen is loaded again.

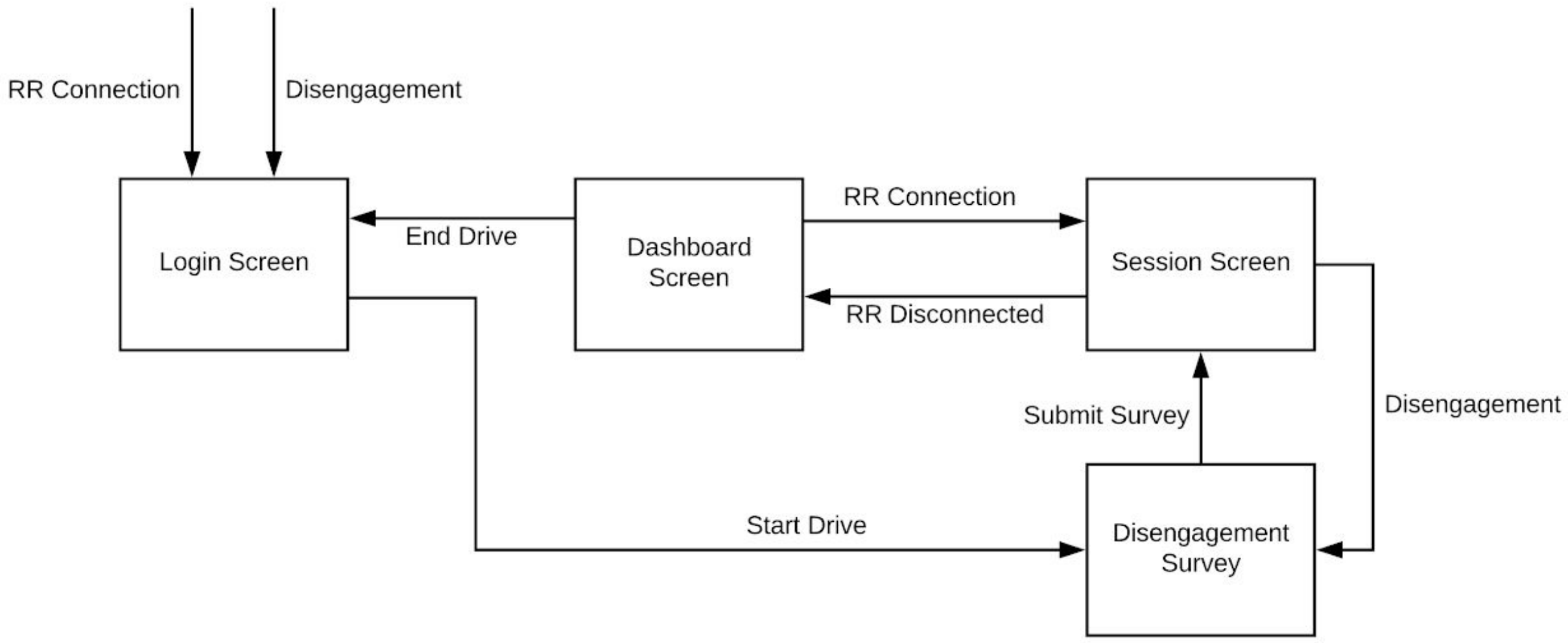

*Figure 24: Feature 1 - Disengagement Edge Case*

The development of this feature revolved around configuring flags for Roadrunner Connection Status and Disengagement Detection. If any of these flags are raised prior to user login as a result of an established roadrunner connection or a triggered disengagement, then the "*Initiate Drive*" button skips the Dashboard screen and directly loads the Session screen or the Disengagement Survey depending upon the type of flag raised. This feature enhances the user interactivity of $\text{IncidentUI}_{\text{droid}}$ by implementing an automated conditional screen flow design for some common scenarios and giving a more intuitive structure to the user interface.

### Feature 2: Save and Edit Event Surveys

In a scenario that involves the AV performing an undesirable maneuver that causes some ride discomfort, the users of $\text{IncidentUI}_{\text{droid}}$ have the ability to trigger an Event by clicking the "New Event" button on the Session tab, which displays the Event Survey on the screen. In the previous iterations of $\text{IncidentUI}_{\text{droid}}$, users have had the ability to either submit the Event Survey and send it to Roadrunner or discard the Event Survey in case the corresponding event was generated unintentionally. The development of this feature involved equipping the users with the ability to "*Save*" a partially filled out Event Survey so that it can be edited, discarded, or submitted later. This feature, as shown in Figure 25, involved the development of a "*Save*" button at the bottom of the Event Survey. Clicking the "*Save*" button preserves the progress for an Event Survey and creates a card associated with that Event Survey in the Session tab.

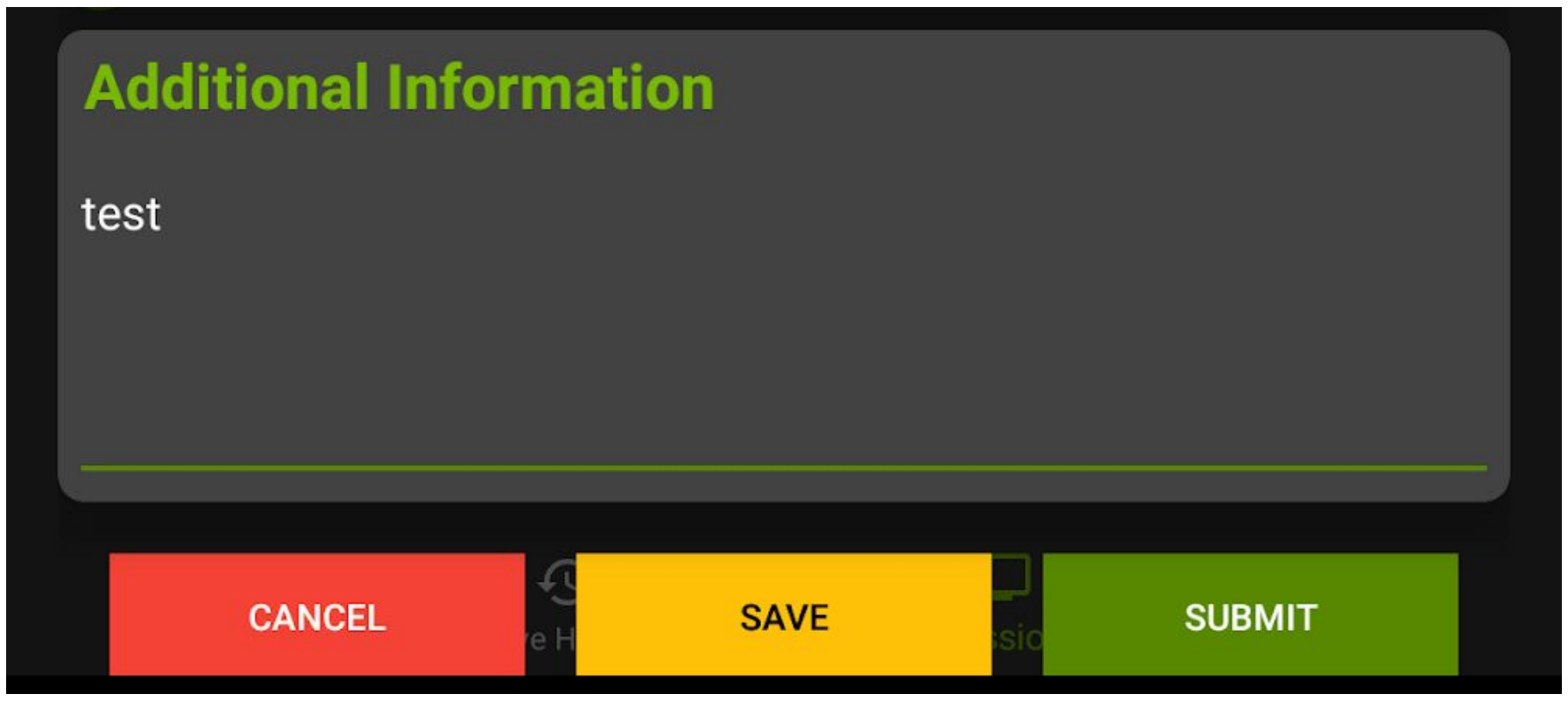

*Figure 25: Event Survey Save Button*

This feature also gives the users the ability to create and save multiple Event Surveys, each of which is stored as an Event Survey Card in a scrollable list. The Event Survey Cards, as shown in Figure 26, display the Event Sequence Numbers for the Event Surveys they are associated with. These cards also include an "*Edit*" button, which can be clicked to access the associated Event Survey filled out with the previously saved progress.

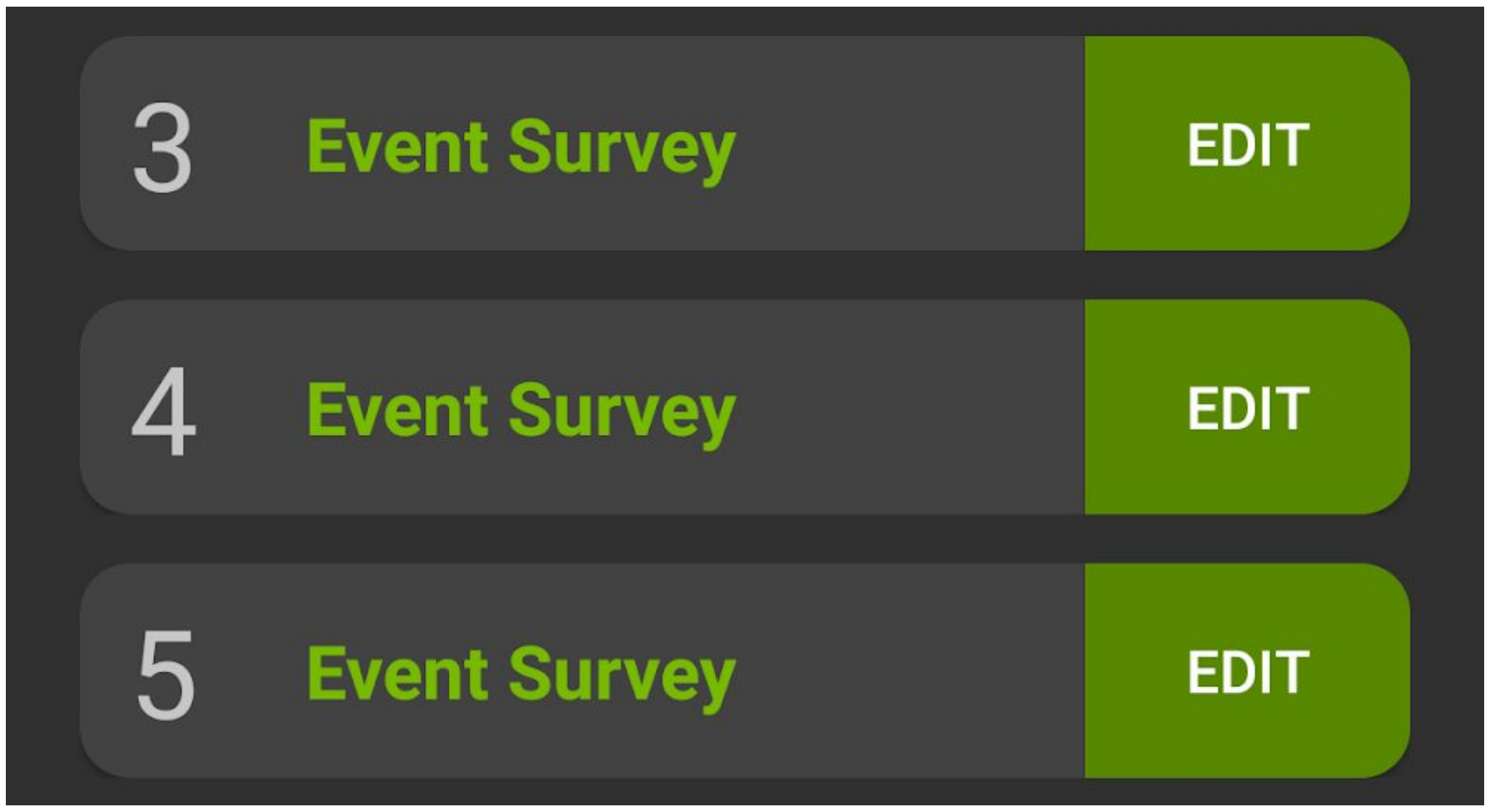

*Figure 26: Event Survey Cards*

After editing the Event Survey, users have the ability to save it again, which preserves the updated progress for the Event Survey in the same associated Event Survey Card. The users also have the ability to discard or submit the Event Survey, both of which result in the associated Event Survey Card being removed from the Session tab. This feature enhances the user interactivity of IncidentUI$_{droid}$ by developing an intuitive user interface that gives the users more control over the ride comfort data collection process.

### Feature 3: Event Comfort Feedback Console

The enhanced portability of the Android Tablet deploying IncidentUI$_{droid}$ and the cross-platform compatibility of the Android environment that allows it to support multiple devices prompted the development of this feature, which involved implementing an Event Comfort Feedback Console, as depicted in Figure 27.

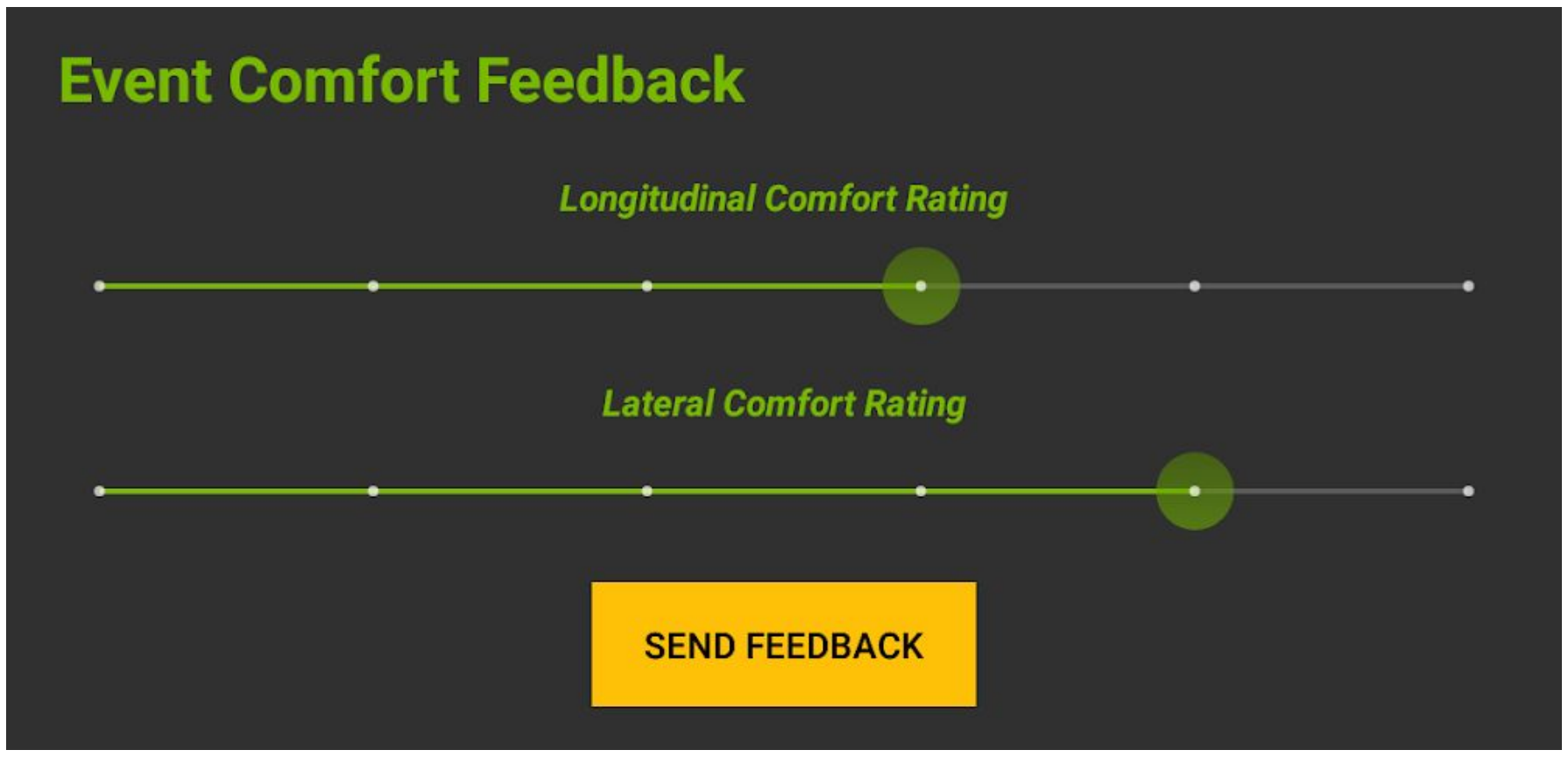

*Figure 27: Event Comfort Feedback Console*

The Event Comfort Feedback Console allows all the passengers in the Ego Vehicle to submit the Longitudinal and Lateral ride comfort rating for an Event generated by the co-pilot. This additional feedback improves the quality of the ride comfort data as it utilizes multifaceted responses from people with varying ride experiences and discomfort perceptions occupying different vehicle seating positions to generate a more comprehensive and well-rounded ride comfort rating for the event. The variations in the ride comfort ratings for different seating positions provide more positional context to the lateral and longitudinal ride discomfort experienced during an event. The multiple varying ride comfort feedback ratings also help eliminate outliers and erroneous responses. Similar to the Event Survey, the Longitudinal and Lateral ride comforts for the event are rated on a scale of 0 to 5. Clicking the "*Send Feedback*" button generates and sends an Event flag and an Event Survey message to Roadrunner with just the Longitudinal and Lateral comfort ratings filled out. This feature enhances the data collection capabilities of IncidentUI$_{droid}$ by developing an intuitive UI that improves the quality of ride comfort data and gives the users more control over the ride comfort data collection process.

### Feature 4: Fill out Pending Event Surveys after End of Session

The development of Feature 2 generated an edge case scenario where a session ends before the user has submitted or discarded all of the pending Event Surveys. During normal operation, a session ends when Roadrunner disconnects from IncidentUI$_{droid}$, which automatically dismisses the Session screen and loads the Dashboard screen. However, in this particular edge case, the demand for the ability to discard or submit an edited pending Event survey after the end of the session, prompted the development of a feature that delays the loading of the Dashboard screen until after all of the pending Event Surveys have been submitted or discarded.

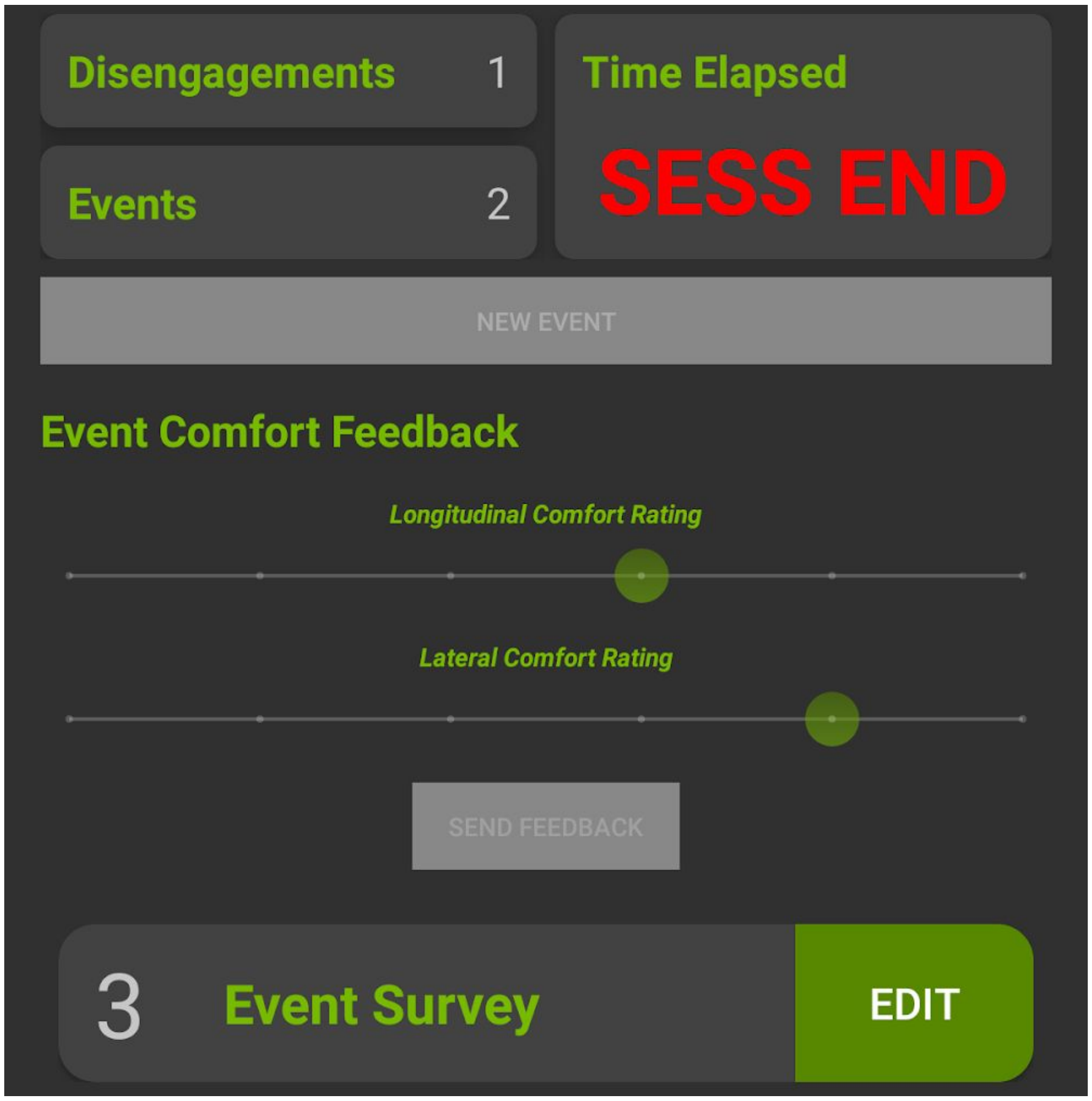

*Figure 28: Pending Event Survey after End of Session*

This feature, as depicted in Figure 28, involved the design and development of a user interface scheme and a conditional screen flow. When a session ends before the user has submitted or discarded all of the pending Event Surveys, this feature modifies the User Interface of the Session screen by displaying “SESS END” and disabling the UI elements that generate a new event or ride comfort feedback. This feature, instead of dismissing the Session screen and launching the Dashboard screen automatically, utilizes the UI elements to indicate the end of the session. If a session ends according to this scenario, a flag is raised to indicate the presence of some pending Event Surveys. Once all of the pending Event Surveys are submitted or discarded, the Session screen is dismissed and the Dashboard screen is launched automatically. This feature enhances the user interactivity of $\text{IncidentUI}_{\text{droid}}$ by implementing an automated conditional screen flow design and utilizing an intuitive user interface scheme for a common scenario.

### 3.6.3 $\text{IncidentUI}_{\text{droid}}$ Hyperion Kit Autonomous Vehicle Test Drive

With the $\text{IncidentUI}_{\text{droid}}$ stable release developed, deployed, and evaluated, the final step in the development of $\text{IncidentUI}_{\text{droid}}$ was initiated. This step involved an AV test drive with the $\text{IncidentUI}_{\text{droid}}$ stable release functioning as a component of the Hyperion Kit in-car architecture. For this test, the Nvidia Shield Tablet running $\text{IncidentUI}_{\text{droid}}$ was integrated with the Pegasus, which is a component of the Hyperion architecture deployed in Nvidia's Ford Fusion Test Autonomous Vehicle. The Pegasus was connected to the Nvidia Shield Tablet via USB Tethering, and modifications were made to the hardware network configurations for the Pegasus and the Nvidia Shield Tablet to integrate the tablet as a part of the local static in-car network and enable communication between Roadrunner and $\text{IncidentUI}_{\text{droid}}$ using the USB Tethering hardware network solution designed previously.

With hardware and network integration achieved, the $\text{IncidentUI}_{\text{droid}}$ in-car testing was initiated. The Test AV was manually driven to Nvidia's private test track where some test laps were conducted to calibrate the perception sensors and the AI computing platform. With the calibration completed and the Pilot and Co-pilot Login information filled out on $\text{IncidentUI}_{\text{droid}}$, Roadrunner was initiated, which immediately started exchanging heartbeat messages with $\text{IncidentUI}_{\text{droid}}$. The exchange of heartbeats indicated an active Roadrunner connection and started the test session by launching the Session screen on $\text{IncidentUI}_{\text{droid}}$. With the session initiated, the pilot engaged the car in self-driving mode, where the actuators assumed control of the test vehicle. In self-driving mode, the AV accelerated, braked, performed lane changes, and turn left and right on the test track.

Next, the $\text{IncidentUI}_{\text{droid}}$ event and disengagement handling systems were tested. When the pilot disengaged the AV, a Disengagement Survey was displayed on the $\text{IncidentUI}_{\text{droid}}$ screen. The AV was not allowed to re-engage until the Disengagement Survey was filled out and submitted. Once the Disengagement Survey was received by Roadrunner, its receipt was acknowledged, the actuators were released, and the AV was allowed to re-engage. The disengagement testing was done multiple times using varying disengagement techniques like manual steering wheel control and manual braking with different longitudinal and lateral sequence numbers to ensure that $\text{IncidentUI}_{\text{droid}}$ functions predictably in every possible scenario. In addition to vigorous disengagement testing, event testing was conducted by generating a new event using the "New Event" button on the Session Tab of $\text{IncidentUI}_{\text{droid}}$. The successful generation of a new event was acknowledged by the display of an Event Survey on the $\text{IncidentUI}_{\text{droid}}$ screen and receipt of an Event Flag by Roadrunner. Once the Event Survey was filled out and submitted, its receipt was also acknowledged by Roadrunner.

Throughout the testing, Roadrunner was disconnected and restarted multiple times to test the integration strength of new architecture design for $\text{IncidentUI}_{\text{droid}}$ and the robustness of $\text{Roadcast}_{\text{Java}}$. The final phase of the evaluation involved testing the UI flow, application functionality, and additional features of $\text{IncidentUI}_{\text{droid}}$, including the four new features developed for the stable release. The stable release for $\text{IncidentUI}_{\text{droid}}$ integrated effortlessly with the Test AV Hyperion architecture. $\text{IncidentUI}_{\text{droid}}$ also functioned faultlessly during the AV test drive and replicated the functionality of IncidentUI accurately whilst greatly improving the user experience, robustness, portability, and compatibility of the original application. This test concluded the project and the development of the $\text{IncidentUI}_{\text{droid}}$ application.

# 4. Results

After eight weeks of design, development, and evaluation of multiple iterations of $IncidentUI_{droid}$, we implemented the final stable release for $IncidentUI_{droid}$. The final stable release is deployed on an Nvidia Shield Android Tablet and utilizes a Java-based system sub-architecture with $Roadcast_{Java}$ and Java implementations of Protobufs. $IncidentUI_{droid}$ greatly improves upon many aspects of IncidentUI with one of the major facets being the enhanced user interactivity of $IncidentUI_{droid}$ that comes with the design of an intuitive user interface. Owing to the multitude of support libraries that come bundled with Android and a comprehensive and intuitive UI development environment included with Android Studio, $IncidentUI_{droid}$ implements a fluid, intuitive, versatile, and readily-modifiable user interface that enhances the AV ride comfort data collection experience.

Another facet of improvement for the $IncidentUI_{droid}$ is its enhanced portability owing to the deployment of $IncidentUI_{droid}$ on an Android tablet, which is in contrast to the fixed in-console Raspberry Pi used for deploying IncidentUI. The Android Tablet running $IncidentUI_{droid}$ is integrated with the in-car Hyperion system architecture and network infrastructure by utilizing one of the previously mentioned hardware network solutions, i.e. USB Tethering, Ethernet LAN, and WiFi. The Android Tablet, owing to its portability, can be passed on to other passengers in the AV which allows multiple passengers to interact with $IncidentUI_{droid}$ and contribute to the ride comfort data collection process. Since the tablet is not fixed to the console and the user interface for $IncidentUI_{droid}$ has been designed to support multiple screen orientations, the tablet can be held and used in any desired orientation, which simplifies the ride comfort data collection experience. Contrary to the IncidentUI deployed on the Raspberry Pi, $IncidentUI_{droid}$ provides support for ride comfort feedback from multiple users. The enhanced cross-platform compatibility of the Android environment and the hardware and software capabilities of an Android tablet provide support for pairing other Android devices to $IncidentUI_{droid}$ to give additional context and feedback and consequently, improve the quality of the ride comfort and disengagement data compiled during a test drive. Since the tablet is not fixed to the console and can be conveniently decoupled from the in-car Hyperion system architecture, the users have the option to readily swap the tablet in and out of the AV. For instance, the tablet running $IncidentUI_{droid}$ can be decoupled from the AV after a test drive and taken away for further analysis of the ride comfort data compiled during the test drive.

Finally, one of the most powerful aspects of $IncidentUI_{droid}$ is the ability to modify and augment the Android front end implementation and the Java-based system sub-architecture for $IncidentUI_{droid}$. The deployment of $IncidentUI_{droid}$ in an Android environment using Android Studio and the redevelopment of the system sub-architecture in pure Java provide a plethora of system and support libraries that support a flexible and robust feature development process. IncidentUI$_{droid}$ also supports prompt modifications to the system sub-architecture, user interface, and application functionality.

In addition to the aforementioned aspects of IncidentUI$_{droid}$ and the extent of their improvement over IncidentUI, this section details the user interface design, system architecture implementation, and the screen and data flow schemes for the final stable release of IncidentUI$_{droid}$.

## 4.1 IncidentUI$_{droid}$ Front End User Interface Design

The user interface for IncidentUI$_{droid}$ went through multiple iterations of design and development with each iteration built upon the previous one. The feedback received from IncidentUI$_{droid}$ demonstrations and testing was utilized to develop new UI elements and enhance the existing ones. The final UI consists of three main screens: Initiate Drive, Dashboard, and Session, with multiple tabs within each screen. The UI is designed and developed using Android XML Layouts, and the functionality of the UI elements is coded in Java. This section elucidates the design and functionality of the final iteration of the user interface and the screen flow scheme implemented for the final IncidentUI$_{droid}$ stable release. Every UI screen and fragment for the IncidentUI$_{droid}$ stable release has been designed and configured to support multiple screen orientations to enhance the user interactivity of IncidentUI$_{droid}$ by enabling effortless application use for any tablet orientation.

### 4.1.1 Initiate Drive Screen

The first screen that is displayed when the application is launched after getting into the ego vehicle is the Initiate Drive screen. A screenshot of the screen is shown in Figure 29. The screen has text fields to enter the Pilot's and Co-pilot's Nvidia IDs. It also has a button "*Start Drive*" to initiate the drive and launch the dashboard screen. The pilot and copilot information are collected for each drive and used to relate AV disengagements, events, ride comfort, session information, and other test data to the test users. The final iteration of the Initiate Drive screen includes the logo for the IncidentUI$_{droid}$ application. The logo along with the other Initiate Drive screen UI elements have been reoriented to support seamless UI function for multiple screen orientations.

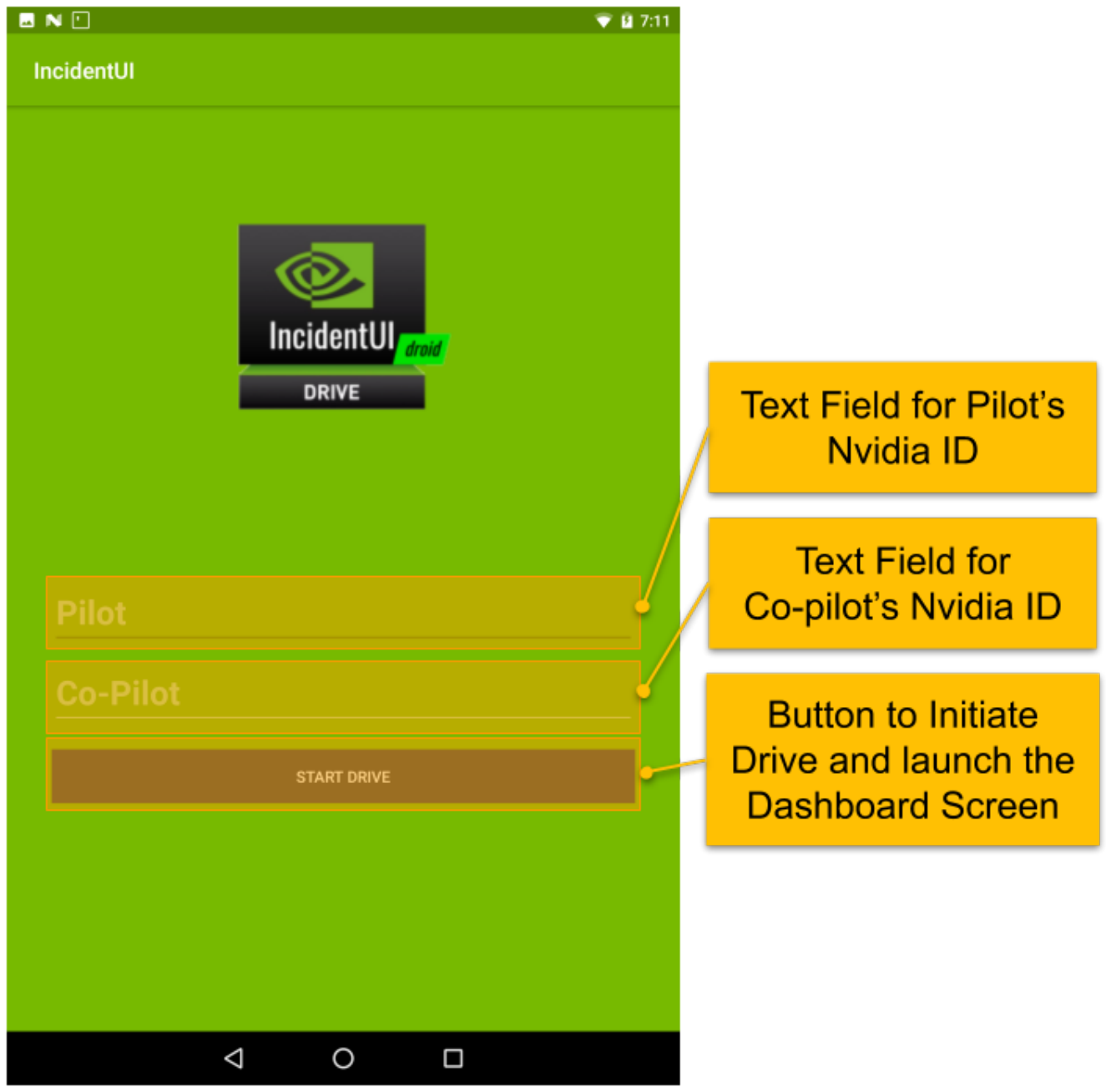

*Figure 29: Initiate Drive Screen*

### 4.1.2 Dashboard Screen

The dashboard screen is displayed when a drive is initiated and in between consecutive active sessions. It has the following three tabs under it.

#### Drive History Tab

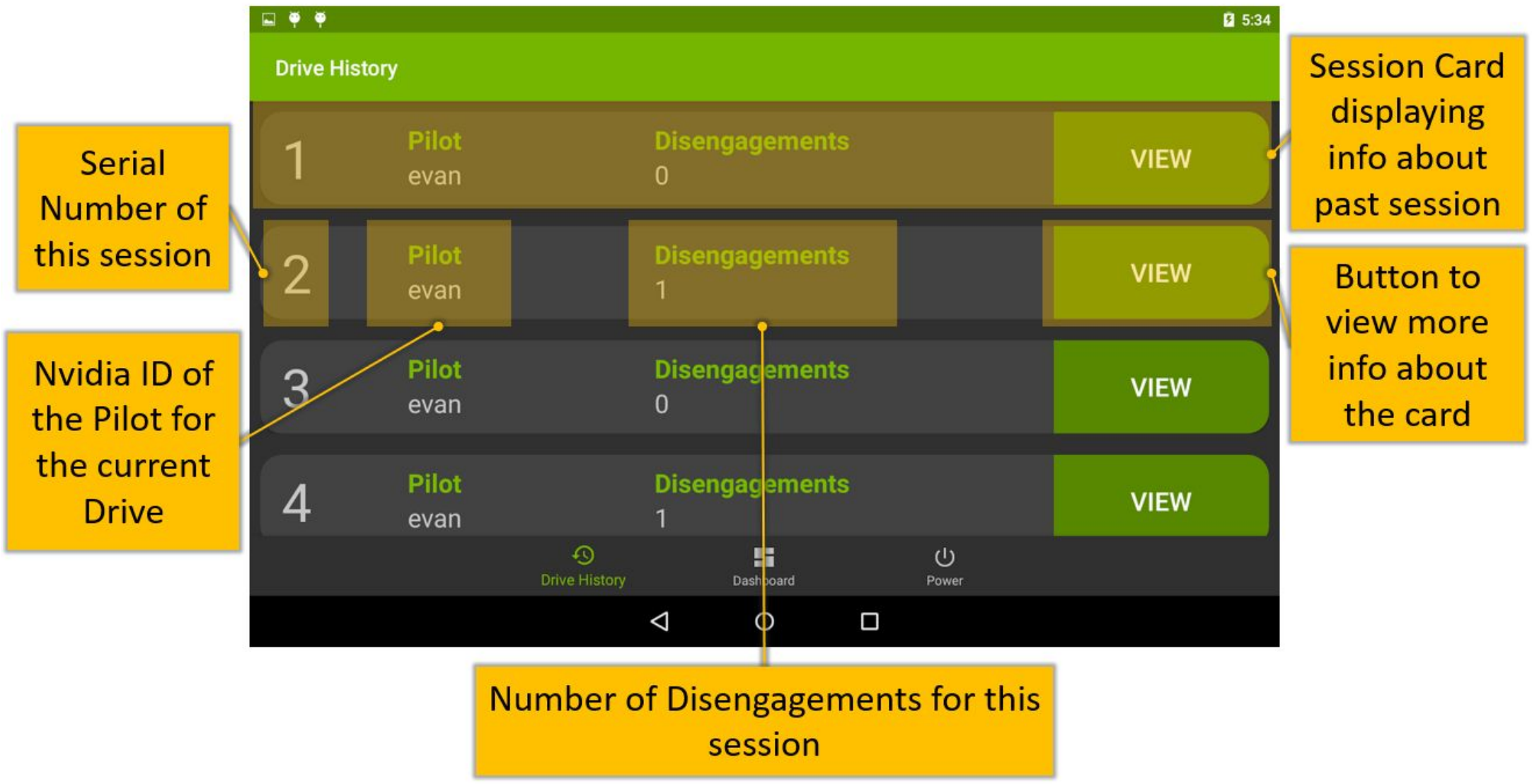

*Figure 30: Drive History Tab*

The Drive History Tab, as shown in Figure 30, stores information about every past session after it has ended. This tab generates a Session Card for every past session and contains information about that session. These cards appear in a scrollable vertical list in this tab. This tab is also available in the Session Screen, allowing the user to view the Drive History during an active session. This tab contains the cards for previous sessions. Each Session Card has the sequence number for the session associated with it. It also displays the Nvidia ID of the Pilot for the current Drive and the number of Disengagements for the associated session.

There is a "*View*" button on each Session Card that, on being clicked, brings up a dialog window that displays more information about that session, as shown in Figure 31. This Session Information dialog window displays the Pilot's and Co-pilot's Nvidia IDs for the current drive. It also displays the number of disengagement and events generated during that session. The dialog window also shows the Session Duration, i.e. the length of time that session lasted. At the bottom of the dialog window, there is a "*Back*" button that dismisses this Session Information dialog and goes back to the Drive History tab.

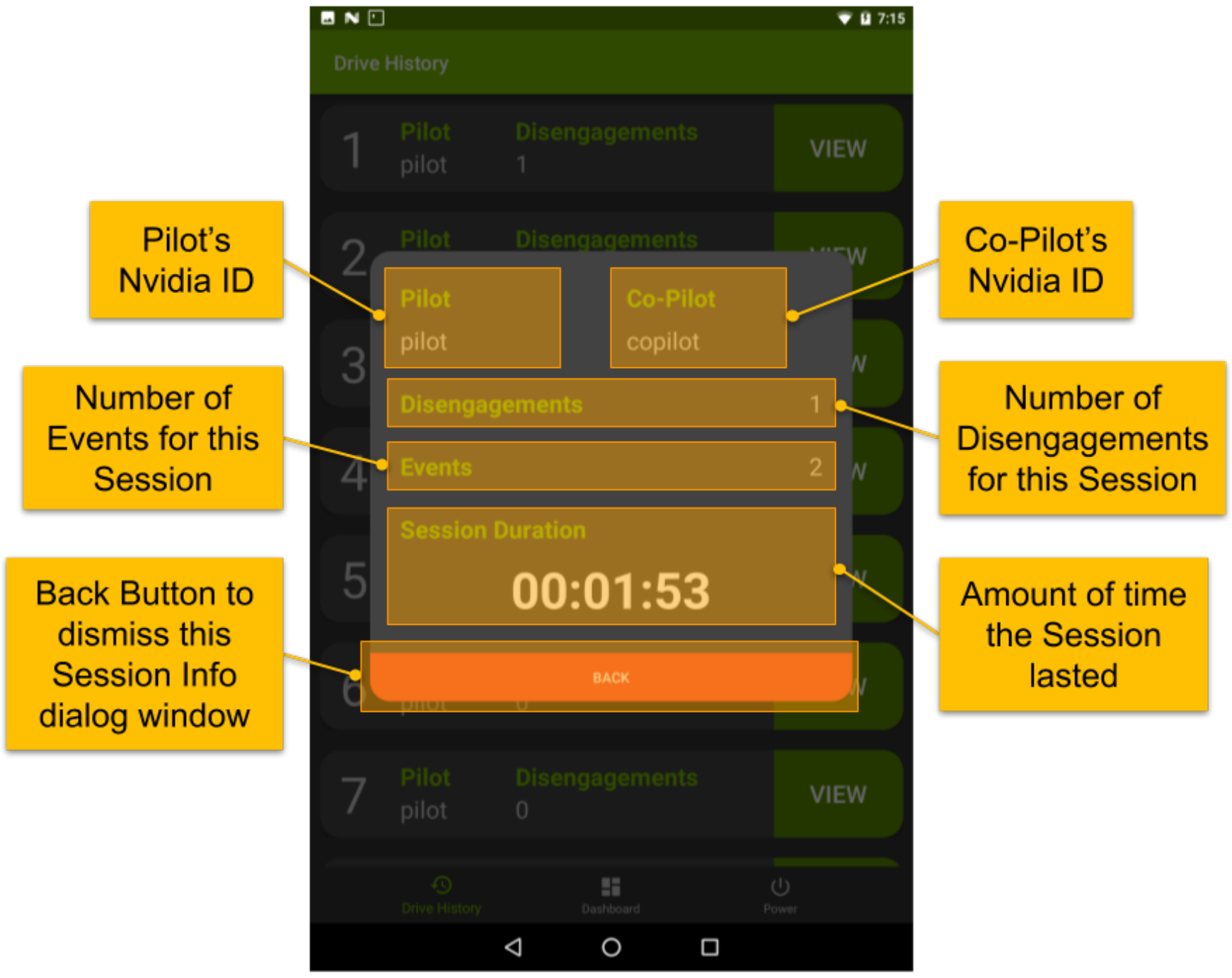

*Figure 31: Session Information Window*

### Dashboard Tab

The dashboard tab, as shown in Figure 32, is displayed by default whenever the dashboard screen is launched. This tab displays the Nvidia IDs of the Pilot and the Co-pilot on the current drive. It has an "*End Drive*" Button that terminates the current drive and takes the user back to the Initiate Drive Screen. At the top right corner of the screen, there is an indicator that shows the current Roadrunner connection status, i.e. if the client running on IncidentUI$_{droid}$ detects a heartbeat from Roadrunner or not.

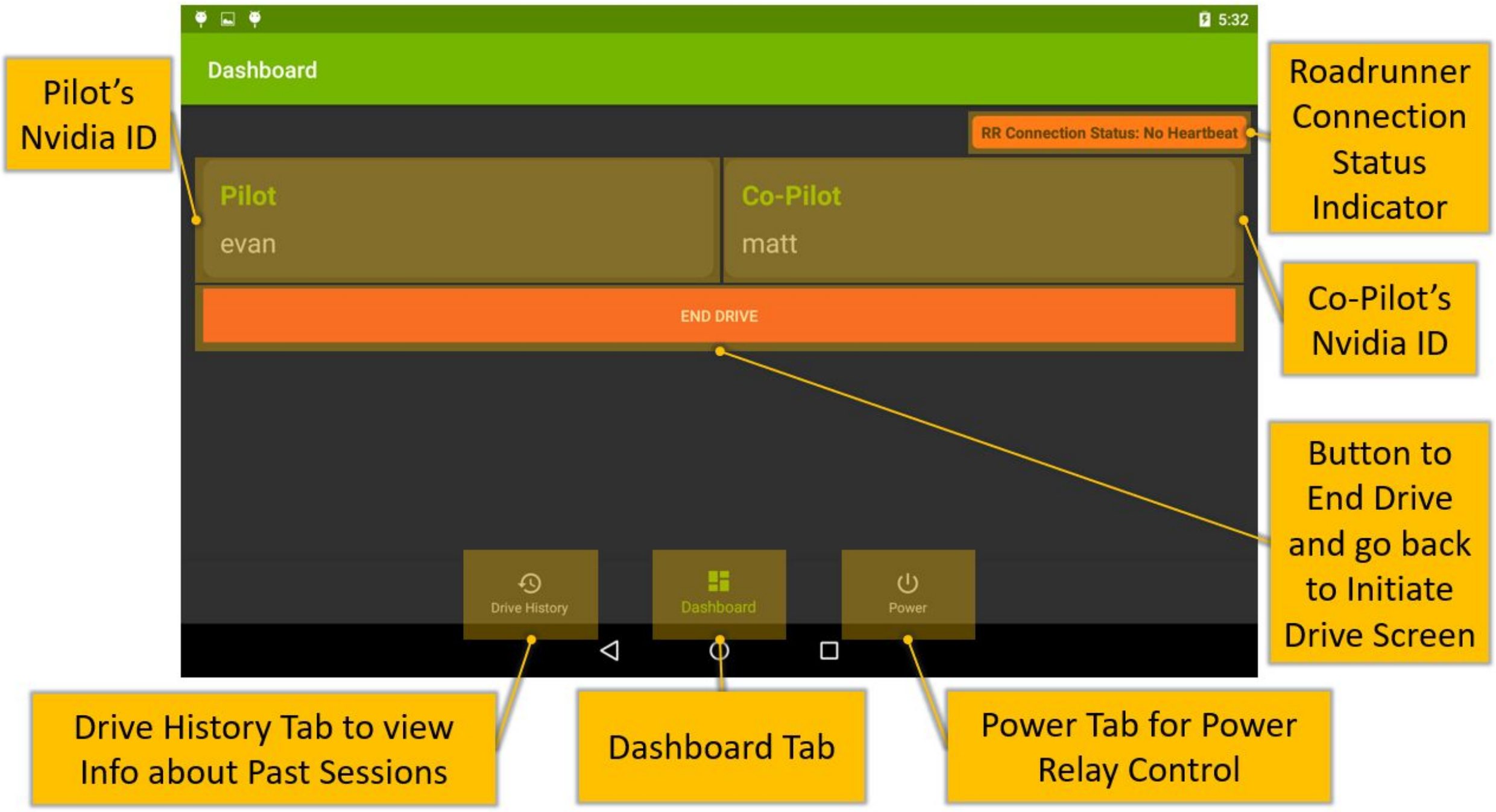

*Figure 32: Dashboard Tab*

At the bottom of the screen, there is a navigation menu that allows users to switch between the three tabs in the dashboard screen, i.e. Drive History Tab, Dashboard Tab, and Power Tab.

#### Power Tab

The power tab is currently empty but has been included as a tab in the UI in case the application is further developed to include the power relay control functionality, thereby completely eliminating the need for the Raspberry Pi in the Hyperion system architecture.

### 4.1.3 Session Screen

During an active session, the session screen is displayed. The ability to initiate or end a session is not under the control of the user; instead, it is controlled by the status of the Roadrunner connection, i.e. the moment Roadrunner connects, a session is initiated, and a session ends when Roadrunner is disconnected. The session screen has the following two tabs:

### Drive History Tab

The Drive History tab is also available in the Session screen to view information about past sessions during an active session. The User Interface elements of this tab and their functions are the same as explained previously for the Dashboard screen.

### Session Tab

The session tab, as shown in Figure 33, is the main tab that is displayed during an active session. It displays the current Pilot's and Co-pilot's Nvidia IDs, along with the number of disengagements and events encountered in this session. There is a "*New Event*" button that records a new event by bringing up the Event Survey. The number of events and disengagements are updated in the session tab in real-time. There is also a chronometer that tracks the amount of time that has elapsed since the current session started. At the top right corner of the screen, there is an indicator that shows the current Roadrunner connection status, i.e. if the client running on $\text{IncidentUI}_{\text{droid}}$ detects a heartbeat from Roadrunner or not.

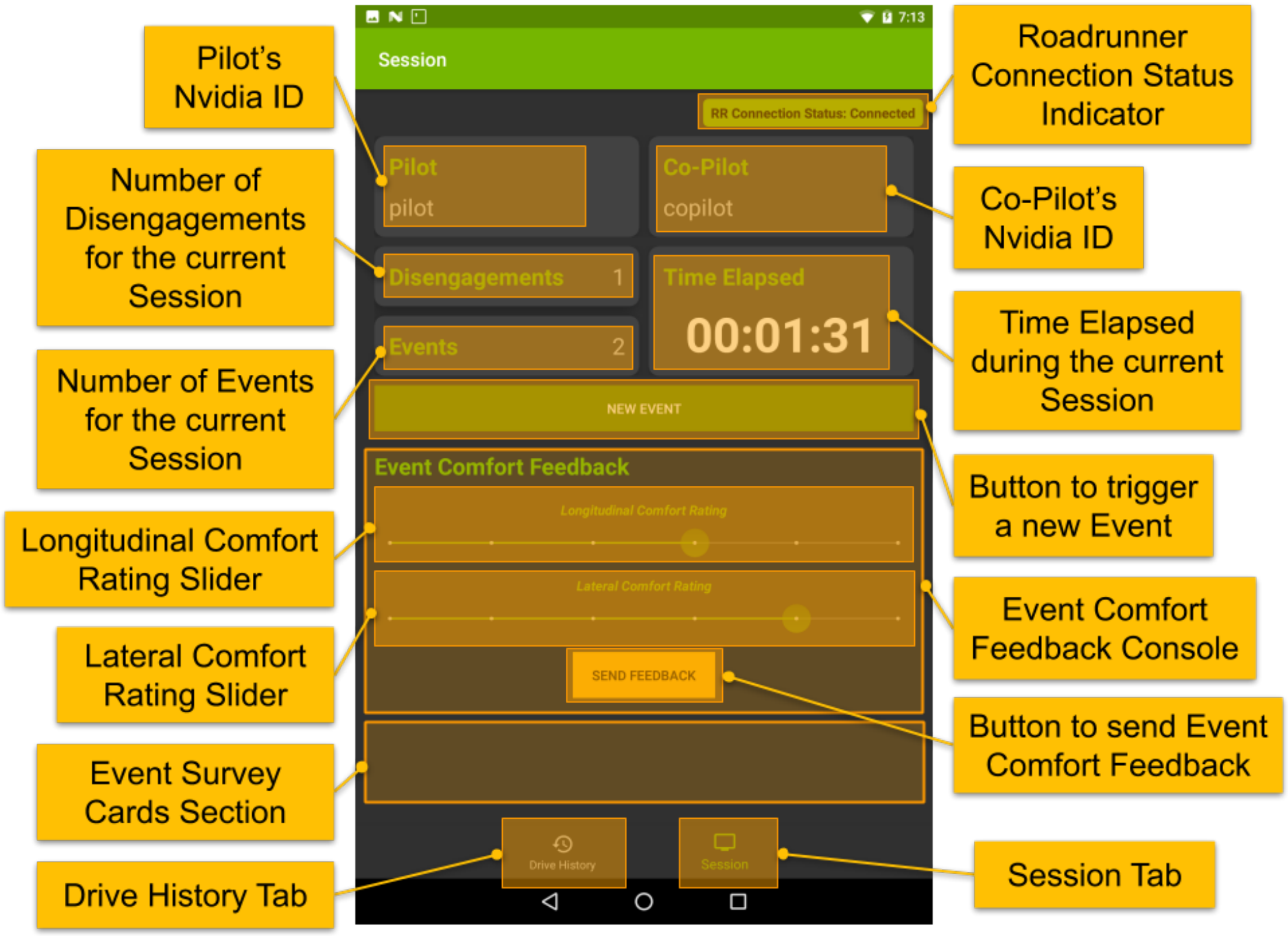

*Figure 33: Session Tab*

The Session Tab also includes an Event Comfort Feedback Console that allows all the passengers in the Ego Vehicle to submit the Longitudinal and Lateral ride comfort rating for an Event generated by the co-pilot. Similar to the Event Survey, the Longitudinal and Lateral ride comforts for the event are rated on a scale of 0 to 5. Clicking the "*Send Feedback*" button generates and sends an Event flag and an Event Survey message to Roadrunner with just the Longitudinal and Lateral comfort ratings filled out. This feature enhances the data collection capabilities of IncidentUI$_{droid}$ by developing an intuitive UI that improves the quality of ride comfort data and gives the users more control over the ride comfort data collection process.

Below the Event Comfort Feedback Console, there is space for a scrollable list of Event Survey Cards, where each card is associated with a pending Event Survey. When an Event Survey is "*Saved*", its progress is saved and a card associated with that Event Survey is created and added to the Event Survey Cards section in the Session tab, as portrayed in Figure 34.

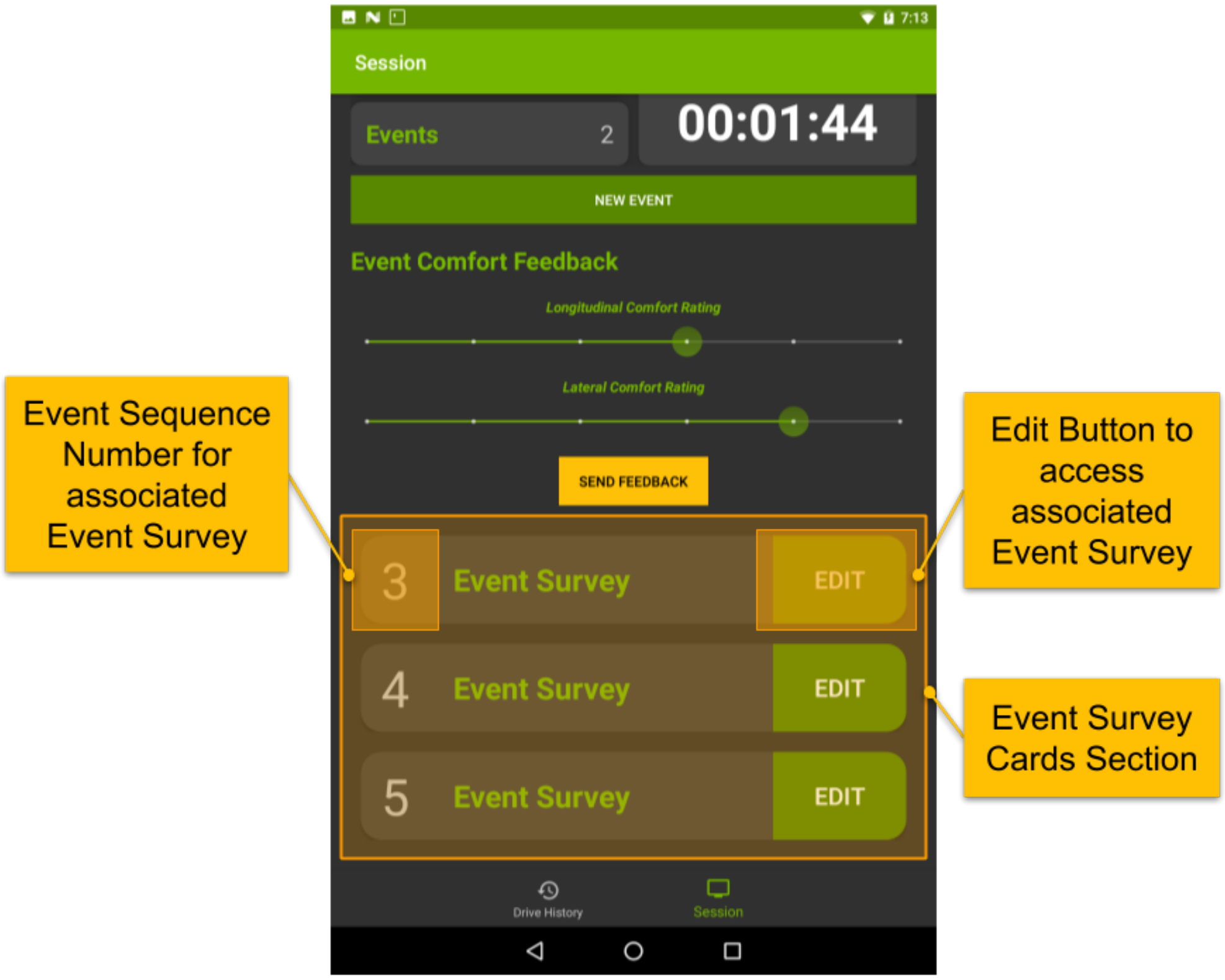

*Figure 34: Event Survey Cards Section*

Users are allowed to create and save multiple Event Surveys, each of which is stored as an Event Survey Card in a scrollable list. The Event Survey Cards, as shown in Figure 34, display the Event Sequence Numbers for the Event Surveys they are associated with. These cards also include an "*Edit*" button, which can be clicked to access the associated Event Survey filled out with the previously saved progress. Editing an Event Survey and saving it again preserves the updated progress for the Event Survey in the same associated Event Survey Card. The users also have the ability to discard or submit the Event Survey, both of which result in the associated Event Survey Card being removed from the Session tab.

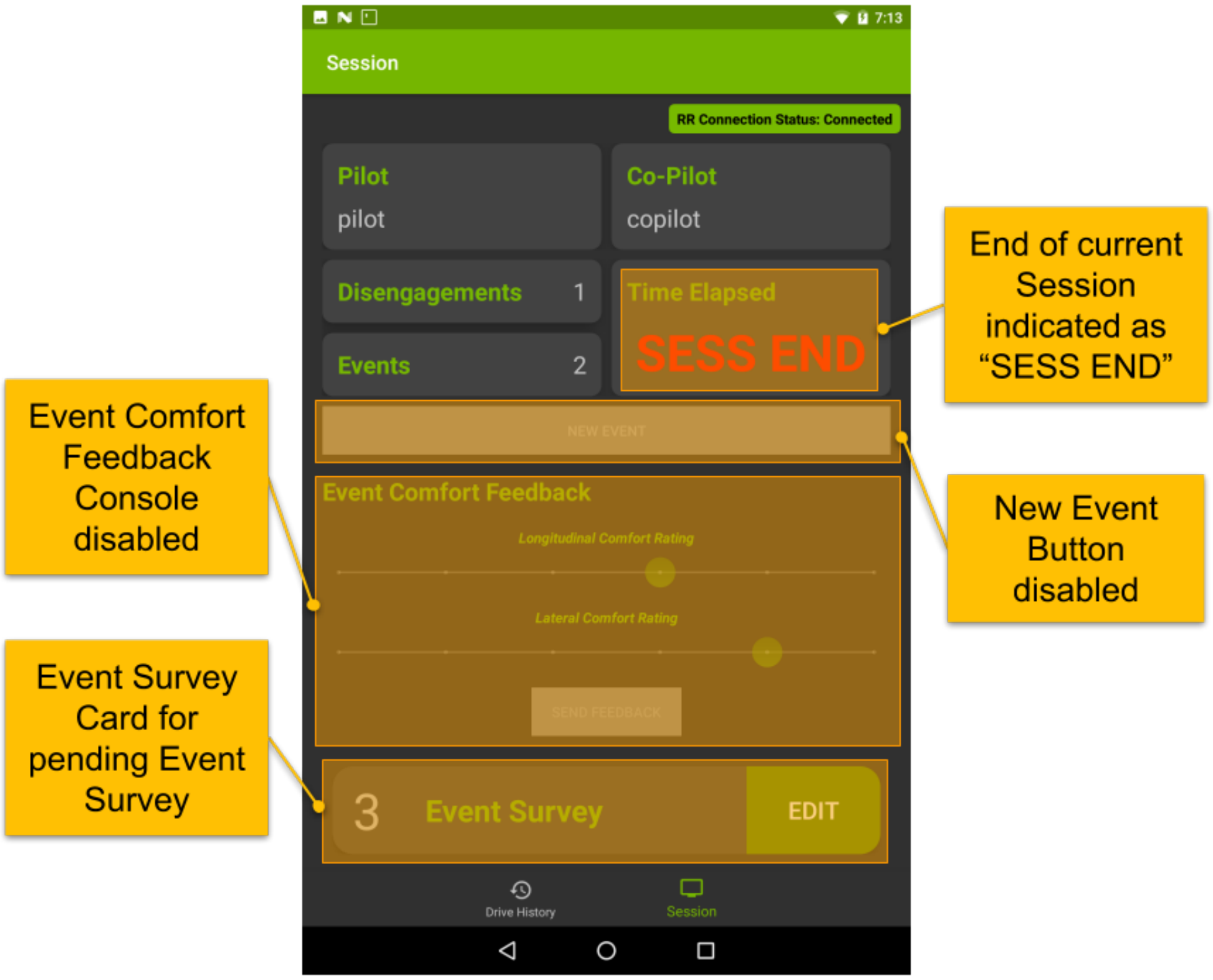

*Figure 35: Session Tab after End of Session with Pending Event Survey*

When a session ends before the user has submitted or discarded all of the pending Event Surveys, the User Interface of the Session screen is modified as shown in Figure 35. The end of the session is indicated by displaying "SESS END" in the Time Elapsed window and disabling the UI elements that generate a new event or ride comfort feedback data, i.e. the New Event Button and the Event Comfort Feedback Console respectively. This application feature, instead of dismissing the Session screen and launching the Dashboard screen automatically, utilizes the UI elements to indicate the end of the session. Once all of the pending Event Surveys are submitted or discarded, the Session screen is dismissed and the Dashboard screen is launched automatically.

The disengagement survey is brought up automatically on the screen when Roadrunner detects an ego vehicle disengagement and sends a disengagement survey request to $\text{IncidentUI}_{\text{droid}}$. At the bottom of the screen, there is a navigation menu that allows users to switch between the two tabs in the session screen, i.e. Drive History Tab and Session Tab.

### 4.1.4 Surveys

The main function of IncidentUI is to track events and disengagements and their causes using surveys. They make use of pre-existing parameters to measure the comfort of the ego vehicle's maneuvers across 2 axes - lateral and longitudinal and use these parameters or other additional information to report the reasoning for the lateral/longitudinal comfort rating assigned by the co-pilot for the event/disengagement. These surveys are displayed and can be filled out only while a session is active, i.e. Roadrunner connection is present. The survey responses are sent to the Roadrunner using Roadcast and then stored in the Pegasus. There are two kinds of surveys used for recording data:

#### Event Survey

The Event Survey is used to track an event. An event is recognized as a scenario where the ego vehicle's maneuver causes lateral and/or longitudinal discomfort without requiring a disengagement from the pilot. Whenever the ego vehicle does a maneuver that the co-pilot believes violates the comfort parameters, they can generate an event. To do so, they can click on the "New Event" on the Session tab, which brings up the "Event Survey" as shown in Figure 36. Ride comfort is measured across 2 axes using sliders on a scale of 0 to 5. Checkboxes let the user choose potential causes for Lateral and Longitudinal Discomfort, with the option to select multiple causes simultaneously. The possible causes of Longitudinal Discomfort are as follows:

- Collision Threat
- Too Fast
- Too Slow
- Jerky Acceleration
- Jerky Break
- False Breaking

The causes of Lateral Discomfort are as follows:

- Collision Threat
- Swerve
- Lateral Jerk
- Too Aggressive
- Too Conservative

The Event Survey also allows the Co-pilot to define the position of the Ego-vehicle at the time of the discomfort event to give locational context to the cause of the event. This locational context improves the quality of the event data, which can then be used to enhance the Drive Software further. The Ego-Vehicle's position is selected from radio buttons for the following options:

- Lane Keep
- Merge
- Lane Change
- Split
- Ramp

The Event Survey also allows the user to provide additional information related to the event that is not covered by the predefined options. The user (usually the co-pilot) can fill out the "*Additional Information*" in a dedicated text box.

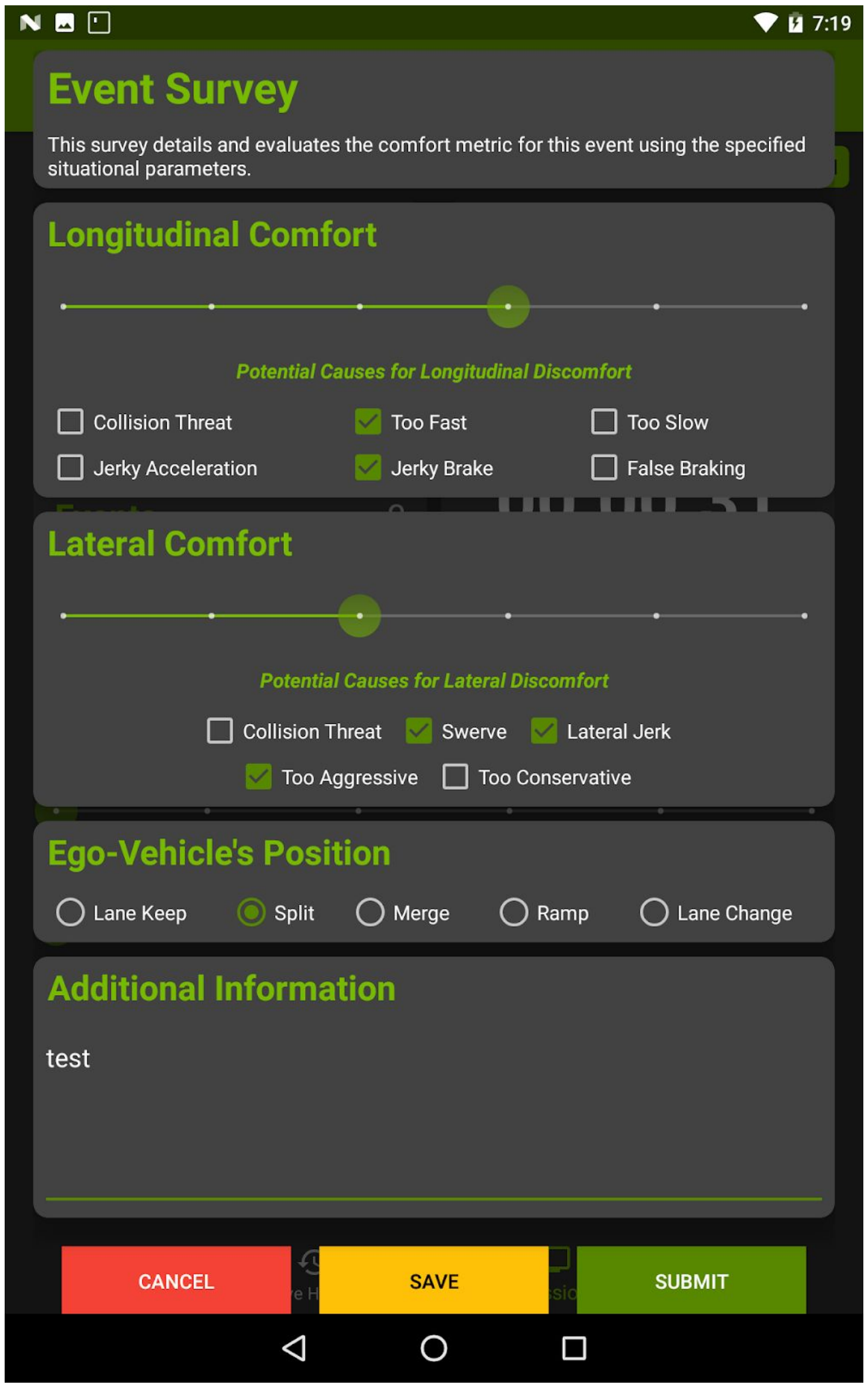

*Figure 36: Event Survey*

Once the Event Survey has been filled out, users have the ability to submit it and send it to Roadrunner. Users can discard the Event Survey by clicking on the "Cancel" button in case the corresponding event was generated unintentionally. The Event Survey also has a "*Save*" button, which on getting clicked saves a partially filled out Event Survey so that it can be edited, discarded, or submitted later. Clicking the "*Save*" button preserves the progress for an Event Survey and creates a card associated with that Event Survey in the Session tab.

### Disengagement Survey

The Disengagement survey, as shown in Figure 37, is used to record disengagement data, i.e. lateral and longitudinal ride comfort, cause of the disengagement, the explanations for the cause of the disengagement, and any additional information describing the disengagement. In case of an ego vehicle disengagement, Roadrunner sends a disengagement survey request to IncidentUI$_{droid}$ using Roadcast, and consequently, the disengagement survey is displayed on the tablet screen.

The ride comfort is measured across 2 axes: longitudinal and lateral, and can be selected using sliders from a range of 0 to 5, where 0 is least comfortable and 5 is most comfortable.

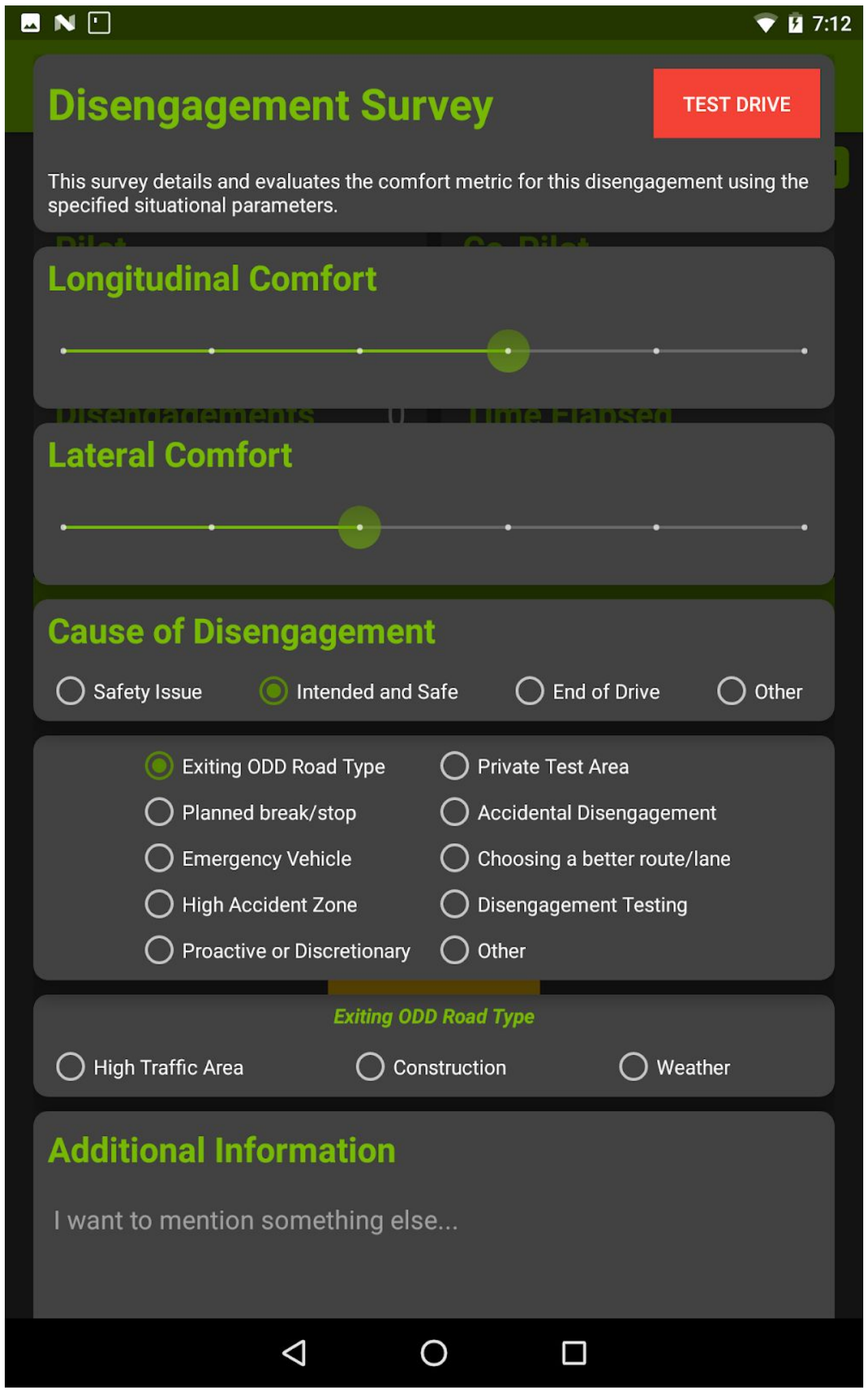

*Figure 37: Disengagement Survey*

The Cause of Disengagement can be selected from four possible options using Radio Buttons. The possible causes of disengagement are as follows :

- Safety Issue
- Intended and safe
- End of Drive
- Other

If the cause of the disengagement is some safety issue caused by the nature of the maneuver performed by the ego vehicle that forced the pilot to take control and trigger a disengagement, then the "*Safety Issue*" option is selected as the Cause of Disengagement on the Disengagement Survey. Selecting this option displays additional fields to provide explanations for the Safety Issue. These fields include potential causes for Lateral and Longitudinal Discomfort and the location of the Ego Vehicle at the time of the disengagement. The layout of the fields is identical to that of the Event Survey, as described in the previous section.

If the disengagement is triggered by the pilot intentionally because of varying reasons and it is not the result of a safety issue, i.e., the disengagement is the result of some reason external to the ego vehicle misbehaving, then the "*Intended and Safe*" option is selected as the Cause of Disengagement on the Disengagement Survey. Choosing the "*Intended and Safe*" option prompts the user to provide more context to the disengagement cause by selecting one of the following explanations using radio buttons:

- Exiting ODD Road Type
- Planned break/stop
- Emergency Vehicle
- High Accident Zone
- Proactive or Discretionary
- Private Test Area
- Accidental Disengagement
- Choosing a better route/lane
- Disengagement Testing
- Other

From these aforementioned options, selecting "*Exiting ODD Road Type*" prompts the user to provide more environmental context to the explanation for the disengagement like a High Traffic Area, Weather Conditions or, a Construction Zone. The rest of the options are self-explanatory and do not require additional context. If none of the predefined options give an accurate explanation for the "*Intended and Safe*" cause of disengagement, then the "*Other*" option can be selected, which displays a text box that can be used to provide an explanation for the disengagement that is not covered by the predefined options.

At the end of an AV drive, the pilot takes control of the ego vehicle to mark the end of the test session and the data collection process, which triggers a disengagement. This scenario can be recorded by selecting the "*End of Drive*" option as the Cause of Disengagement on the Disengagement Survey. If none of the aforementioned options accurately define the cause of disengagement, then the "*Other*" option can be selected on the Disengagement Survey, which displays a text box that can be used to provide a cause for the disengagement that is not covered by the predefined options. The Disengagement Survey also allows the user to provide additional information related to the disengagement that is not covered by the predefined options. The user (usually the co-pilot) can fill out the "*Additional Information*" in a dedicated text box.

During an AV test drive at Nvidia's private test track, multiple disengagements and events are triggered to test and tweak the Hyperion Software and Hardware. The Disengagement Survey has a "*Test Drive*" button on the top-right corner that automatically submits a disengagement survey with the following information filled out:

- Longitudinal Comfort: 0
- Lateral Comfort: 0
- Cause of Disengagement: Intended and Safe
- Explanation: Private Test Area
- Additional Information: None

This feature is set up to ensure that the test users are not bogged down by having to fill out the disengagement survey with duplicate information every time an intentional disengagement is triggered as part of the test session on the private test track.

## 4.2 IncidentUI$_{droid}$ Screen Flow Scheme

The user interface for IncidentUI$_{droid}$ consists of three main screens: Initiate Drive, Dashboard, and Session, with multiple tabs within each screen. Users can interact with the user interface by switching between the screens or modifying the UI elements. The screen flow scheme for IncidentUI$_{droid}$ is represented in Figure 38. The first screen that is displayed when the application is launched after getting into the ego vehicle is the Initiate Drive screen. The Initiate Drive screen has text boxes to fill out the Pilot's and Co-pilot's Nvidia IDs, and an "*Initiate Drive*" button, which on getting clicked initiates a new drive and launches the Dashboard Screen.

The Dashboard Screen has three Tabs: Dashboard Tab, Power Tab, and Drive History Tab, with the Dashboard Tab displayed by default. Each of these tabs can be displayed by selecting them from the Tab Navigator at the bottom of the Dashboard screen. The Drive History Tab consists of a scrollable list of Session Cards that contain information about the associated sessions. The session cards also have a "*View*" button, which on being clicked expands the corresponding session card and displays a Session Info dialog window with more information for the associated session. The expanded Session Card dialog window has a "*Back*" button, which on getting clicked dismisses the dialog window and displays the Drive History Tab.

When a Roadrunner connection is detected by IncidentUI$_{droid}$, the Session Screen is loaded automatically. The Session screen has two tabs: the Drive History Tab and the Session Tab, with the Session Tab displayed by default. Either of these tabs can be displayed by selecting them from the Tab Navigator at the bottom of the Session screen.

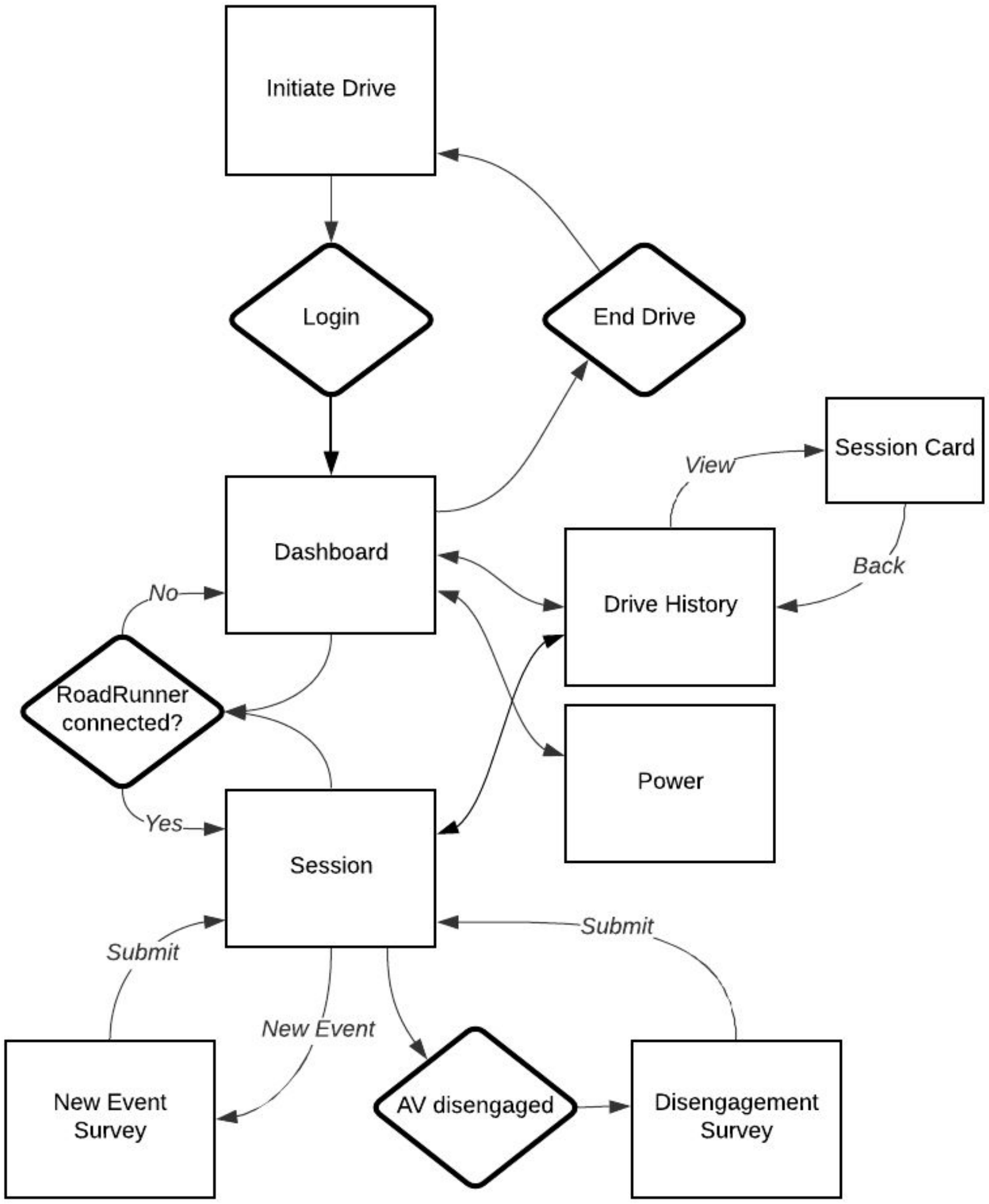

*Figure 38: IncidentUI$_{droid}$ Screen Flow Scheme*

The Drive History tab for the Session screen follows the same screen flow scheme as the one for the Dashboard screen. When an AV disengagement is detected by IncidentUI$_{droid}$, the Disengagement Survey is displayed on the screen as a pop-up dialog window. Once the Disengagement Survey has been filled out, it can be submitted or discarded by clicking the "*Submit*" and "*Cancel*" buttons respectively, both of which dismiss the survey and display the Session screen. The users can click the "*New Event*" button on the Session tab to trigger a new Event. Triggering a new Event displays the Event Survey on the screen. Similar to the Disengagement Survey, the Event Survey can be submitted or discarded by clicking the "*Submit*" and "*Cancel*" buttons respectively. The Event Survey can also be saved for editing later by clicking the "Save" button. Saving, discarding or, submitting the Event Survey dismisses the Event Survey dialog window and displays the Session screen.

When the active Roadrunner connection is terminated, the Session screen is dismissed and the Dashboard screen is launched again. If another Roadrunner connection is detected, a new session is initiated and the Session screen is launched. The Dashboard screen has an "*End Drive*" button, which, on getting clicked, ends the current test drive, and launches the Initiate Drive screen again. The aforementioned intuitive screen flow scheme is designed to enhance the user experience for IncidentUI$_{droid}$'s user interface and to implement a screen flow that complements and enhances IncidentUI$_{droid}$'s features.

## 4.3 IncidentUI$_{droid}$ System Architecture Design

The system architecture for IncidentUI$_{droid}$ went through multiple system integration design plans and numerous iterations of development with each iteration built upon the previous one. System integration testing using the Roadrunner Emulator was used to evaluate the efficacy of the system architecture design. The evaluation results were used to modify and enhance the system architecture design and implementation. The final system architecture design, as represented in Figure 39, is developed using Architecture Design Plan B, which involves the development of Roadcast$_{Java}$: a new communications protocol coded purely in Java. Roadcast$_{Java}$ replicates the functionality of the C++-coded Roadcast and utilizes Java implementations of Protobufs to serialize and deserialize data on the IncidentUI$_{droid}$ end. Similar to Roadcast, Roadcast$_{Java}$ implements a Client and a Server to regulate the exchange of data with Roadrunner. The final system architecture involves a Java-based IncidentUI$_{droid}$ system sub-architecture integrated with the Roadrunner system architecture.

The IncidentUI$_{droid}$ System Architecture consists of the Roadcast$_{Java}$ Client and Server that communicate with the Roadrunner Server and Client respectively. Roadcast$_{Java}$ utilizes Java implementations of Protobufs on the IncidentUI$_{droid}$ side, while Roadcast utilizes C++ implementations of the same Protobufs definitions on the Roadrunner end to maintain compatibility between the data structures and serialization schemes on either side of the system architecture. The Roadcast$_{Java}$ Client and Server run on two separate background worker threads and constantly listen for messages in a loop. The interactive user interface for IncidentUI$_{droid}$ runs on the main User Interface thread, which is independent of and runs parallel to the Client and Server background threads. The separate UI thread enables the user interface to remain interactive throughout the operation of the $Roadcast_{Java}$ Client and Server.

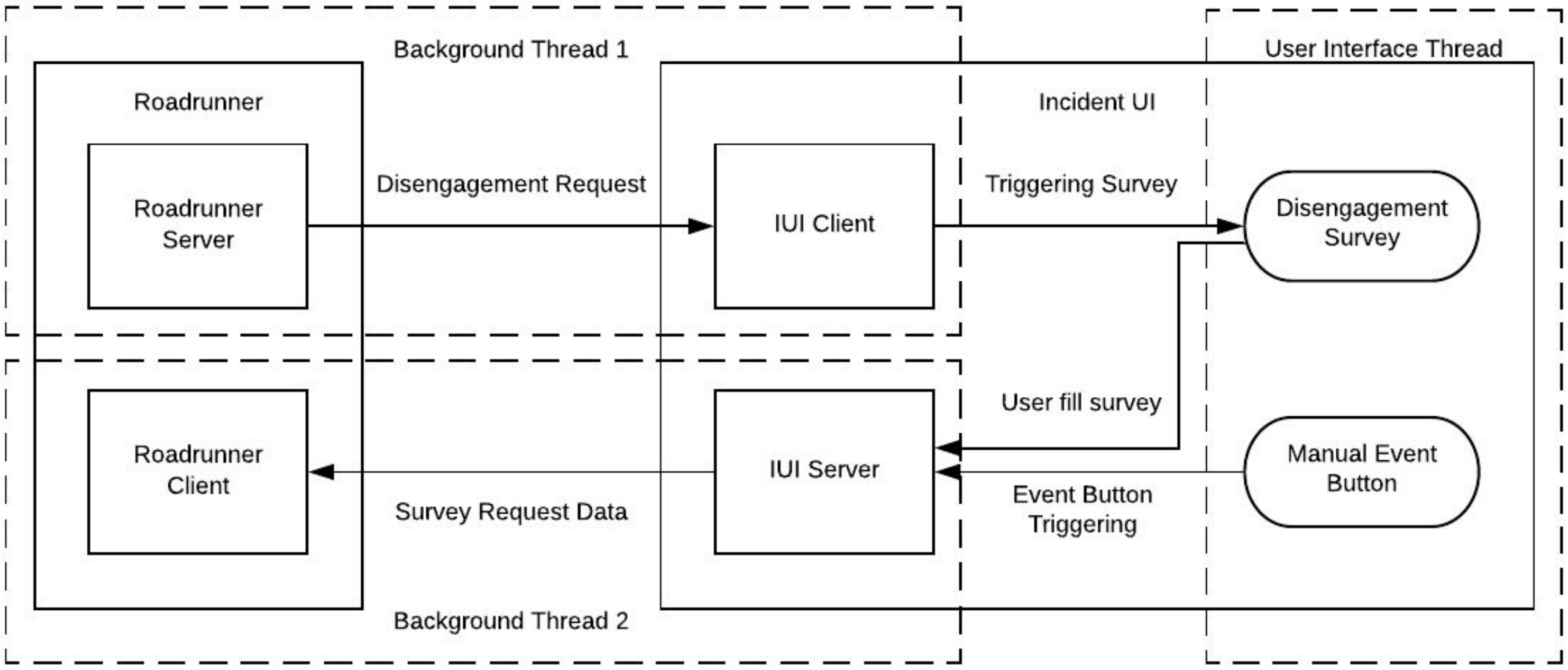

*Figure 39: $IncidentUI_{droid}$ System Architecture Design*

The $Roadcast_{Java}$ Client is responsible for receiving and processing survey requests and heartbeat messages sent by Roadrunner. The Client runs on a background worker thread as an "*IntentService*'' and listens for incoming data from Roadrunner in a loop. The $Roadcast_{Java}$ Client is configured to listen for messages from Roadrunner on a specific port and from a specific Server IP address that the Roadcast Server running on Roadrunner is bound to. The $Roadcast_{Java}$ Client, on receiving a message from the Roadcast Server running on Roadrunner, deserializes the message data directly into a Java object using the Java implementations of Protobufs. The Client then sends this object or a corresponding signal to the UI thread using a "*Handler*". This "*Handler*" is utilized by the Android front end to access the data sent by Roadrunner and execute UI changes accordingly.

$Roadcast_{Java}$ utilizes a server to generate and send survey data and heartbeat messages to Roadrunner. The $Roadcast_{Java}$ Server runs on another background worker thread as an "*IntentService*'' and sends generated messages to Roadrunner in a loop. The Server is bound to a specific IP address and configured to accept client connections on a specific port. The Server also utilizes this port to send serialized messages to the Roadcast Client running on Roadrunner. When a survey is filled out and submitted on the main UI thread, the survey data is packaged into a Java object. The Server then sends this object or a corresponding signal to the background worker thread using a "*Handler*". On receiving the Java survey object and/or the signal from the "*Handler*", the $\text{Roadcast}_{\text{Java}}$ Server utilizes the Java class serialization methods to serialize the Java object and generate a message, which is then transmitted to the Roadcast Client running on Roadrunner.

On application startup, the $\text{Roadcast}_{\text{Java}}$ client attempts to connect to the Roadcast server in a loop using a specific port and Server IP address. On a successful connection, the $\text{Roadcast}_{\text{Java}}$ client receives and processes the heartbeat messages sent by the Roadcast server. Similarly, the $\text{Roadcast}_{\text{Java}}$ server gets initialized and bound to a specific IP address and starts listening for client connections on a specific port. On a successful Roadcast client connection, the $\text{Roadcast}_{\text{Java}}$ server generates and sends heartbeat messages to the connected Roadcast client. After a successful and robust two-way connection is established between Roadrunner and $\text{IncidentUI}_{\text{droid}}$, varying survey request messages and corresponding survey response messages are sent back and forth during an active test session. At the start of every active session, a Login Survey Request is sent by Roadrunner and the corresponding Login Survey Response containing information about the current Pilot and Copilot is sent back by $\text{IncidentUI}_{\text{droid}}$. This exchange is followed by a series of Event and Disengagement survey requests and responses sent back and forth between Roadrunner and $\text{IncidentUI}_{\text{droid}}$.

### Disengagement Survey Flow

During an active Session, if Roadrunner detects an AV disengagement, the Roadrunner Server creates and sends a Disengagement Survey Request message to $\text{IncidentUI}_{\text{droid}}$. When the $\text{Roadcast}_{\text{Java}}$ Client receives the message, it reads the included Disengagement Lateral and Longitudinal Sequence Numbers and sends a signal to the UI thread using the "*Handler*" to display the Disengagement Survey.

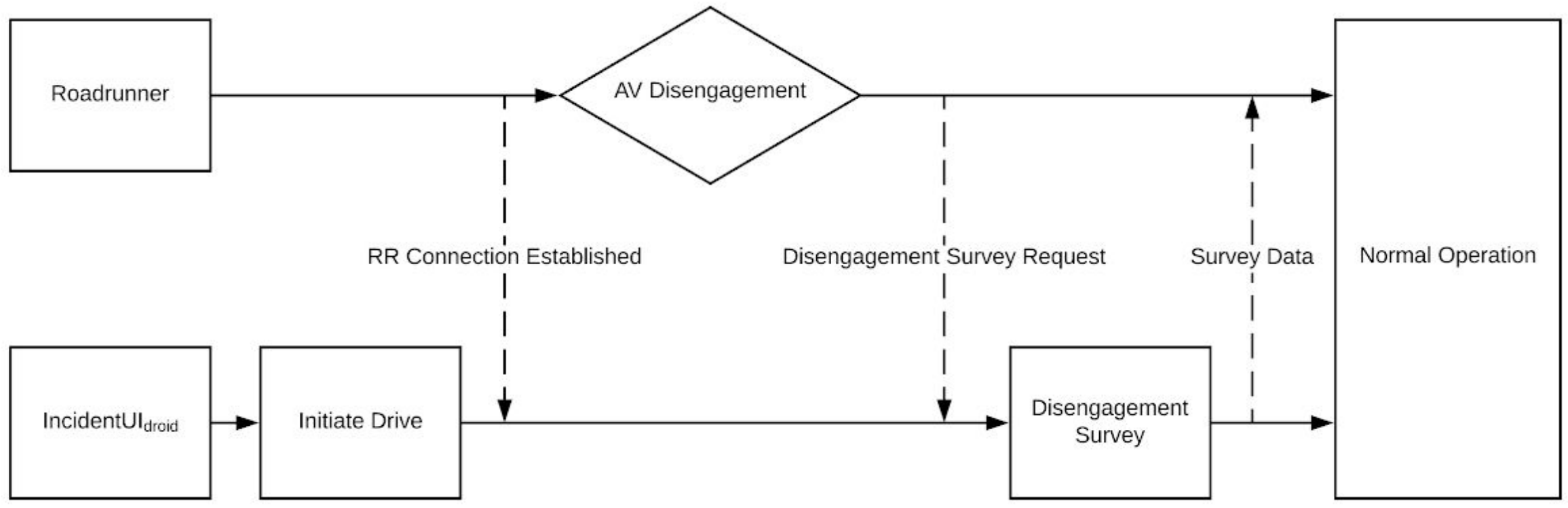

*Figure 40: Disengagement Survey Flow*

After the Disengagement Survey is filled out, a signal is sent back to the $\text{Roadcast}_{\text{Java}}$ Server using the "*Handler*". The $\text{Roadcast}_{\text{Java}}$ Server packages the Disengagement Survey data into a Disengagement Survey Response message and sends it to the Roadrunner Client, which, on receiving the message, deserializes it into a C++ structure. The Lateral and Longitudinal Sequence Numbers packaged with the Disengagement Survey Response message match the ones received with the Disengagement Survey Request message to ensure that the disengagement data is sequentially organized and can be related to the corresponding Survey Request message. After the message is received, a full cycle of operation is complete and normal operation resumes. The data and logic workflow for the Disengagement Survey is represented in Figure 40.

### Event Survey Flow

During an active session, if the user triggers an event by clicking on the "New Event" button, the Event Survey is displayed on the screen, and a signal is sent to the $\text{Roadcast}_{\text{Java}}$ Server via the "*Handler*". The $\text{Roadcast}_{\text{Java}}$ Server, on receiving this signal, creates an Event Flag with an Event Sequence Number attached to it and sends the Event Flag to the Roadrunner Client. After the Event Survey is filled out, another signal is to the $\text{Roadcast}_{\text{Java}}$ Server using the "*Handler*". The $\text{Roadcast}_{\text{Java}}$ Server packages the Event Survey data into an Event Survey Response message and sends it to the Roadrunner Client, which, on receiving the message, deserializes it into a C++ structure. The Event Sequence Number packaged with the Event Survey Response message matches the one sent with the Event Flag to ensure that the event data is sequentially organized and can be related to the corresponding Event Flag that records the time of the Event. After the message is received, a full cycle of operation is complete and normal operation resumes. The data and logic workflow for the Event Survey is represented in Figure 41.

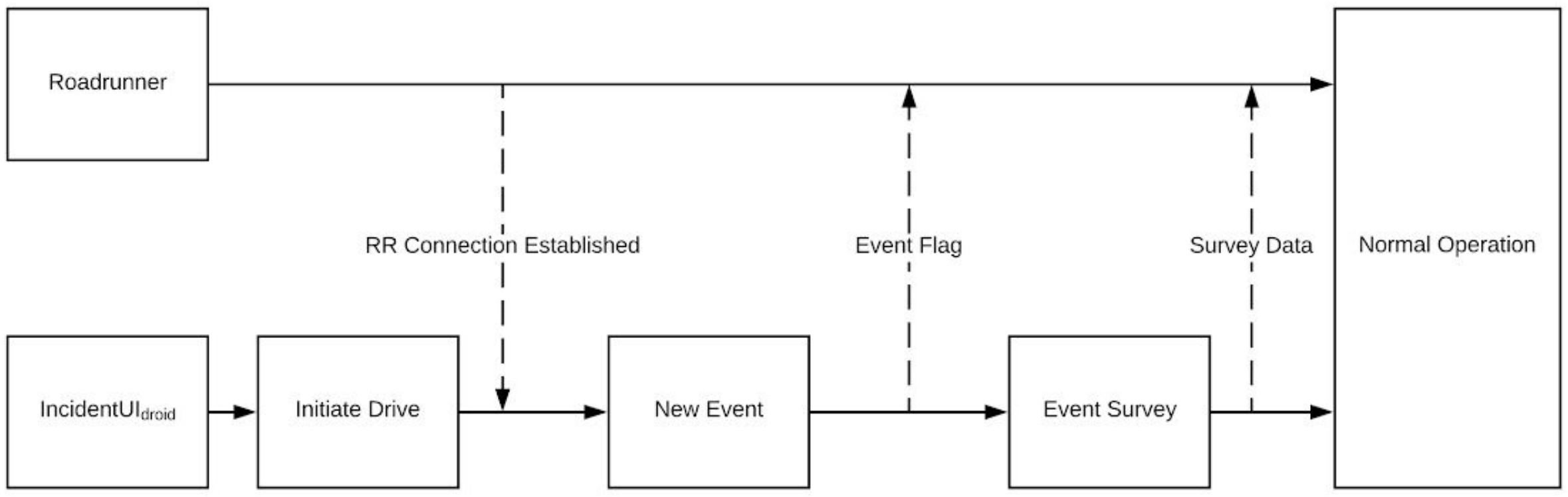

*Figure 41: Event Survey Flow*

Throughout the span of an active session, heartbeats are exchanged between Roadrunner and $\text{IncidentUI}_{\text{droid}}$ to indicate an active roadrunner connection status. When the Roadrunner connection terminates, the test session is concluded.

# 5. Conclusion

For the final deliverable, we submitted a fully functional IncidentUI$_{droid}$ prototype set up on an Nvidia Shield Tablet, which, in addition to replicating and enhancing the functionality of the existing IncidentUI, demonstrated an intuitive user interface and further practical features. Over the span of the development process, we transplanted the IncidentUI from an in-console Raspberry Pi Auxiliary Display to an Android Tablet Platform by developing a new Android front end, redesigning the system architecture and integration scheme, and testing the application functionality iteratively and comprehensively.

The main sticking point we tackled and resolved during the span of this project was the lack of portability, user interactivity, and cross-platform compatibility of the existing IncidentUI application. The development of IncidentUI$_{droid}$ in an Android environment solves every aspect of the aforementioned sticking point. Android application development through Android Studio provides the tools and features needed to develop an intuitive, lightweight, and practical User Interface coded in XML. The use of an Android tablet to deploy IncidentUI$_{droid}$ enhances the portability of the application and provides support for ride comfort feedback from multiple users. The enhanced cross-platform compatibility of the Android environment and the hardware and software capabilities of an Android tablet provide support for pairing other Android devices to IncidentUI$_{droid}$ to give additional context and feedback and consequently, improve the quality of the ride comfort and disengagement data compiled during a test drive. The deployment of IncidentUI$_{droid}$ in an Android environment using Android Studio and the redevelopment of the system sub-architecture in pure Java provide a plethora of system and support libraries that support a flexible and robust feature development process. IncidentUI$_{droid}$ also supports prompt modifications to the system sub-architecture, user interface, and application functionality.

The user interface for IncidentUI$_{droid}$ went through multiple iterations of design and development with each iteration built upon the previous one. The feedback received from IncidentUI$_{droid}$ demonstrations and testing was utilized to develop new UI elements and enhance the existing ones. IncidentUI$_{droid}$'s UI is designed to enhance the ride comfort data collection experience by implementing an intuitive user interface that complements and enhances IncidentUI$_{droid}$'s features. IncidentUI$_{droid}$ runs on a peripheral Android tablet that is integrated with the in-car Hyperion architecture using a hardware network solution like USB Tethering, Ethernet LAN, or WiFi. Since the Android tablet running IncidentUI$_{droid}$ is not fixed to the central console, it can be used to any screen orientation and passed around to other passengers in the ego vehicle to collect ride comfort feedback. This additional feedback improves the quality of the ride comfort data as it utilizes multifaceted responses from people with varying ride experiences and discomfort perceptions occupying different vehicle seating positions to generate a more comprehensive and well-rounded ride comfort rating.

The enhanced cross-platform compatibility of the Android environment and the hardware and software capabilities of an Android tablet provide support for pairing other Android devices to $\text{IncidentUI}_{droid}$ to give additional context and feedback and consequently, improve the quality of the ride comfort and disengagement data compiled during a test drive. Since the tablet is not fixed to the console and can be conveniently decoupled from the in-car Hyperion system architecture, the users have the option to readily swap the tablet in and out of the AV.

In a nutshell, the transition of IncidentUI from a fixed Raspberry Pi display to a portable Android tablet enhances the user experience for the disengagement and ride comfort data collection process. $\text{IncidentUI}_{droid}$ enhances the portability, intuitive user interactivity, and cross-platform compatibility of the existing IncidentUI application. The deployment of $\text{IncidentUI}_{droid}$ in an Android environment and the development of a purely java-based system sub-architecture facilitates the maintenance and modification of the user interface, application features, and system architecture of $\text{IncidentUI}_{droid}$.

# 6. Future Work Recommendations

With the deployment of the stable release for $IncidentUI_{droid}$, we laid the groundwork for developing an Android-based AV disengagement and ride comfort evaluation application, but owing to the 8 week-long duration of the project, we couldn't implement every possible feature and so, there is still a lot of room for improvement. The Android-based $IncidentUI_{droid}$ provides a versatile environment for future improvement and extension. In this section, we mention some ideas for future development and new features that would enhance $IncidentUI_{droid}$.

## Multiple Devices Connection

The portability of the tablet deploying $IncidentUI_{droid}$ allows it to be passed around to other passengers in the ego vehicle to collect ride comfort feedback. This additional feedback improves the quality of the ride comfort data. Another way to improve the quality of the ride comfort data and gather ride comfort feedback from multiple passengers is to enable multiple devices to connect to the $IncidentUI_{droid}$ running on the Co-pilot's tablet. The enhanced cross-platform compatibility of the Android environment and the hardware and software capabilities of an Android tablet provide support for pairing other Android devices to $IncidentUI_{droid}$. The Event Comfort Feedback Console is designed to work as the front end for this feature.

The development of this feature, which is supported by the Event Comfort Feedback Console and Android's integration versatility, would involve setting up the co-pilot's $IncidentUI_{droid}$ as the master and other passengers' devices as the slaves. When the master $IncidentUI_{droid}$ generates an event, all the slave $IncidentUI_{droid}$s are notified to utilize the Event Comfort Feedback Console to submit the longitudinal and lateral comfort ratings for the event. These event comfort feedback ratings can be sent directly to Roadrunner with the same Event Sequence number as the Event, which would require unidirectional master-to-slave sync. The ratings can also be sent to the master $IncidentUI_{droid}$, where they are compiled into one Event Survey Response message and sent to Roadrunner; this design would require a bidirectional master and slave sync. The development of this feature improves the quality of the ride comfort data as it utilizes multifaceted responses from people with varying ride experiences and discomfort perceptions occupying different vehicle seating positions to generate a more comprehensive and well-rounded ride comfort rating for the event.

## App Store or Cloud Support

Currently, the Android Application Package (apk) for $\text{IncidentUI}_{\text{droid}}$ is modified and installed directly from Android Studio. Publishing $\text{IncidentUI}_{\text{droid}}$ on an app store or cloud would allow multiple devices to run $\text{IncidentUI}_{\text{droid}}$ and make the application updates more readily available to all devices running $\text{IncidentUI}_{\text{droid}}$. Instead of requiring access to the development machines to update the version of $\text{IncidentUI}_{\text{droid}}$, all of the devices running $\text{IncidentUI}_{\text{droid}}$ could be simultaneously updated by installing the latest version of $\text{IncidentUI}_{\text{droid}}$ from the cloud. The development of this feature would improve the accessibility of $\text{IncidentUI}_{\text{droid}}$ and also facilitate its development. The increased availability of $\text{IncidentUI}_{\text{droid}}$ would also enhance the ride comfort data collection experience and improve the quality of the ride comfort data.

## Dedicated Web Application and Remote Database

The AV team uses an online spreadsheet to track which employees are using which AV during a test drive. This spreadsheet is filled out manually for every test drive. Developing a web application that replicates the functionality of the spreadsheet and offers some additional $\text{IncidentUI}_{\text{droid}}$ sync features would enhance the user experience for $\text{IncidentUI}_{\text{droid}}$. A suggested feature involves the Pilot and Co-pilot Login information on the $\text{IncidentUI}_{\text{droid}}$ being automatically filled out once it is pulled from the web application, which stores the Pilot and Co-pilot information prior to the test drive. Attaching a Remote Real-Time Database to the web application would also enable unified data management and analysis. This feature involves configuring $\text{IncidentUI}_{\text{droid}}$ to push all the ride data to the remote database after every test drive. This ride data is accessible from the attached web application. Storing the data on a remote database in addition to the Pegasus SSD creates a copy of the data for comparison in case the SSD encounters data loss. The web application can also be utilized to run scripts to analyze the data stored in the remote database. The development of this feature would enhance the user experience for $\text{IncidentUI}_{\text{droid}}$ and also improve the accessibility and quality of the ride comfort data.

# 7. Discussions

After eight weeks of design, development, and evaluation, we delivered the final stable release for $\text{IncidentUI}_{\text{droid}}$ that yielded the desired functionality while testing on an AV test drive and received positive feedback from Nvidia's AV Developers. $\text{IncidentUI}_{\text{droid}}$ enhances the portability, intuitive user interactivity, cross-platform compatibility, and feature development flexibility of the existing IncidentUI application. Our contribution towards laying the groundwork for developing an intuitive and robust Android-based AV disengagement and ride comfort evaluation application paved the way for Nvidia's AV Developers to further modify, scale, and enhance the IncidentUI application.

Over the duration of this project, we successfully tackled our goal to develop an Android application that runs on a peripheral tablet and communicates with the Pegasus to detect AV disengagements and report ride comfort. We designed and developed an Android XML-based intuitive user interface for $\text{IncidentUI}_{\text{droid}}$. We also redesigned the system architecture by implementing the system communications protocol: Roadcast, and the data serialization scheme, i.e. Protobufs, in Java. In addition to contributing to Nvidia's AV infrastructure, we also improved our technical skills and gained invaluable industry experience working for Nvidia. Working on a team software development project utilizing the agile methodology taught us the importance of collaborative development and introduced us to the industry guidelines and standards for project planning and code design. From analyzing the existing IncidentUI and system architecture design to designing, developing, and implementing the Android-based $\text{IncidentUI}_{\text{droid}}$ and the Java-based system sub-architecture, to ultimately evaluating and modifying various aspects of $\text{IncidentUI}_{\text{droid}}$ to develop a final stable release, every step of our Major Qualifying Project equipped us with a more comprehensive understanding of the software development process.

Facing obstacles during the development process, like failure to integrate $\text{IncidentUI}_{\text{droid}}$ with the system architecture using an NDK interface, was also valuable and taught us how to adapt and improvise to devise a new solution and steer development in a viable direction. The collaborative development environment we established during the project was based around not only dividing up tasks and features but also around sharing resources and technical knowledge to improve work efficiency. This project experience was very informative and allowed us to grow personally and professionally. Working with Nvidia prepared us for the technology industry and served as a perfect conclusion to our undergraduate studies in Computer Science. We would again like to extend our gratitude to everyone who assisted us in the development of our project.